\documentclass[%
superscriptaddress,
preprint,
nolongbibliography,
 amsmath,amssymb,
 aps,
 prd
]{revtex4-2}

\usepackage{graphicx}
\usepackage{dcolumn}
\usepackage{bm}
\usepackage[hidelinks]{hyperref}

\usepackage{braket}
\usepackage{upgreek}
\usepackage{bbm}
\usepackage{xcolor}
\newcommand{\rev}[1]{{#1}}

\newcommand{\eq}[1]{\begin{equation}
\begin{aligned}
#1
\end{aligned}
\end{equation}}

\begin{document}

\preprint{}

\title{Collective response and noise of a levitated ferromagnet lattice for ultralight dark matter detection}

\author{Dongyi Yang}
\affiliation{School of Physics, Peking University, Beijing 100871, China}

\author{Xiao Yang}
\affiliation{Beijing Key Laboratory of Quantum Sensing and Precision Measurement, and Center for Quantum Information Technology, and Institute of Quantum Electronics, and School of Electronics, Peking University, Beijing 100871, China}

\author{Chenxi Sun}
\email{scx@pku.edu.cn}
\affiliation{Beijing Key Laboratory of Quantum Sensing and Precision Measurement, and Center for Quantum Information Technology, and Institute of Quantum Electronics, and School of Electronics, Peking University, Beijing 100871, China}

\author{Jianwei Zhang}
\email{james@pku.edu.cn}
\affiliation{School of Physics, Peking University, Beijing 100871, China}




\date{\today}

\begin{abstract}

Ultralight dark matter can induce weak oscillating magnetic-like signals and can therefore be searched for with precision magnetometry. Levitated ferromagnets provide a sensitive platform for such searches, but a single ferromagnet is limited in total polarized spin and readout performance. \rev{We investigate a levitated ferromagnet lattice as a scalable detector architecture for ultralight dark matter.} We develop a theoretical description of the collective lattice response in the fully trapped regime, incorporating dipole-dipole interactions, finite-size effects, and boundary-induced mode mixing. We further analyze the collective noise budget and show that interaction effects mainly produce a narrow blind zone through thermal-noise amplification, while away from this region, the lattice preserves favorable collective noise scaling. \rev{We then derive projected sensitivities to axion-electron, dark-photon, and axion-photon couplings for a reference theoretical benchmark. Under the benchmark assumptions, the lattice improves the projected reach in all three channels relative to a single-ferromagnet detector. The axion-photon channel receives an additional coherent enhancement at the level of an order-of-magnitude estimate, whose coefficient depends on the electromagnetic geometry of the lattice and shield.}

\end{abstract}

\maketitle


\section{Introduction}

Ultralight dark matter (ULDM) candidates, including axions, axion-like particles, and dark photons, can be described as classical, coherently oscillating fields on laboratory scales~\cite{preskill_cosmology_1983,abbott_cosmological_1983,dine_not-so-harmless_1983}. Their oscillation frequencies are set by their masses, typically spanning a wide range from sub-Hz to kHz and beyond. In this regime, ULDM interacts with standard-model particles through effective couplings that can be interpreted as oscillating, spatially coherent forces or fields, providing a promising target for precision measurement experiments~\cite{graham_experimental_2015,arias_wispy_2012,stadnik_searching_2015}.

A number of experimental approaches have been proposed and implemented to probe such signals, including nuclear magnetic resonance, atomic magnetometry, torsion balances, and resonant electromagnetic detectors~\cite{budkerProposalCosmicAxion2014,graham_new_2013,jackson_kimball_search_2023}. Among these, levitated ferromagnets have recently emerged as a sensitive platform for detecting spin-dependent interactions of ULDM~\cite{kaliaUltralightDarkMatter2024}. In these systems, the intrinsic spin of a ferromagnet couples to external magnetic-like fields induced by dark matter, and the resulting mechanical motion can be read out with high precision. The sensitivity of such detectors is determined by the interplay between the intrinsic susceptibility of the ferromagnet and the noise properties of the readout system. 

A natural direction for improving the sensitivity of such detectors is to extend from a single ferromagnet to an array of many identical elements. The basic idea and its leading-order sensitivity gain were introduced in our recent letter \cite{yang_ultralight_2026}. In this work, we investigate a ferromagnet lattice as a platform for ULDM detection. The key idea is to perform a collective readout of a large number of ferromagnets, thereby enhancing the effective signal-to-noise ratio while preserving the intrinsic dynamics of each individual element. Unlike simply increasing the size of a single magnet, this approach allows one to scale up the total polarized spin without significantly altering the single-particle response, providing a route to improved sensitivity that is compatible with existing levitation and readout techniques.

The transition from a single particle to a lattice, however, introduces several new physical ingredients that must be treated carefully. First, dipole-dipole interactions between ferromagnets modify the dynamical response of the system, leading to a momentum-dependent susceptibility and the emergence of collective modes. Second, the finite size of the lattice breaks translational invariance, resulting in boundary-induced mode mixing that couples the coherent ($k=0$) mode to other modes. Third, the noise properties of the detector are altered in a nontrivial way: different noise sources scale differently with the number of particles, and interaction-induced mode mixing can significantly modify the effective noise spectrum. A consistent treatment of these effects is necessary in order to reliably assess the performance of the lattice detector.

In this paper, we develop a theoretical framework that incorporates these ingredients in a unified manner. We first describe the single-particle dynamics in terms of a susceptibility matrix in the fully trapped regime. We then generalize this description to a periodic lattice, introducing the interaction kernel in momentum space and identifying the coherent mode that couples to spatially uniform signals. The effects of finite size are captured by a boundary-induced self-energy, which modifies the response of the coherent channel and gives rise to characteristic features such as near-singular modes and blind zones. Building on this dynamical description, we analyze the noise properties of the lattice, including thermal, imprecision, and backaction contributions, and determine how they combine to set the final sensitivity.

\rev{Using this framework, we evaluate the projected sensitivities of the ferromagnet lattice for three representative ULDM couplings: axion-electron, dark-photon, and axion-photon interactions, with sensitivity curves based on a reference theoretical configuration. Under these benchmark assumptions, the lattice configuration improves the projected reach in all three cases compared to a single-ferromagnet detector, primarily through the collective reduction of the effective noise floor. In the axion-photon channel, the signal may receive an additional coherent enhancement associated with the electromagnetic response of the lattice. Its order-$N$ scaling follows from the long-wavelength overlap argument developed below, whereas its numerical coefficient is geometry dependent. Overall, the lattice provides a systematic and scalable detector architecture for enhancing sensitivity across multiple ULDM scenarios.}

The remainder of this paper is organized as follows. In Sec.~II, we review the dynamics of a single levitated ferromagnet and introduce the susceptibility formalism. In Sec.~III, we describe the lattice configuration and relevant experimental considerations. In Sec.~IV, we develop the theoretical framework for the lattice response, including interaction effects and boundary-induced mode mixing. In Sec.~V, we analyze the noise properties of the lattice detector. In Sec.~VI, we apply the framework to ULDM detection and present projected sensitivities. Finally, we summarize our results and discuss future directions in Sec.~VII.

\section{Single levitated ferromagnet}

We consider a single levitated ferromagnetic particle with a large, macroscopically polarized spin. The internal magnetization is assumed to be rigidly locked to the body of the particle, such that the spin degrees of freedom can be described by a single unit vector
\begin{equation}
\hat{\mathbf n} = \frac{\mathbf S}{|\mathbf S|},
\end{equation}
where $\mathbf S$ denotes the total spin angular momentum of the ferromagnet. The magnetic moment of the ferromagnet is given by
\eq{
    \boldsymbol{\mu} = - \gamma_e \mathbf S = \mu \hat{\mathbf n},
}
where $\gamma_e$ is the electron gyromagnetic ratio, and $\mu = -\gamma_\mathrm e |\mathbf S|$. In addition to its intrinsic spin, the particle possesses mechanical angular momentum $\mathbf L$, associated with rigid-body rotation. The total angular momentum is therefore
\begin{equation}
\mathbf J = \mathbf L + \mathbf S .
\end{equation}

The ferromagnet is assumed to be levitated in a trapping potential that fixes its center-of-mass position. Throughout this work, we focus on the rotational dynamics and neglect translational degrees of freedom. The orientation of the ferromagnet is fully specified by the direction of $\hat{\mathbf n}$, which we parametrize using spherical angles $(\theta,\phi)$ with respect to a fixed laboratory frame. The particle is subject to an orientation-dependent potential $V(\hat{\mathbf n})$, which denotes the total potential energy relevant to the rotational dynamics. It may include contributions from the trapping potential, residual magnetic fields, gravity-induced torques, or other externally engineered fields. This potential defines one or more equilibrium orientations of the magnetization. In addition, weak time-dependent magnetic fields may act on the spin degree of freedom and serve either as probe signals or as external drives. Dissipation is assumed to be negligible on the timescales of interest, consistent with the high mechanical quality factors achievable in levitated systems.

The equations of motion of the ferromagnet can be written as
\begin{equation}
\partial_t \mathbf J = - \hat{\mathbf n} \times \nabla_{\hat{\mathbf n}} V ,
\qquad
\partial_t \hat{\mathbf n} = \boldsymbol{\Omega} \times \hat{\mathbf n},
\end{equation}
where $\boldsymbol{\Omega}$ is the angular velocity of the rigid body and the mechanical angular momentum satisfies $\mathbf L = I \boldsymbol{\Omega}$, with $I$ the moment of inertia of the ferromagnet. In the following, we assume that the ferromagnets are spherically symmetric, so that $I$ is a scalar. In that case, one can show that
\begin{equation}
j_n \equiv \frac{\mathbf J\cdot \hat{\mathbf n}}{S}
\end{equation}
is a constant of motion. Physically, $j_n$ characterizes the projection of the total angular momentum onto the magnetization direction. In the generic situations considered in this work, it is an $\mathcal O(1)$ dynamical invariant and is typically taken to be $j_n=1$. For the equilibrium configuration around which we linearize, we separate the weak time-dependent probe field from the static orientational potential. The equilibrium orientation $\hat{\mathbf n}_0$ is therefore determined by the static trapping part of the potential,
\begin{equation}
\nabla_{\hat{\mathbf n}} V_\mathrm{trap} \big|_{\hat{\mathbf n}_0} = 0 .
\end{equation}
We study the response of the system to weak perturbations about this equilibrium. Writing $\theta(t)=\theta_0+\delta\theta(t)$ and $\phi(t)=\phi_0+\delta\phi(t)$, and assuming small angular deviations, the equations of motion can be linearized in $\delta\theta$ and $\delta\phi$.

To leading order, the dynamics are fully captured by a set of coupled linear equations for the angular fluctuations, reading \cite{kaliaUltralightDarkMatter2024}
\eq{
& \left[\partial_t^2\left(\begin{array}{cc}
1 & 0 \\
0 & \sin ^2 \theta_0
\end{array}\right)+j_n \omega_I \sin \theta_0 \partial_t\left(\begin{array}{cc}
0 & 1 \\
-1 & 0
\end{array}\right)
+\omega_I\left(\begin{array}{cc}
v_{\theta \theta} & v_{\theta \phi} \\
v_{\phi \theta} & v_{\phi \phi}
\end{array}\right)\right]\binom{\delta \theta}{\delta \phi}=0,
\label{eq:EOM}
}
where
\begin{equation}
\omega_{\mathrm I}=\frac{|\mu|}{\gamma_{\mathrm e} I}
\end{equation}
is the Einstein-de Haas frequency, and
\begin{equation}
v_{ij} =
\frac{\gamma_{\mathrm e}}{|\mu|}
\frac{\partial^2 V_\mathrm{trap}}{\partial i\,\partial j}
\bigg|_{\hat{\mathbf n}_0},
\qquad i,j\in\{\theta,\phi\},
\end{equation}
encode the local curvature of the trapping potential at equilibrium. 
External time-dependent magnetic fields enter as driving terms through their coupling to the magnetic moment. In the frequency domain, the linearized equations define a susceptibility matrix that relates the angular response of the ferromagnet to an applied oscillatory magnetic field.

Rather than reproducing the full derivation, which has been presented in detail in ref.~\cite{kaliaUltralightDarkMatter2024}, we summarize the resulting linear response in terms of the susceptibility matrix. For a weak oscillating magnetic field $\mathbf B(t)=\mathbf B_0 e^{-i\omega t}$, the angular response of the ferromagnet takes the form
\begin{equation}
\begin{pmatrix}
\delta\theta(\omega) \\
\delta\phi(\omega)
\end{pmatrix}
=
-\mu B_0
\chi(\omega)
\begin{pmatrix}
b_\theta \\
b_\phi
\end{pmatrix},
\label{eq:response}
\end{equation}
where $(b_\theta,b_\phi)$ denote the components of the magnetic field in the local angular basis. The inverse susceptibility matrix is given by
\eq{
\chi(\omega)^{-1}= I\left[-\omega^2\left(\begin{array}{ll}
1 & 0 \\
0 & 1
\end{array}\right)-i j_n \omega_I \omega\left(\begin{array}{cc}
0 & 1 \\
-1 & 0
\end{array}\right)+\omega_I\left(\begin{array}{cc}
v_{\theta \theta} & 0 \\
0 & v_{\phi \phi}
\end{array}\right)\right],
\label{eq:chi}
}
where we choose local coordinates such that the equilibrium orientation is represented by $\theta_0=\pi/2$ and $\phi_0=0$. We further diagonalize the curvature matrix at the equilibrium point, so that
\begin{equation}
v_{ij} = \mathrm{diag}(v_{\theta\theta},\, v_{\phi\phi}),
\qquad
v_{\theta\theta} > 0,\; v_{\phi\phi} > 0 .
\end{equation}
Equation~(\ref{eq:chi}) is the general linearized susceptibility matrix near the equilibrium orientation. It contains both oscillatory and precessional responses: the diagonal terms encode trap-induced libration, while the antisymmetric term proportional to $j_n\omega_{\mathrm I}\omega$ describes precessional mixing associated with angular-momentum conservation. Different limits of Eq.~(\ref{eq:chi}) correspond to distinct dynamical regimes, including fully trapped, partially trapped, and gyroscopic motion. In the present work, however, our main focus is the fully trapped regime, in which the orientational dynamics are dominated by the restoring torques of the trap.

In the fully trapped regime, the trapping potential provides restoring torques in both angular directions transverse to the equilibrium orientation. This is the primary operating regime considered in this work. Although the general susceptibility in Eq.~(\ref{eq:chi}) contains a precessional mixing term proportional to $j_n\omega_{\mathrm I}\omega$, the response in the parameter range of interest is dominated by trap-induced libration. Accordingly, in the following analysis we adopt the oscillation-dominated approximation and use the simplified inverse susceptibility
\begin{equation}
\chi^{-1}(\omega)
=
I
\begin{pmatrix}
-\omega^2 + \omega_{\mathrm I} v_{\theta\theta} & 0 \\
0 & -\omega^2 + \omega_{\mathrm I} v_{\phi\phi}
\end{pmatrix}.
\end{equation}
The two eigenfrequencies are therefore
\begin{equation}
\omega_{\theta,\phi}=\sqrt{\omega_{\mathrm I}v_{\theta\theta,\phi\phi}}.
\end{equation}
These two modes correspond to small-angle librations about the equilibrium orientation. For frequencies well below resonance the response is quasi-static and set mainly by the trap stiffness. Very close to resonance, the susceptibility is enhanced and the neglected precessional mixing can modify the detailed lineshape. At frequencies above the larger libration frequency, the response is rapidly suppressed, so that the largest resonance frequency also roughly defines the upper frequency limit of the efficient frequency band. The system can be optimized for maximal sensitivity in a narrow frequency range around a target resonance. In the present work, however, our primary interest is broadband sensitivity away from resonance. This fully trapped, oscillation-dominated regime provides the controlled starting point for the lattice analysis in the following sections.

In realistic experimental implementations, the trapping potential of a levitated ferromagnet generally includes contributions from residual magnetic fields inside the magnetic shielding. Even in multi-layer $\mu$-metal shields, leakage fields at the level of $B_\mathrm{ leak} \sim 10^{-9}$--$10^{-8}\ \mathrm{T}$ are common. Such residual fields generate an orientational potential of the form
\begin{equation}
V(\hat{\mathbf n}) \simeq - \boldsymbol{\mu} \cdot \mathbf B_\mathrm{ leak},
\end{equation}
which produces a restoring torque for small angular deviations about the alignment direction. Expanding this potential to quadratic order around the alignment direction gives a trap curvature of order $|\mu|B_\mathrm{ leak}$, which implies the parametric estimate
\begin{equation}
\omega_\mathrm{ trap}^2 \sim \frac{|\mu| B_\mathrm{ leak}}{I}.
\end{equation}
For typical parameters of levitated micron-scale ferromagnets, even nanotesla-level leakage fields yield trapping frequencies well above the characteristic rotational frequency $\omega_I$. As a result, complete rotational freedom is generally difficult to realize in present-day experiments, and the fully trapped regime is typically the most natural and experimentally relevant dynamical configuration.

\rev{The trapping curvature adopted in the benchmark below should not, however, be identified only with this residual leakage field. It represents an engineered orientational confinement that is assumed to be approximately uniform throughout the lattice. For example, a controlled bias field aligned with the equilibrium magnetization gives
\begin{equation}
V_{\theta\theta}\simeq V_{\phi\phi}\simeq |\mu|B_{\mathrm{bias}}.
\label{eq:bias_curvature}
\end{equation}
With $V_{\theta\theta}=V_{\phi\phi}=10^{-13}\ \mathrm{J}$ and $|\mu|=2.3\times10^{-8}\ \mathrm{A\,m^2}$, the benchmark corresponds to
\begin{equation}
B_{\mathrm{bias}}\simeq 4.3\ \upmu\mathrm{T}.
\label{eq:benchmark_bias}
\end{equation}
This field is parametrically distinct from the nanotesla-level residual leakage discussed above. A controlled weak bias field is one possible realization; an equivalent nonmagnetic orientational trapping mechanism would lead to the same linear response provided that it supplies the curvature in Eq.~(\ref{eq:bias_curvature}). The uniformity and technical noise associated with this confinement are architecture-dependent requirements of an experimental implementation.}

For completeness, Eq.~(\ref{eq:chi}) also covers partially trapped and gyroscopic limits. In the partially trapped case, only one angular direction has finite restoring curvature, leading to a hybrid response with one librational and one nearly free rotational mode. In the gyroscopic limit, the restoring curvature vanishes in both directions and the dynamics are governed mainly by angular-momentum conservation. Since neither limit is central to the lattice design studied here, we do not discuss them further in detail. The partially trapped regime may nevertheless arise in situations where the residual field is effectively uniaxial and provides curvature only along one angular direction, for example in the presence of approximate axial symmetry. The gyroscopic regime requires that the orientational curvature be suppressed in both angular directions, which could occur in exceptionally well-shielded environments, in actively compensated field configurations, or outside the influence of the geomagnetic field in space. While such conditions are challenging and are not the primary focus of current implementations, they are not excluded in principle. The gyroscopic limit is therefore of conceptual interest and may be relevant for alternative measurement strategies that exploit long coherence times and slow precession rather than harmonic resonance.

\section{Ferromagnet Lattice: Geometry and Levitation}

\subsection{Lattice geometry}
We consider a lattice composed of $N$ identical levitated ferromagnets, each of radius $R$, mass density $\rho$, magnetization $M$, and magnetic moment $\boldsymbol{\mu}_i$, where $\hat{\mathbf n}_i$ denotes the orientation of the intrinsic spin of the $i$th ferromagnet. The ferromagnets are positioned at fixed equilibrium locations $\mathbf r_i$, forming a three-dimensional crystal-like array with lattice constant $a_0$. The overall linear size of the lattice is $L_\mathrm{lat} \sim N^{1/3} a_0$, and the lattice is enclosed within a magnetic shield of characteristic size $L_\mathrm{sh}$. Throughout this work we assume the hierarchy
\begin{equation}
R \ll a_0 \ll  L_\mathrm{lat} \ll L_\mathrm{sh} ,
\end{equation}
which ensures that (i) each ferromagnet behaves as a point magnetic dipole at inter-particle distances, (ii) the lattice occupies a region small compared with the shield size, and (iii) cavity modes of the shield vary slowly across the lattice volume. We also assume that each ferromagnet is sufficiently far from the shield, so that the direct influence of the shield on the lattice dynamics can be neglected. The influence of the approximation of this hierarchy will be discussed in section~V.C.

Each ferromagnet retains its own rotational degrees of freedom, described by the angular variables $(\theta_i,\phi_i)$ or equivalently by the unit vector $\hat{\mathbf n}_i$. The moment of inertia of an individual spherical ferromagnet is
\begin{equation}
I = \frac{2}{5} m R^2 ,
\end{equation}
with $m = \frac{4}{3}\pi R^3 \rho$. Importantly, the lattice construction does not alter the intrinsic dynamical parameters $(\mu, I)$ of each constituent particle. The extension from a single ferromagnet to a lattice increases the total polarized spin while preserving the intrinsic single-particle dynamical parameters of each constituent element. To make this separation explicit, it is useful to introduce collective angular variables corresponding to the spatially uniform ($k=0$) mode, together with orthogonal combinations describing nonuniform ($k\neq 0$) modes. ULDM induces an effective magnetic-like field that is effectively uniform across the lattice scale. Consequently, the signal couples dominantly to the collective $k=0$ mode. In contrast, inter-particle magnetic dipole–dipole interactions generate spatially varying internal fields that couple different sites and primarily excite nonuniform modes. This separation between collective and internal degrees of freedom underlies the possibility of coherent signal enhancement in the lattice while controlling many-body effects.

A ferromagnet lattice is not equivalent to a single larger ferromagnet containing the same total number of polarized spins. For a single spherical ferromagnet of radius $R_\mathrm{ big}$, the magnetic moment and moment of inertia scale as
\begin{equation}
\mu_\mathrm{ big} \propto R_\mathrm{ big}^3,
\qquad
I_\mathrm{ big} \propto R_\mathrm{ big}^5 .
\end{equation}
Increasing the size therefore leads to a rapid growth of inertia, which suppresses the high-frequency dynamical response and limits the accessible bandwidth. By contrast, for a uniformly polarized configuration, the total magnetic moment magnitude scales as
\begin{equation}
|\mu_\mathrm{ tot}| \sim N |\mu| ,
\end{equation}
while each constituent retains its original moment of inertia $I$. The intrinsic susceptibility of each ferromagnet therefore remains unchanged, and the intrinsic single-particle dynamical bandwidth is not reduced simply by increasing $N$. The lattice enhances the total polarized spin participating in the measurement without introducing the dynamical penalties associated with scaling up a single rigid body.

An alternative architecture would be to allow the ferromagnetic particles to move freely in space, forming a gas-like ensemble rather than a fixed lattice. In principle, such a configuration could be stabilized against magnetic attraction by introducing additional electrostatic repulsion, for example by charging the particles to produce mutual Coulomb forces. While this approach can also prevent clustering, it introduces substantial dynamical complications. In a gas-like configuration, the relative positions of the ferromagnets fluctuate in time, leading to time-dependent dipole–dipole couplings. Collisions or close encounters between particles generate stochastic torques and momentum exchange, producing additional relaxation channels beyond those present in a fixed lattice. These processes result in collisional noise and dephasing, which degrade the coherence of collective motion. Moreover, in the absence of fixed equilibrium positions, the uniform ($k=0$) collective mode is no longer dynamically protected. The center-of-mass and relative coordinates become coupled through interactions and external trapping forces, and the collective angular response mixes with translational and collisional degrees of freedom. As a consequence, the simple separation between collective signal enhancement and internal many-body dynamics, which is central to the lattice configuration, is lost. A fixed lattice architecture suppresses collisional processes, eliminates translational diffusion, and stabilizes the collective degrees of freedom. These features make the lattice configuration significantly more favorable than a gas-like ensemble for maximizing coherent signal accumulation.

\subsection{Levitation and mechanical stability of the lattice}

A central experimental requirement for realizing a ferromagnet lattice is the ability to stably levitate and spatially separate these magnetic particles to prevent mutual attraction that causes clustering. In this section we discuss several possible levitation architectures and their suitability for constructing a stable lattice configuration.

\paragraph*{Acoustic trapping.}

A natural and experimentally mature approach is acoustic levitation using ultrasonic standing-wave fields or phased-array transducers \cite{chen_acoustic_2019,foresti_acoustophoretic_2013,marzo_holographic_2015,sukhanov_three-dimensional_2020}. In such systems, pressure nodes provide restoring forces that confine small particles in three dimensions. By engineering multiple pressure nodes, one can generate a three-dimensional array of stable trapping sites with lattice constant $a_0$. For micron-scale ferromagnets, the acoustic trapping force can readily exceed gravitational forces while remaining tunable independently of magnetic interactions. The restoring force scale can be written parametrically as
\begin{equation}
F_\mathrm{ trap} \sim k_\mathrm{ trap} \, \delta x ,
\end{equation}
where $k_\mathrm{ trap}$ is the effective mechanical stiffness of the acoustic node. The magnetic dipole–dipole force between neighboring ferromagnets scales as
\begin{equation}
F_\mathrm{ dd} \sim \frac{\mu_0 |\mu|^2}{4\pi a_0^4} .
\end{equation}
For experimentally realistic parameters, one can design $k_\mathrm{ trap}$ such that $F_\mathrm{ trap} \gg F_\mathrm{ dd}$ over the relevant displacement range, ensuring mechanical stability of the lattice against dipolar attraction. Acoustic trapping therefore provides a flexible and scalable platform for arranging $N$ ferromagnets over a spatial extent $L_\mathrm{lat}$, while maintaining fixed equilibrium positions and suppressing translational diffusion. This architecture is well suited for room-temperature implementations.

\paragraph*{Superconducting magnetic levitation.}

An alternative levitation strategy employs superconducting magnetic levitation, where Meissner screening currents provide vertical support \cite{kaliaUltralightDarkMatter2024}. While this method can achieve extremely low mechanical dissipation, it strongly constrains the geometry of the system. In particular, superconducting levitation typically stabilizes motion primarily along the vertical direction and restricts the allowable vertical separation between particles and the superconducting surface. The equilibrium levitation height is \cite{kaliaUltralightDarkMatter2024} 
\eq{
z_0=\left(\frac{\mu_0 M^2 R^3}{16 \rho g}\right)^{\frac{1}{4}},
}
which is normally of the same order as $R$, and strongly limits the vertical extent of the lattice. Also, the magnetic field created by the ferromagnets at the surface of the superconductor is limited by the critical magnetic field of the superconductor, which limits the enlargement of the lattice. For a lattice configuration extending over a macroscopic size $L_\mathrm{lat}$, these geometric constraints make it difficult to realize a fully three-dimensional array of well-separated trapping sites. Furthermore, magnetic field gradients associated with superconducting screening may introduce spatial inhomogeneities in the local magnetic environment across the lattice. For these reasons, although superconducting levitation is highly attractive for single-particle experiments, it is less suitable for constructing an extended ferromagnet lattice and is not the preferred architecture in the present context.

\paragraph*{Optical levitation and vacuum operation.}

Optical trapping provides another viable route to constructing a ferromagnet lattice. Optical tweezers or optical standing-wave traps can generate arrays of stable trapping sites with precise spatial control \cite{tseng_search_2025}. In contrast to acoustic trapping, optical levitation is naturally compatible with high-vacuum environments that offer several advantages. The absence of air eliminates collisional damping, reduces thermal noise acting on the rotational degrees of freedom, and provides an ideal mechanical isolation from the environment. Reduced thermal load improves long-term stability and may enable long-term integration. In practice, this approach faces substantial technical challenges. Ferromagnetic materials generally exhibit strong optical absorption and scattering at typical trapping wavelengths, leading to enhanced radiation-pressure imbalance and significant internal heating. These effects compromise trap stability and can degrade magnetic properties, posing a limitation for low-noise operation \cite{millen_optomechanics_2020,romero-isart_large_2011,ranjit_zeptonewton_2016}. Nevertheless, recent developments in optical manipulation of mesoscopic objects suggest that lattice-scale implementations are feasible in principle \cite{millen_optomechanics_2020,frimmer_cooling_2016,chang_cavity_2010}.

\paragraph*{Conventional magnetic levitation.}

Conventional magnetic levitation using static magnetic field gradients may also be considered. Earnshaw’s theorem prohibits stable levitation of a purely ferromagnetic particle in static magnetic fields alone; however, dynamic stabilization techniques or hybrid magnetic–mechanical schemes can provide effective confinement. In practice, magnetic levitation of multiple ferromagnets arranged in a regular lattice would require carefully engineered field configurations to prevent cross-coupling between sites. Moreover, magnetic gradients used for translational confinement may interfere with the controlled magnetic environment required for precision ULDM searches. As a result, while magnetic levitation is feasible for small numbers of particles, scaling to large $N$ with uniform trapping properties across a large size is technically demanding.

\rev{The alternatives above illustrate possible ingredients of a future apparatus, but the numerical benchmark used in this work is not assigned to any one of them. In particular, it should not be interpreted as an already specified cryogenic acoustic, optical, or magnetic lattice. It is a reference theoretical configuration chosen to quantify the collective scaling and the limitations introduced by dipolar interactions and finite-size mode mixing. Realizing all of its assumptions simultaneously would require an architecture-specific analysis of cryogenic compatibility, shielding, trapping uniformity, SQUID integration, and the associated technical noise.}

Regardless of the specific levitation architecture, the essential requirement is that translational restoring forces dominate magnetic dipole–dipole attraction. Provided that
\begin{equation}
k_\mathrm{ trap} a_0 \gg \frac{\mu_0 |\mu|^2}{4\pi a_0^4},
\end{equation}
the equilibrium lattice remains mechanically stable and resistant to collapse. Since the trapping stiffness can be tuned independently of the magnetic properties of the ferromagnets, there exists a broad experimental parameter window in which a stable lattice of $N$ levitated ferromagnets can be realized over a macroscopic scale.

\section{Interaction-induced modification of the lattice response}
We now turn to the modification of the coherent lattice response induced by magnetic dipole-dipole interactions. For a single levitated ferromagnet, the linear response to a weak magnetic field is fully characterized by the susceptibility matrix $\chi(\omega)$ derived in Sec.~II. In a lattice, however, each ferromagnet also experiences the magnetic field generated by all the others. The resulting dipole-dipole interaction couples different sites and turns the problem into a genuine many-body response problem. Even when the interaction is sufficiently weak not to qualitatively alter the order of magnitude of the response, it can still reshape the spectral structure by shifting the collective modes and mixing the uniform response with nonuniform lattice modes. A controlled treatment of these effects is therefore necessary before discussing the final detector noise.

The analysis naturally proceeds in two steps. We first consider an ideal periodic lattice, for which translational invariance allows the dynamics to be diagonalized in reciprocal space. This provides the unperturbed collective-mode structure and shows explicitly that a spatially uniform external field couples only to the $\mathbf k=0$ mode. We then incorporate finite-boundary effects, which break translational invariance and induce mode mixing between the coherent channel and modes with $\mathbf k\neq 0$. This boundary-induced mixing will be the key mechanism responsible for the interaction correction to the observable response.

\subsection{Lattice dynamics in reciprocal space}

We begin from the linear response of an individual ferromagnet in the fully trapped regime. For compactness, we denote by $\delta\mathbf n_i(\omega)$ the two-component angular response of the $i$th ferromagnet in the transverse plane. In the absence of inter-particle coupling, the response is simply
\begin{equation}
\delta\mathbf n_i(\omega)=-\mu\,\chi(\omega)\,\mathbf B_{\mathrm{ext}}(\omega),
\label{eq:lattice_response_single_site}
\end{equation}
where $\mu$ is the magnetic moment and $\mathbf B_{\mathrm{ext}}$ is assumed to be approximately uniform across the lattice.

In the lattice, each ferromagnet is additionally driven by the magnetic field produced by all other ferromagnets. For a pair of sites $i$ and $j$ separated by $\mathbf r_{ij}=\mathbf r_i-\mathbf r_j$, the dipole field can be written in terms of the standard dipolar tensor. Restricting to the two transverse angular components relevant to the linearized dynamics, we write the interaction-induced field on site $i$ as
\begin{equation}
\mathbf B_i^{\mathrm{int}}(\omega)
=
B_{\mathrm{dd}}
\sum_{j\neq i}
T_{ij}\,\delta\mathbf n_j(\omega),
\label{eq:interaction_field_realspace}
\end{equation}
where
\begin{equation}
B_{\mathrm{dd}}
\equiv
\frac{\mu_0|\mu|}{4\pi a_0^3}
\label{eq:Bdd_def}
\end{equation}
sets the characteristic dipolar field scale for lattice spacing $a_0$, and the dimensionless interaction kernel is
\begin{equation}
T_{ij}
=
\frac{a_0^3}{r_{ij}^3}
\begin{pmatrix}
1-3\hat r_{ij,x}^2 & -3\hat r_{ij,x}\hat r_{ij,y}\\[4pt]
-3\hat r_{ij,x}\hat r_{ij,y} & 1-3\hat r_{ij,y}^2
\end{pmatrix}.
\label{eq:dipolar_kernel_realspace}
\end{equation}
Here $\hat r_{ij,a}\equiv r_{ij,a}/r_{ij}$ denotes the $a$th component of the unit vector along the bond, and we set $T_{ii} = 0$ to exclude self-interactions. The equation of motion of the lattice then becomes
\begin{equation}
\delta\mathbf n_i(\omega)
=
-\mu\,\chi(\omega)\left[
\mathbf B_{\mathrm{ext}}(\omega)+\mathbf B_i^{\mathrm{int}}(\omega)
\right].
\label{eq:lattice_eom_realspace_1}
\end{equation}
Substituting Eq.~(\ref{eq:interaction_field_realspace}) into Eq.~(\ref{eq:lattice_eom_realspace_1}) and rearranging gives the coupled linear system
\begin{equation}
\sum_j
\left[
\chi^{-1}(\omega)\delta_{ij}
+
\mu B_{\mathrm{dd}} T_{ij}
\right]
\delta\mathbf n_j(\omega)
=
-\mu\,\mathbf B_{\mathrm{ext}}(\omega),
\label{eq:lattice_master_realspace}
\end{equation}
where the external field is assumed uniform, so the source term is site independent. This equation is exact within the linear-response regime. Compared with the single-ferromagnet problem, the only new ingredient is the dipolar kernel $T_{ij}$, which couples the responses of different lattice sites.

For an ideal periodic lattice, the interaction kernel depends only on the relative displacement between sites, and the problem can therefore be diagonalized in reciprocal space. \rev{We use the Fourier convention}
\rev{\begin{equation}
\delta\mathbf n_i(\omega)
=
\sum_{\mathbf k}
e^{i\mathbf k\cdot\mathbf r_i}\,
\mathbf u_{\mathbf k}(\omega),
\qquad
\mathbf u_{\mathbf k}(\omega)
=
\frac{1}{N}\sum_i
e^{-i\mathbf k\cdot\mathbf r_i}\,
\delta\mathbf n_i(\omega).
\label{eq:uk_fourier_def}
\end{equation}}
\rev{Here $N$ is the total number of ferromagnets and $\mathbf k$ runs over the reciprocal lattice. In particular, a spatially uniform angular displacement satisfies $\mathbf u_0=\delta\mathbf n$.} The Fourier transform of the interaction kernel is defined by
\begin{equation}
T(\mathbf k)
=
\sum_{j\neq i}
e^{-i\mathbf k\cdot(\mathbf r_i-\mathbf r_j)}\,T_{ij},
\label{eq:Tk_def}
\end{equation}
which depends only on the relative displacement and is therefore diagonal in reciprocal space. Equation~(\ref{eq:lattice_master_realspace}) then reduces to
\begin{equation}
\left[
\chi^{-1}(\omega)
+
\mu B_{\mathrm{dd}}\,T(\mathbf k)
\right]
\mathbf u_{\mathbf k}(\omega)
=
-\mu\,\mathbf B_{\mathrm{ext}}(\omega)\,\delta_{\mathbf k,0}.
\label{eq:lattice_master_kspace}
\end{equation}
It is therefore natural to define the interaction-modified susceptibility matrix
\begin{equation}
\chi_{\mathrm{mod}}^{-1}(\omega,\mathbf k)
\equiv
\chi^{-1}(\omega)
+
\mu B_{\mathrm{dd}}\,T(\mathbf k).
\label{eq:chi_mod_def}
\end{equation}
The many-body problem in the ideal periodic lattice is thus reduced to a set of independent $2\times 2$ response problems labeled by $\mathbf k$. The role of the dipole-dipole interaction is to renormalize the susceptibility differently for different lattice modes, thereby shifting and reshaping the collective spectrum.

A crucial consequence of Eq.~(\ref{eq:lattice_master_kspace}) is that a spatially uniform external field couples only to the $\mathbf k=0$ mode. In the ideal periodic limit, all modes with $\mathbf k\neq 0$ remain unexcited under uniform driving, and the observable response is entirely governed by the coherent channel,
\begin{equation}
\mathbf u_0(\omega)
=
-\mu\,\chi_{\mathrm{mod}}(\omega,\mathbf 0)\,
\mathbf B_{\mathrm{ext}}(\omega).
\label{eq:u0_ideal_lattice}
\end{equation}
At this stage, the lattice behaves as a coherent sum of identical ferromagnets with a susceptibility renormalized by the interaction. No mode mixing occurs, and the response is fully governed by the coherent channel. The situation changes once translational invariance is broken.
In a finite lattice, the boundary introduces a nonperiodic correction to the dipolar kernel, so that the reciprocal-space modes are no longer exact eigenmodes of the dynamics. As a result, the coherent $\mathbf k=0$ response is mixed with modes at $\mathbf k\neq 0$, and the observable susceptibility acquires an additional correction beyond the ideal periodic result. We analyze this boundary-induced mode mixing in the next subsection.

\subsection{Boundary-induced mode mixing and effective coherent response}

The simplification obtained in the previous subsection relies entirely on translational invariance. In a finite lattice, this symmetry is broken by the boundary, and the reciprocal-lattice modes are no longer exact eigenmodes of the dynamics. Physically, this means that even under a spatially uniform drive, the coherent $\mathbf k=0$ response can leak into modes with $\mathbf k\neq 0$, which then feed back into the observable response of the lattice. The boundary effect therefore does not merely shift the collective spectrum but mixes the coherent channel with nonuniform lattice modes.

To make this structure explicit, we decompose the full interaction kernel into a translationally invariant part and a boundary correction,
\begin{equation}
T_{\mathrm{tot}} = T + T_{\mathrm{bnd}}.
\label{eq:T_tot_split}
\end{equation}
Here $T$ denotes the part that can still be diagonalized in reciprocal space, while $T_{\mathrm{bnd}}$ contains the nonperiodic correction induced by the finite boundary. The diagonal part of $T_\mathrm{bnd}$ in reciprocal space can be absorbed into the definition of the interaction-modified susceptibility. After this redefinition, $T_{\mathrm{bnd}}$ is understood as the off-diagonal component responsible for mode mixing.

\rev{Since the experimentally relevant signal is the coherent response under a nearly uniform external field, we define the coherent channel directly as the spatial average}
\rev{\begin{equation}
\mathbf u_{\mathrm{coh}}(\omega)
\equiv
\frac{1}{N}\sum_i\delta\mathbf n_i(\omega)
=\mathbf u_0(\omega).
\label{eq:u_coh_formal}
\end{equation}}
\rev{With the Fourier convention in Eq.~(\ref{eq:uk_fourier_def}), the boundary-induced mode-mixing matrix is}
\rev{\begin{equation}
V_{\mathbf k\mathbf q}
\equiv
\frac{1}{N}
\sum_{i,j}
e^{-i\mathbf k\cdot\mathbf r_i}
(T_{\mathrm{bnd}})_{ij}
e^{i\mathbf q\cdot\mathbf r_j}.
\label{eq:Vkq_def}
\end{equation}}
In the absence of $T_{\mathrm{bnd}}$, one has $V_{\mathbf k\mathbf q}=0$ and the modes remain decoupled, recovering the ideal periodic result derived in Sec.~IV.A. The entire effect of the finite boundary is therefore encoded in the off-diagonal matrix $V_{\mathbf k\mathbf q}$. \rev{The mode-space equation of motion is}
\begin{equation}
\chi_{\mathrm{mod}}^{-1}(\omega,\mathbf k)\,
\mathbf u_{\mathbf k}
+
\mu B_{\mathrm{dd}}
\sum_{\mathbf q\neq \mathbf k}
V_{\mathbf k\mathbf q}\,
\mathbf u_{\mathbf q}
=
-\mu\,\mathbf B_{\mathrm{ext}}\,\delta_{\mathbf k,0}.
\label{eq:eom_mode_projected}
\end{equation}
For $\mathbf k\neq 0$, these modes are not directly driven by the uniform external field; they are populated only through boundary-induced coupling to other modes, ultimately originating from the coherent mode. Their response may therefore be viewed as virtual leakage channels that renormalize the $\mathbf k=0$ dynamics.

When the leakage into nonuniform modes remains perturbative, so that the dynamics is still governed primarily by the coherent channel, one may solve Eq.~(\ref{eq:eom_mode_projected}) iteratively. To leading nontrivial order, the $\mathbf k\neq 0$ modes are induced by the coherent mode according to
\begin{equation}
\mathbf u_{\mathbf k}
\simeq
- \mu B_{\mathrm{dd}}\,
\chi_{\mathrm{mod}}(\omega,\mathbf k)\,
V_{\mathbf k 0}\,
\mathbf u_0,
\qquad
\mathbf k\neq 0.
\label{eq:uk_from_u0}
\end{equation}
Substituting this back into the $\mathbf k=0$ equation yields an effective equation of motion for the coherent response,
\begin{equation}
\left[
\chi_{\mathrm{mod}}^{-1}(\omega,\mathbf 0)
-
\Sigma_{\mathrm{bnd}}(\omega)
\right]
\mathbf u_0
=
-\mu\,\mathbf B_{\mathrm{ext}}.
\label{eq:u0_effective_eom}
\end{equation}
Here
\begin{equation}
\Sigma_{\mathrm{bnd}}(\omega)
\equiv
|\mu|^2 B_{\mathrm{dd}}^2
\sum_{\mathbf k\neq 0}
V_{0\mathbf k}\,
\chi_{\mathrm{mod}}(\omega,\mathbf k)\,
V_{\mathbf k 0}
\label{eq:Sigma_bnd_def}
\end{equation}
is the self-energy induced by the boundary correction. 

Equation~(\ref{eq:u0_effective_eom}) is the central result of this subsection. It shows that, after integrating out the nonuniform modes, the observable coherent response retains the same form as in the ideal lattice, but with an additional self-energy correction generated by virtual excursions into $\mathbf k\neq 0$ modes. The boundary therefore plays the role of a mode-mixing perturbation: it does not create an independent collective mode by itself, but modifies the coherent response through coupling to the existing spectrum of nonuniform lattice modes. The corrected coherent susceptibility can thus be written as
\begin{equation}
\chi_{\mathrm{coh}}^{-1}(\omega)
=
\chi_{\mathrm{mod}}^{-1}(\omega,\mathbf 0)
-
\Sigma_{\mathrm{bnd}}(\omega),
\label{eq:chi_coh_def}
\end{equation}
so that
\begin{equation}
\mathbf u_0(\omega)
=
-\mu\,
\chi_{\mathrm{coh}}(\omega)\,
\mathbf B_{\mathrm{ext}}(\omega).
\label{eq:u0_chicoh}
\end{equation}
This form reduces the full finite-lattice problem to an effective response theory for the coherent channel alone. All boundary effects are encoded in the frequency-dependent matrix $\Sigma_{\mathrm{bnd}}(\omega)$.

The physical meaning of Eq.~(\ref{eq:Sigma_bnd_def}) is transparent. The factor $V_{0\mathbf k}$ describes the boundary-induced leakage of the coherent mode into a nonuniform mode $\mathbf k$, while $\chi_{\mathrm{mod}}(\omega,\mathbf k)$ governs the response of that intermediate mode under the interaction-modified lattice dynamics. The factor $V_{\mathbf k0}$ then describes the coupling of this amplitude back into the coherent channel. The boundary correction is therefore controlled both by the strength of the mode mixing and by the susceptibility of the intermediate nonuniform modes. Even when the boundary itself is weak, its effect can be enhanced if some $\mathbf k\neq 0$ modes are close to singular response. This structure will be important for identifying the parameter regime in which the perturbative treatment remains valid.

For later convenience, it is useful to introduce a dimensionless form of the self-energy. Writing
\begin{equation}
\chi_{\mathrm{mod}}^{-1}(\omega,\mathbf k)
=
X_0\,\xi(\omega,\mathbf k),
\label{eq:xi_def_general}
\end{equation}
where $X_0 =  I\omega_I v_{\theta\theta}$ sets the overall susceptibility scale and $\xi(\omega,\mathbf k)$ is dimensionless. The matrix $\xi$ is normalized so that its entries are simply $1$ and $v_{\phi\phi}/v_{\theta\theta}$ in the absence of interaction. One obtains
\begin{equation}
\sigma_{\mathrm{bnd}}(\omega)\equiv \frac1{X_0}\Sigma_\mathrm{bnd}(\omega)
=
\lambda_\mathrm{int}^2
\sum_{\mathbf k\neq 0}
V_{0\mathbf k}\,
\xi^{-1}(\omega,\mathbf k)\,
V_{\mathbf k0},
\label{eq:Sigma_bnd_dimensionless}
\end{equation}
with 
\eq{
\lambda_\mathrm{int} = \frac{R^3}{3 a_0^3} \frac{\omega_s}{v_{\theta\theta}}
}
a dimensionless parameter characterizing the interaction strength. Here 
\eq{
\omega_s = \gamma_e \mu_0 M
}
is a material-dependent frequency scale set by the magnetization $M$ of the ferromagnet. In this form, the boundary correction is naturally separated into three ingredients: the overall interaction scale $\lambda_\mathrm{int}^2$, the geometric mode-mixing matrix elements $V_{0\mathbf k}$, and the intrinsic lattice response $\xi^{-1}(\omega,\mathbf k)$. 

\subsection{Perturbative regime, near-singular modes, and beyond}

The effective coherent susceptibility derived above provides a compact description of the boundary correction. However, this description is valid only when the leakage from the coherent channel into nonuniform modes remains perturbative. We now examine the regime in which this assumption is valid, and identify the physical mechanism that can enhance the boundary effect.

To obtain Eq.~(\ref{eq:uk_from_u0}) from Eq.~(\ref{eq:eom_mode_projected}), one assumes that, on the right-hand side of the formal solution, the dominant source term is the direct boundary-induced mixing with the coherent mode, while the feedback from other nonuniform modes remains subleading. In other words, the nonuniform modes are assumed to be slaved to the coherent response rather than forming an independent strongly mixed sector. Under this approximation,
\begin{equation}
\mathbf u_{\mathbf k}
\simeq
-\mu B_{\mathrm{dd}}\,
\chi_{\mathrm{mod}}(\omega,\mathbf k)\,
V_{\mathbf k0}\,
\mathbf u_0,
\qquad
\mathbf k\neq 0.
\label{eq:uk_from_u0_repeated}
\end{equation}
The validity of this approximation should be checked self-consistently. Substituting the leading-order expression Eq.~(\ref{eq:uk_from_u0_repeated}) back into Eq.~(\ref{eq:eom_mode_projected}), one finds that the contribution generated by mode-to-mode feedback among the nonuniform sector is small only if
\begin{equation}
\left\|
-\mu B_{\mathrm{dd}}
\sum_{\mathbf q\neq 0,\mathbf k}
V_{\mathbf k\mathbf q}\,
\chi_{\mathrm{mod}}(\omega,\mathbf q)\,
V_{\mathbf q0}
\right\|
\ll
\left\|
V_{\mathbf k0}
\right\|,
\qquad
\mathbf k\neq 0.
\label{eq:self_consistency_condition}
\end{equation}
This is the direct self-consistency condition for the coherent-mode-dominance approximation. It states that the secondary leakage path
$
0\rightarrow \mathbf q \rightarrow \mathbf k
$
must remain small compared with the primary leakage path
$
0\rightarrow \mathbf k
$.
Only in this regime can the nonuniform modes be integrated out perturbatively to produce the boundary self-energy of the coherent channel.

This condition has two distinct physical implications. The first is the intuitive one: the interaction should not be too strong overall. If the dipolar interaction scale becomes too large, then the boundary-induced couplings are no longer a small perturbation on top of the unperturbed lattice modes, and the entire nonuniform sector can become strongly hybridized. In that regime, the response is no longer organized around a dominant coherent mode, and the perturbative elimination of $\mathbf k\neq 0$ modes breaks down globally. At a qualitative level, one therefore expects the coherent-mode-dominance approximation to remain viable when $\lambda_\mathrm{int}$ is not too large, typically $\lambda_\mathrm{int}\lesssim 0.1$ up to geometry- and frequency-dependent corrections.

The second implication is more subtle but more important for the present system. Even when the overall interaction strength is moderate, the self-consistency condition can still fail if some nonuniform modes acquire an anomalously large susceptibility. This occurs when the matrix $\chi_{\mathrm{mod}}^{-1}(\omega,\mathbf k)$ develops a small eigenvalue, or equivalently when the dimensionless matrix $\xi(\omega,\mathbf k)$ introduced in Eq.~(\ref{eq:xi_def_general}) becomes nearly singular. In that case, the response of the intermediate mode $\mathbf q$ is strongly enhanced, and the secondary leakage process in Eq.~(\ref{eq:self_consistency_condition}) is no longer suppressed. To make this point explicit, let $\nu_\alpha(\omega,\mathbf k)$ denote the eigenvalues of $\xi(\omega,\mathbf k)$. Then the induced amplitude of a nonuniform mode contains contributions weighted by $1/\nu_\alpha(\omega,\mathbf k)$. Therefore, even if the geometric mixing matrix elements $V_{\mathbf k\mathbf q}$ are small, the corresponding correction can become large when one or more $\nu_\alpha(\omega,\mathbf k)$ approach zero. The breakdown of the approximation is then not driven by a parametrically large boundary term, but by the strong response of a subset of near-singular modes. Physically, the boundary opens a leakage channel from the coherent mode into the nonuniform sector, and the near-singular modes amplify this leakage by providing almost resonant intermediate states.

It is useful to distinguish these two mechanisms conceptually. The first is a global strong-coupling effect: the interaction strength itself is so large that no perturbative treatment around the coherent mode is justified. The second is a local spectral effect: the interaction is still moderate in a bulk sense, but the lattice contains exceptional modes at specific frequencies for which the response is strongly enhanced. In the present problem, the latter is the more relevant mechanism. Over much of the spectrum, the coherent-mode picture remains valid and the interaction correction is smooth. The main source of irregular behavior is instead the presence of near-singular modes, which distort the coherent susceptibility over a restricted frequency range through enhanced boundary-induced mixing.

When the approximation fails, the relevant near-singular modes must be retained explicitly rather than integrated out perturbatively. A convenient way to proceed is to divide the full momentum space into two parts: a subspace $\mathcal S$ containing the near-singular modes and the coherent mode, and the remaining regular modes. The regular modes continue to contribute perturbatively through the same self-energy structure as before, while the modes in $\mathcal S$ are treated on equal footing with the coherent mode. After excluding the modes in $\mathcal S$ from the perturbative treatment, the coherent-mode-dominance condition is replaced by
\begin{equation}
\left\|\sum_{\mathbf q \notin \mathcal S}  V_{\mathbf k\mathbf q}\mathbf u_{\mathbf q} \right\|\ll \left\| V_{\mathbf k 0} \mathbf u_0 \right\|.
\label{eq:near_singular_condition_refined}
\end{equation}
That is, once the near-singular modes have been separated out, the residual modes outside this subset must still remain weakly coupled in the same sense as before. The subset $\mathcal S$ therefore plays a role analogous to a near-degenerate subspace: instead of solving directly for $\rev{\mathbf u_0}$ alone, one solves the coupled equations within this subspace. Nevertheless, the choice of the subspace $\mathcal S$ is not unique. In principle, one may enlarge $\mathcal S$ to improve the accuracy of the treatment, at the cost of solving a correspondingly larger set of coupled linear equations involving the matrix elements $V_{\mathbf k\mathbf q}$.

The amplitudes of modes outside the subset can then be fully expressed in terms of the amplitudes of the modes inside the subset,
\begin{equation}
\mathbf u_{\mathbf k}
=
\xi^{-1}(\omega,\mathbf k)
\sum_{\mathbf s\in\mathcal S}
\lambda_{\mathrm{int}} V_{\mathbf k\mathbf s} \mathbf u_{\mathbf s},
\qquad
\mathbf k\notin\mathcal S.
\label{eq:uk_outside_subset}
\end{equation}
Substituting this relation back into the equations of motion, one obtains
\rev{\begin{equation}
\sum_{\mathbf s'\in\mathcal S}
\left[
\xi(\omega,\mathbf s)\delta_{\mathbf s,\mathbf s'}
-\lambda_{\mathrm{int}} V_{\mathbf s,\mathbf s'}
\right]
\mathbf u_{\mathbf s'}
=
0,
\qquad
\mathbf s\in\mathcal S\setminus\{0\},
\label{eq:subset_eq_nonzero}
\end{equation}}
and
\rev{\begin{equation}
\left[
\xi(\omega,0)
-\lambda_{\mathrm{int}}^2
\sum_{\mathbf k\notin\mathcal S}
V_{0\mathbf k}\xi^{-1}(\omega,\mathbf k)V_{\mathbf k0}
\right]\mathbf u_0
-\lambda_{\mathrm{int}}
\sum_{\mathbf s\in\mathcal S\setminus\{0\}}
\left[
V_{0\mathbf s}
+\lambda_{\mathrm{int}}
\sum_{\mathbf k\notin\mathcal S}
V_{0\mathbf k}\xi^{-1}(\omega,\mathbf k)V_{\mathbf k\mathbf s}
\right]
\mathbf u_{\mathbf s}
=
-\frac{\mu B_{\mathrm{ext}}}{X_0}.
\label{eq:subset_eq_zero_full}
\end{equation}}
The modes outside the subset $\mathcal S$ also generate a self-energy correction for each mode inside $\mathcal S$. However, by construction, all potentially singular modes have already been removed from this residual self-energy. The remaining contribution is therefore expected to remain small compared with the direct coupling term $V_{0\mathbf s}$ and may be neglected to leading order. Treating this as a partitioned matrix equation, the solution for the coherent response becomes
\begin{equation}
\chi'_\mathrm{coh}(\omega) = \frac1{X_0}\left[\xi(\omega,0)
-\sigma_\mathrm{bnd}'(\omega)\right]^{-1},
\label{eq:u0_near_singular}
\end{equation}
where the dimensionless self-energy is
\begin{equation}
\sigma_{\mathrm{bnd}}'(\omega)
=
\lambda_{\mathrm{int}}^2\left[
\sum_{\mathbf s,\mathbf s'\in\mathcal S\setminus \{0\}}
V_{0\mathbf s}
G_{\mathbf s,\mathbf s'}(\omega)
V_{\mathbf s'0} 
+ 
\sum_{\mathbf k\notin\mathcal S}
V_{0\mathbf k}\xi^{-1}(\omega,\mathbf k)V_{\mathbf k 0}
\right],
\label{eq:Sigma_bnd_near_singular}
\end{equation}
and
\rev{\begin{equation}
G_{\mathbf s,\mathbf s'}^{-1}(\omega)
=
\xi(\omega,\mathbf s)\delta_{\mathbf s,\mathbf s'}
-\lambda_{\mathrm{int}} V_{\mathbf s,\mathbf s'},
\qquad
\mathbf s,\mathbf s'\in\mathcal S\setminus\{0\},
\label{eq:G_subset_def}
\end{equation}}
is the response matrix of the near-singular subset excluding the coherent mode. In other words, once the near-singular modes are retained explicitly, the effect of the boundary has the same formal structure as before, except that the simple susceptibility of an isolated intermediate mode is replaced by the full response matrix of the near-singular subset. \rev{This expression reduces to the perturbative result in Sec.~IV.B when $\mathcal S\setminus\{0\}$ is empty or when the couplings among the retained modes may be neglected.}

\subsection{\texorpdfstring{Small-$\mathbf k$ structure of the interaction kernel\ \rev{and static angular stability}}{Small-k structure of the interaction kernel and static angular stability}}

The analysis above shows that the boundary correction is most strongly enhanced when the coherent mode couples to nonuniform modes with anomalously large susceptibility. \rev{The zero-frequency limit of the same mode spectrum also determines whether the aligned configuration is a local minimum of the static angular energy.} To identify where such near-singular modes arise, we now examine the structure of the interaction kernel $T(\mathbf k)$ for a simple cubic lattice. Its reciprocal lattice is again simple cubic, with a minimum wave number
\begin{equation}
k_0 = \frac{2\pi}{L_\mathrm{lat}},
\label{eq:k0_def}
\end{equation}
where $L_\mathrm{lat}$ is the linear size of the lattice. There is no simple closed-form expression for the full lattice interaction kernel in reciprocal space. We focus on its small-$\mathbf k$ behavior, which is the most relevant regime for the boundary correction. Physically, this is because the boundary correction is concentrated near the edge of the lattice in real space, and therefore couples predominantly to long-wavelength modes in reciprocal space.

According to the Ewald summation method, the dipolar interaction of the discrete lattice can be decomposed into a long-range continuum contribution and a short-range discrete correction \cite{de_leeuw_simulation_1980,stamm_coherent_2018}. The long-range part can be Fourier transformed analytically as in continuum space, while the short-range part contributes only regular corrections. In the limit $k a_0\ll 1$, the interaction kernel takes the form
\begin{equation}
T(\mathbf k)
=
4\pi
\begin{pmatrix}
\hat k_x^2-\frac13 & \hat k_x\hat k_y\\[4pt]
\hat k_x\hat k_y & \hat k_y^2-\frac13
\end{pmatrix}
+
\mathcal O\!\left[(k a_0)^2\right],
\label{eq:Tk_smallk}
\end{equation}
where $\hat{\mathbf k}=\mathbf k/k$ is the unit vector along $\mathbf k$. The leading term in Eq.~(\ref{eq:Tk_smallk}) is nonanalytic: it depends only on the direction of $\mathbf k$, but not on its magnitude. This is a direct consequence of the long-range nature of the dipole-dipole interaction. As a result, a family of modes with different $|\mathbf k|$ but nearly the same direction share the same leading interaction energy, and therefore nearly the same self-energy correction. The near-singular modes therefore appear in directional groups rather than as isolated points in momentum space. Consequently, once a near-singular direction exists, a whole set of nearby long-wavelength modes can become strongly enhanced, making the response particularly sensitive to the boundary-induced mode mixing.

Using Eq.~(\ref{eq:Tk_smallk}), one can obtain the eigenvalues of the dimensionless modified susceptibility matrix $\xi$ in the small-$\mathbf k$ approximation,
\begin{equation}
\nu_1
=
1-\bar\omega^2+\frac{4\pi}{3}\lambda_{\mathrm{int}},
\label{eq:lambda1_smallk}
\end{equation}
\begin{equation}
\nu_2
=
1-\bar\omega^2
-
4\pi\lambda_{\mathrm{int}}
\left(
\frac23-\hat k_z^2
\right),
\label{eq:lambda2_smallk}
\end{equation}
where
\begin{equation}
\bar\omega
=
\frac{\omega}{\sqrt{\omega_I v_{\theta\theta}}}
\label{eq:omega_bar_def}
\end{equation}
is the dimensionless frequency, with $\bar\omega=1$ corresponding to the reference single-particle resonance frequency set by $v_{\theta\theta}$. Here we specify the fully trapped regime where the single-particle susceptibility is approximately diagonal. The first eigenvalue $\nu_1$ is independent of $\mathbf k$, whereas the second one $\nu_2$ retains the directional dependence through $\hat k_z$. The first branch therefore produces a uniform interaction shift in the small-$\mathbf k$ limit, while the second branch is anisotropic and can approach zero along particular directions. The anisotropic near-singular structure discussed in the previous subsection is therefore expected to originate primarily from this second branch.

\rev{At $\omega=0$, Eq.~(\ref{eq:chi_mod_def}) is the Fourier representation of the static angular Hessian, normalized in $\xi$ by $X_0=I\omega_Iv_{\theta\theta}$. Since $X_0>0$, the uniformly aligned state is a local minimum only if $\xi(0,\mathbf k)$ is positive definite for every wave vector. Equivalently, defining the minimum over both eigenvalue branches and all wave vectors as
\begin{equation}
\nu_{\min}(0)\equiv\min_{\mathbf k,\,a=1,2}\nu_a(0,\mathbf k),
\label{eq:angular_stability_hessian}
\end{equation}
static angular stability requires $\nu_{\min}(0)>0$. For isotropic trapping curvature, the most restrictive long-wavelength bulk value follows from $\nu_2(0,\hat k_z=0)$,
\begin{equation}
\nu_{\min}(0)\simeq 1-\frac{8\pi}{3}\lambda_{\mathrm{int}}.
\label{eq:bulk_angular_stability}
\end{equation}
The benchmark value $\lambda_{\mathrm{int}}=0.0665$ gives $\nu_{\min}(0)=0.44289>0$ in this approximation. We also construct the full static Hessian for the finite $100^3$ open-boundary lattice. The boundary generates site-dependent first-order torques, so the uniformly aligned configuration is defined as an engineered stationary reference state after local linear compensation of these torques. The compensation changes the first derivative of the energy but not its second derivative and therefore leaves the angular Hessian unchanged. The matrix-free calculation described in Appendix~\ref{app:static_angular_hessian} gives
\begin{equation}
\nu_{\min}^{\mathrm{fin}}=0.44818,
\label{eq:finite_angular_stability}
\end{equation}
corresponding to a minimum energy curvature of $4.48\times10^{-14}\,\mathrm{J}$ for the benchmark trapping scale. This positive finite-lattice minimum lies $1.19\%$ above the bulk estimate and directly verifies static angular stability of the compensated benchmark configuration. The finite-size sequence and numerical validation are given in Appendix~\ref{app:static_angular_hessian}.}

\rev{At finite frequency, the same eigenvalue spectrum determines the onset of the near-singular dynamical response.} In particular, once the frequency is tuned such that $\nu_2$ becomes small for some direction $\hat{\mathbf k}$, all sufficiently small-$k$ modes along nearby directions acquire similarly large \rev{responses} and contribute coherently to the boundary self-energy. As $\hat k_z \in [0,1]$, these two eigenvalues together determine a frequency band
\eq{
\bar\omega \in \left[\sqrt{
\max\left({1-\frac{8\pi}3\lambda_\mathrm{int},0}\right)
}, \sqrt{1+\frac{4\pi}3\lambda_\mathrm{int}}\right],
\label{eq:freq_interval}
}
and near-singular modes can only exist within this frequency band.

\subsection{Boundary correction formalism and numerical results}

The ideal interaction kernel $T(\mathbf k)$ discussed above corresponds to an infinite lattice with periodic boundary conditions. For the numerical treatment of a finite lattice, we adopt an operational decomposition: the diagonal part of the finite-lattice kernel in reciprocal space is identified as $T(\mathbf k)$, while the residual off-diagonal part is treated as the boundary-induced mixing term. This definition coincides with the ideal bulk kernel only in the infinite-lattice limit. For a finite lattice, the translational symmetry is broken by the boundary, and the actual interaction kernel is obtained by restricting the ideal kernel to the occupied region of the lattice. Let the finite lattice occupy a spatial region $D$, and define the occupation function
\begin{equation}
s(\mathbf r)=
\left\{
\begin{array}{ll}
1, & \mathbf r\in D,\\
0, & \mathbf r\notin D.
\end{array}
\right.
\label{eq:occupation_function}
\end{equation}
Correspondingly, we introduce the projection operator
\begin{equation}
S=\sum_{\mathbf r}s(\mathbf r)\ket{\mathbf r}\bra{\mathbf r}.
\label{eq:S_projection}
\end{equation}
The interaction kernel of the finite lattice is then
\begin{equation}
T^{\mathrm{finite}}=STS.
\label{eq:Tfinite_def}
\end{equation}
In reciprocal space, this becomes
\begin{equation}
T^{\mathrm{finite}}_{\mathbf k\mathbf q}
=
\sum_{\mathbf p}
\tilde S(\mathbf p-\mathbf k)\,
T(\mathbf p)\,
\tilde S(\mathbf q-\mathbf p),
\label{eq:Tfinite_kspace}
\end{equation}
where
\begin{equation}
\tilde S(\mathbf q-\mathbf k)
\equiv
\bra{\mathbf k}S\ket{\mathbf q}
=
\frac{1}{N}
\sum_{\mathbf r}
s(\mathbf r)e^{i(\mathbf q-\mathbf k)\cdot \mathbf r}
\label{eq:S_tilde_def}
\end{equation}
is the Fourier transform of the occupation function. The corresponding boundary correction is therefore
\begin{equation}
V_{\mathbf k\mathbf q}
=
\sum_{\mathbf p}
\tilde S(\mathbf p-\mathbf k)\,
T(\mathbf p)\,
\tilde S(\mathbf q-\mathbf p)
-
\delta_{\mathbf k\mathbf q}T(\mathbf k).
\label{eq:Vkq_boundary_finite}
\end{equation}
In an infinite periodic lattice, $\tilde S$ reduces to a Kronecker delta and the kernel remains diagonal in $\mathbf k$ space. In a finite lattice, however, the window function $\tilde S$ broadens the momentum support and couples different reciprocal-space modes. The boundary correction is precisely the off-diagonal part generated by this finite-size projection. Note that, in general, the diagonal kernel $T(\mathbf k)$ defined here is not identical to that of the infinite lattice, since the boundary correction also contributes nonzero diagonal terms that have been absorbed into the definition of $T(\mathbf k)$.

\begin{figure}
    \centering
    \includegraphics[width=0.75\linewidth]{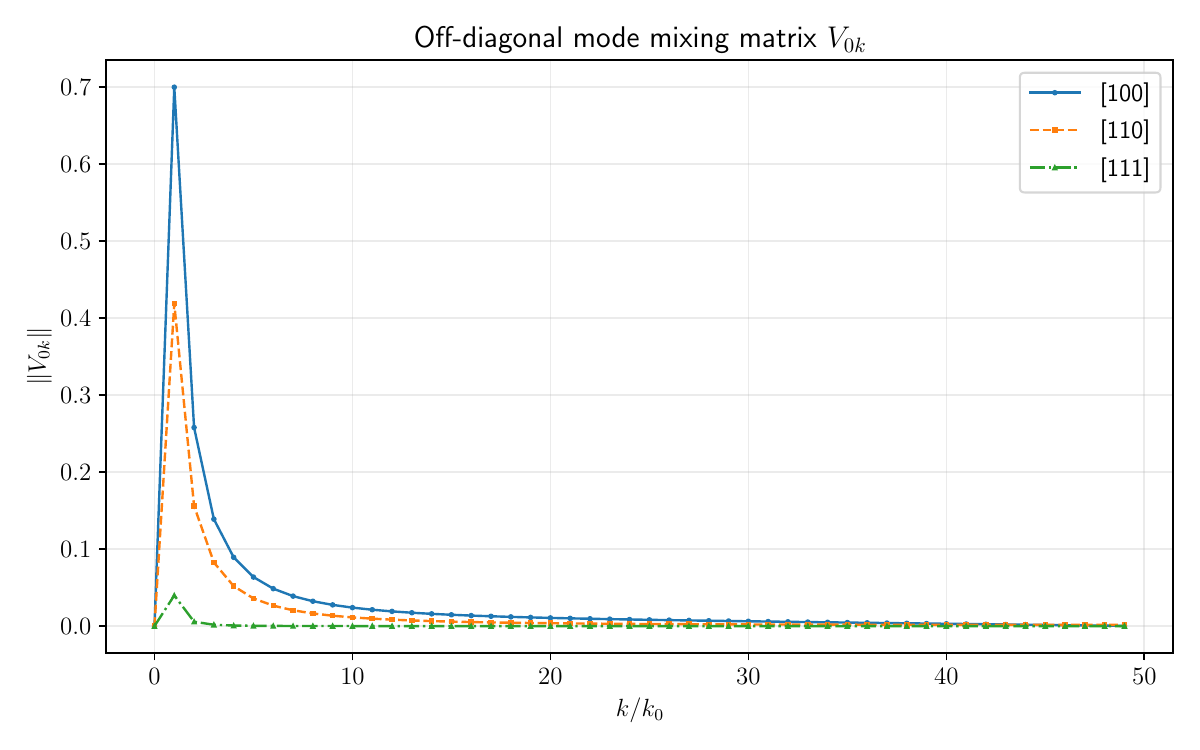}
    \caption{Boundary-induced mode-mixing amplitude $V_{0\mathbf k}$ along the three high-symmetry directions of the simple cubic lattice. The dominant weight is concentrated in the small-$k$ region, showing that the coherent mode couples most strongly to long-wavelength nonuniform modes.}
    \label{fig:v0k_norm_cuts}
\end{figure}

\rev{To evaluate the boundary correction quantitatively, we numerically construct a $100\times100\times100$ simple-cubic lattice with open boundaries and separate the finite-lattice kernel into its reciprocal-space diagonal and off-diagonal parts. The diagonal part defines $T(\mathbf k)$, while the remaining mode-mixing contribution defines $T_{\mathrm{bnd}}$, so that by construction}
\begin{equation}
V_{\mathbf k\mathbf k}=0.
\label{eq:Vkk_zero}
\end{equation}
\rev{Details of the numerical implementation and the treatment of the near-singular subspace are given in Appendix~\ref{app:finite_lattice_numerics}. Fig.~\ref{fig:Tper_diag_components_100} shows the diagonal components of the numerically extracted kernel along the $\mathbf k \parallel(1,0,0)$ direction.} In the small-$k$ region, the result agrees well with the analytical expression obtained for the ideal lattice in Sec.~IV.D. \rev{The sharp tendency toward zero at the smallest accessible $k$ is produced by the finite-size window function $\tilde S(\mathbf q)$; the analytical expression in Sec.~IV.D describes the corresponding bulk long-wavelength behavior. Fig.~\ref{fig:v0k_norm_cuts} shows the numerical value of $V_{0\mathbf k}$ along the three high-symmetry directions. The dominant contribution is concentrated at small $k$, confirming that the boundary couples the coherent mode most strongly to long-wavelength nonuniform modes.}

\begin{figure}
    \centering
    \includegraphics[width=0.75\linewidth]{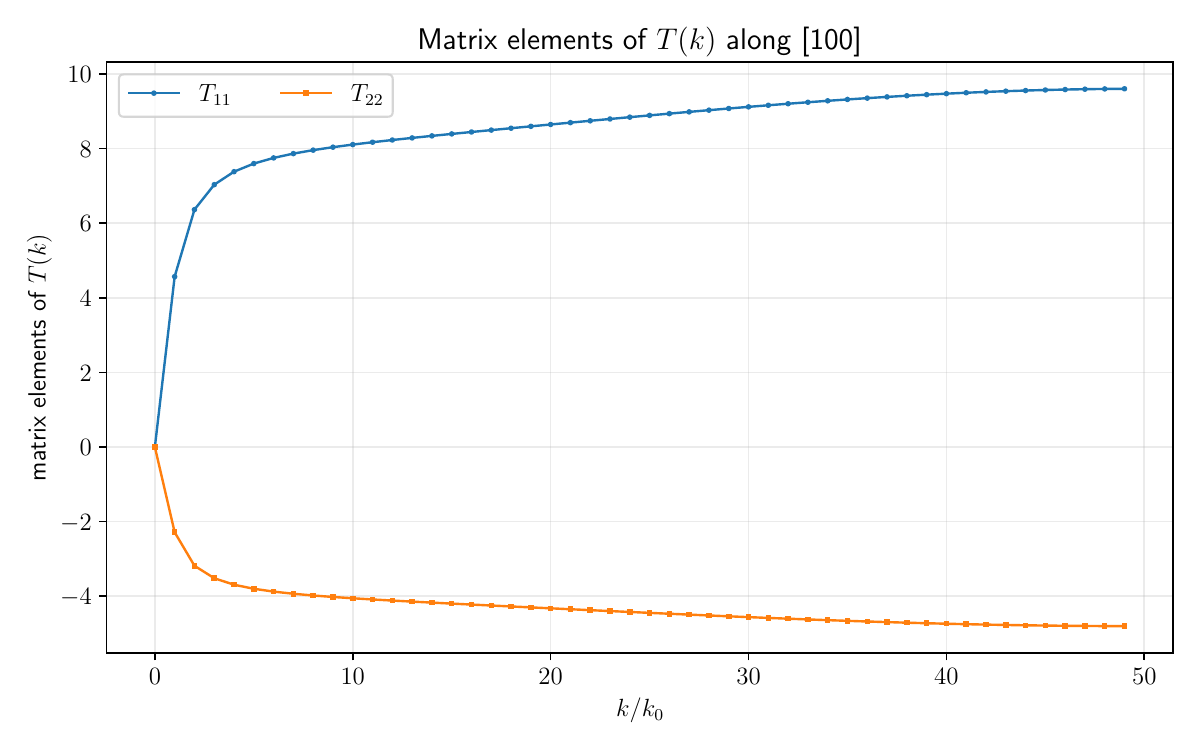}
    \caption{Diagonal components of the interaction kernel $T(\mathbf k)$ extracted numerically for a $100\times100\times100$ simple cubic lattice, shown along the $\mathbf k\parallel(1,0,0)$ direction. In the small-$k$ regime, the result agrees with the analytical long-wavelength expression of the ideal lattice. The sharp tendency toward zero at the smallest $k$ is caused by the finite-size window function rather than by the bulk interaction itself.}
    \label{fig:Tper_diag_components_100}
\end{figure}

\begin{figure}
    \centering
    \includegraphics[width=0.75\linewidth]{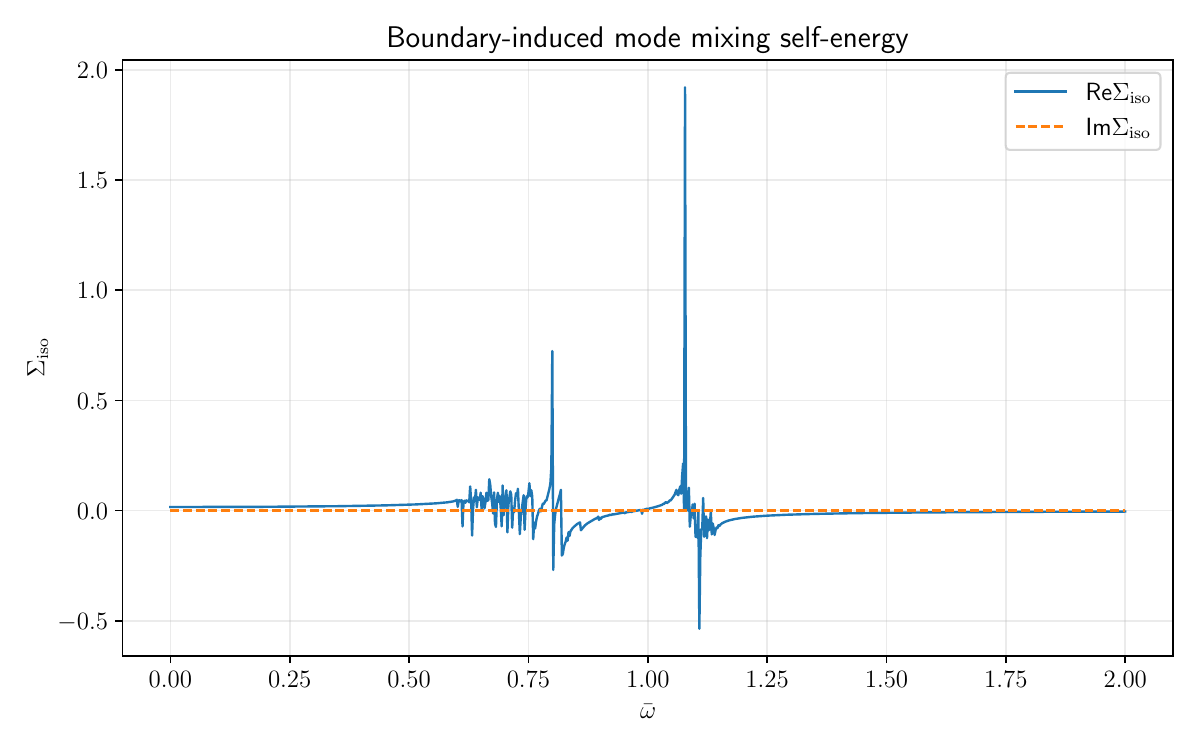}
    \caption{Boundary-induced self-energy of the coherent mode, calculated for the finite simple cubic lattice. The dominant contribution comes from long-wavelength nonuniform modes coupled to the coherent channel by the boundary. The vertical axis is shown on a symmetric logarithmic scale: it is linear in the interval $[-1,1]$ and logarithmic outside this range, allowing the positive and negative parts and the overall large-scale structure to be displayed in a single plot. The resulting structure is strongly frequency dependent, reflecting the propagation of the intermediate lattice modes.}
    \label{fig:self_energy_iso}
\end{figure}

With the finite-lattice kernel determined numerically, we next evaluate the boundary self-energy and the resulting coherent susceptibility. In the following calculations, we take $\lambda_\mathrm{int} = 0.0665$. In the perturbative regime, the self-energy is obtained from Eq.~(\ref{eq:Sigma_bnd_def}), while in the presence of near-singular modes the corresponding generalized expression Eq.~(\ref{eq:Sigma_bnd_near_singular}) is used. In both cases, the physical content is the same: the coherent response is renormalized by virtual leakage into nonuniform modes and the subsequent return to the coherent channel. The numerical result for the self-energy is shown in \rev{Fig.}~\ref{fig:self_energy_iso}. Within the frequency interval Eq.~(\ref{eq:freq_interval}), the self-energy develops a sequence of small peaks associated with individual discrete $\mathbf k$ modes. Each of these peaks is approximately antisymmetric about its center, with one positive and one negative side. Among them, two dominant peaks stand out as the main deviations from zero. The first one is associated with the near-singular point of the $\mathbf k=(k_0,0,0)$ mode. Since this mode has the largest boundary mixing amplitude $V_{0\mathbf k}$, it gives rise to the highest individual peak. The second dominant structure is associated approximately with the point where the eigenvalue $\nu_1$, which is constant in the small-$k$ approximation, approaches zero. In this case, a large number of momentum modes become nearly singular simultaneously, producing the strongest enhancement over the entire frequency range.

\begin{figure}
    \centering
    \includegraphics[width=0.75\linewidth]{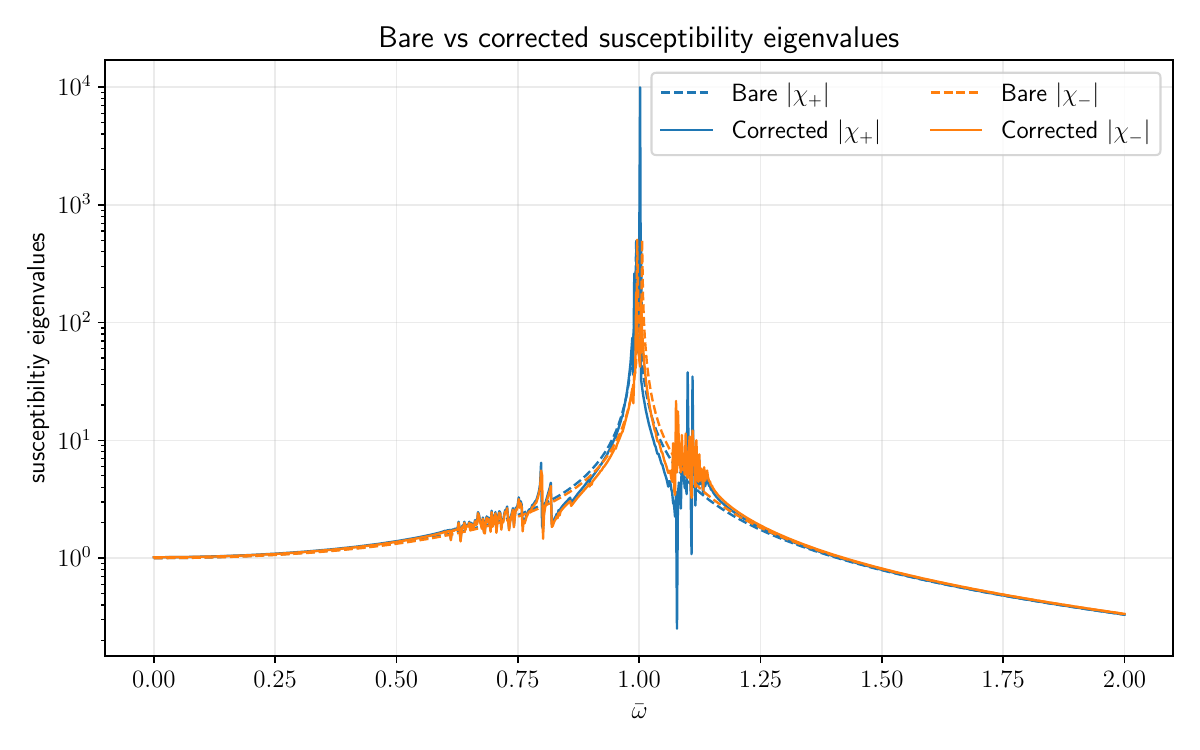}
    \caption{Comparison between the bare coherent susceptibility and the susceptibility corrected by the boundary-induced self-energy. The finite-size interaction does not simply suppress the response uniformly, but reshapes its spectral structure through enhanced mixing with long-wavelength nonuniform modes, especially near the frequencies associated with the near-singular sector.}
    \label{fig:susceptibility_bare_vs_corrected_eigenvalues}
\end{figure}

The resulting corrected coherent susceptibility is compared with the bare coherent susceptibility in \rev{Fig.}~\ref{fig:susceptibility_bare_vs_corrected_eigenvalues}.
As anticipated from the preceding analysis, the boundary correction does not produce a simple overall suppression of the response. Instead, it introduces a structured frequency dependence through the coupling to long-wavelength nonuniform modes. Over most of the spectrum the correction remains moderate, but near frequencies where the small-$k$ sector approaches a singular response, the susceptibility is visibly distorted by the enhanced mode mixing. This behavior is precisely the finite-size manifestation of the near-singular directional structure discussed in Sec.~IV.D.

\rev{Once the interaction kernel $T(\mathbf k)$ is determined, Eq.~(\ref{eq:chi_mod_def}) fixes the eigenfrequencies at which the nonuniform modes become nearly singular. These modes generate the strong thermal-noise amplification discussed in Sec.~V and therefore determine the frequency range of the blind zone. Numerical precision and frequency-grid resolution control how well the individual discrete peaks and line shapes inside this range are resolved, while the underlying near-singular spectrum is fixed by the interaction kernel. The present $1001$-point scan consequently provides a qualitative representation of the internal blind-zone structure. Its detailed peaks depend on the finite-lattice geometry and boundary coupling, whereas the dominant long-wavelength sector admits the analytical estimate in Eq.~(\ref{eq:freq_interval}). The agreement of the full numerical calculation with this estimate confirms that the small-$\mathbf k$ analysis captures the modes that set the leading blind-zone range.}

These results complete the deterministic part of the interaction analysis. The boundary correction modifies the coherent susceptibility through a frequency-dependent self-energy, whose dominant contribution originates from long-wavelength modes. The same structure will play an important role in the noise problem, because thermal fluctuations can also enter the coherent readout channel through these interaction-induced mode-mixing processes. We will turn to this question in the next section.

\section{Noise Analysis of the Lattice}

In this section, we analyze the noise properties of the ferromagnet lattice. Under collective readout, different noise sources scale differently with the number of ferromagnets $N$, depending on whether they are independent at each site or correlated across the lattice. The central advantage of the lattice architecture is that thermal and imprecision-related contributions can be parametrically reduced by collective operation, while backaction can be rebalanced through the readout circuit. The simple noninteracting scaling must be refined once interaction-induced mode mixing is included. Taking all these effects into account, we show that an effective magnetic-field sensitivity scaling that follows a power law of $1/\sqrt{N}$ can be achieved.

\subsection{Single-ferromagnet noise budget}

We begin by reviewing the noise properties of a single levitated
ferromagnet, which forms the basis for the lattice analysis. A detailed discussion of the single ferromagnet noise budget can be found in \rev{Refs.}~\cite{kaliaUltralightDarkMatter2024,ankel_noise_2025}. A weak oscillatory magnetic field induces an angular response of the ferromagnet according to Eq.~(\ref{eq:response}). The angular motion of the ferromagnet produces a magnetic flux signal in the pickup coil, which further produces an induced current as a detection signal. For small angular deviations, the flux can be written as
\begin{equation}
\Phi(\omega)=  -\mathbf x^{\mathsf T} \eta \, \hat{\mathbf n}(\omega),
\label{eq:flux_def}
\end{equation}
where $\eta$ is the flux coupling factor between the
ferromagnet and the pickup coil, and $\mathbf x$ is the
coil's normal vector. Assuming that the ferromagnet sits at the center of the coil and that one of its angular directions $\theta,\phi$ is aligned with the coil normal, the flux coupling factor can be derived from direct electromagnetism calculation as 
\eq{
\eta = \frac{N_p\mu_0|\mu|}{2R_p},
}
where $N_p$ is the number of loops and $R_p$ is the radius of the coil. More generally, for an arbitrary coil geometry, the coupling is described by a matrix that mixes the two angular directions. For simplicity, we restrict attention to symmetrized setups in which the two angular directions are read out independently. Combining Eqs.~(\ref{eq:response})
and (\ref{eq:flux_def}), the transfer function from magnetic field to measured flux is
\begin{equation}
\mathcal T(\omega)
\equiv
\frac{\Phi(\omega)}{B(\omega)}
=
\eta\,\mu\,\mathbf x^{\mathsf T}\chi(\omega)\mathbf b .
\label{eq:transfer_function}
\end{equation}
Noise in the measured magnetic field can be traced back to fluctuations in either the mechanical dynamics of the ferromagnet or the readout circuitry. The magnetic field noise spectral density $S_B(\omega)$ is defined by
\begin{equation}
S_B(\omega)
=
\frac{S_\Phi(\omega)}{|\mathcal T(\omega)|^2},
\label{eq:SB_def}
\end{equation}
where $S_\Phi(\omega)$ denotes the magnetic flux noise spectrum of the
measurement. Specifically, in the fully trapped regime, the susceptibility matrix is diagonal, and in a trivial setup, we align the coils onto the two angular directions $\theta,\phi$, making the transfer function to be diagonal as well. The system thus introduces no correlation between the two angular directions, as long as the noise spectrum of the readout coils is diagonal itself. We utilize this assumption in the following discussions, as it is generally possible to approximate this condition.

The noise budget receives contributions from three principal physical mechanisms. The first is thermal noise arising from environmental
dissipation of the rotational motion. The coupling of the ferromagnet to its surrounding atmosphere introduces a damping rate $\gamma$, which, according to the fluctuation-dissipation theorem, is accompanied by a random torque with spectral density
\begin{equation}
S_{\tau,1}^\mathrm{ th}
=
4k_B  T I \gamma .
\end{equation}
Here and in the following, we use the subscript $1$ to denote that the term represents a single ferromagnet noise explicitly. This stochastic torque drives angular fluctuations in a similar way that a magnetic-field-induced torque drives angular motions. The thermal noise can thus be expressed as an equivalent magnetic field noise, which reads
\begin{equation}
S_{B,1}^\mathrm{ th}(\omega)
=
\frac{1}{|\mu|^2}S_{\tau,1}^\mathrm{ th}.
\label{eq:thermal_noise}
\end{equation}
In this simplified description, the thermal torque noise is white, and the corresponding equivalent magnetic-field noise is therefore frequency independent.

The second contribution arises from the readout imprecision noise. The SQUID detector possesses an intrinsic flux noise floor characterized by a spectral density $S_{\Phi,1}^\mathrm{ imp}$. Since this noise enters directly at the readout stage and is independent of the ferromagnet dynamics, the corresponding equivalent magnetic-field noise follows straightforwardly from Eq.~(\ref{eq:SB_def}). Because the transfer function depends on the magnetic susceptibility, the imprecision noise acquires a strong frequency dependence. In particular, it exhibits a pronounced dip at the resonance frequency and increases at large frequencies. At sufficiently high frequencies, the susceptibility approaches zero, causing the imprecision noise to dominate the total noise budget and effectively setting an upper frequency limit for the ferromagnet-based sensor.

The third contribution is the measurement backaction. Fluctuations of the SQUID current generate a fluctuating magnetic field in the pickup circuit, which acts back on the ferromagnet and drives additional motion. Denoting the SQUID current noise by $S_I^\mathrm{ SQUID}$, the resulting magnetic field noise can be written as
\begin{equation}
S_{B,1}^\mathrm{ back}
=
\frac{\eta^2}{|\mu|^2} S_I^\mathrm{ SQUID}.
\label{eq:backaction_noise}
\end{equation}
In contrast to imprecision noise, backaction represents a real fluctuating force acting on the system rather than a limitation of the measurement resolution. The backaction noise is therefore also independent of ferromagnet dynamics or susceptibility matrix, and its spectral form depends solely on $S_I^\mathrm{ SQUID}$.

The imprecision and backaction noise discussed above originate from the same physical source, namely the intrinsic quantum fluctuations of the SQUID detector. In the readout process, the SQUID converts the magnetic flux induced by the ferromagnet into a measurable output signal, while its current fluctuations are fed back into the pickup circuit and generate a fluctuating magnetic field acting on the ferromagnet. These two effects define the two complementary noise channels of the readout: the flux imprecision noise $S_{\Phi}^\mathrm{ imp}$ and the SQUID current noise $S_I^\mathrm{ SQUID}$. For a linear quantum detector, these two noise channels are not independent. Following Ref.~\cite{kaliaUltralightDarkMatter2024}, it is convenient to characterize them by the two energy-resolution parameters
\begin{equation}
\kappa_\Phi \equiv \frac{S_{\Phi}^\mathrm{ imp}}{L_S},
\qquad
\kappa_I \equiv S_I^\mathrm{ SQUID} L_S,
\end{equation}
where $L_S$ is the SQUID inductance. The quantum limit constrains their product as \(\kappa_\Phi \kappa_I \gtrsim \hbar^2\), up to convention-dependent factors of order unity \cite{clerk_introduction_2010,kaliaUltralightDarkMatter2024}. In the present work, we follow Ref.~\cite{kaliaUltralightDarkMatter2024} and adopt the symmetric choice
\begin{equation}
\kappa_\Phi=\kappa_I=\kappa ,
\end{equation}
so that the flux and current noise are taken to contribute equally in the sense of energy resolution. Under this assumption,
\begin{equation}
S_{\Phi}^\mathrm{ imp}=L_S \kappa,
\qquad
S_I^\mathrm{ SQUID}=\frac{\kappa}{L_S},
\qquad
\frac{S_{\Phi}^\mathrm{ imp}}{S_I^\mathrm{ SQUID}}=L_S^2 .
\end{equation}
For the numerical estimates, we take $\kappa=1000\,\hbar$ for the current setup following Ref.~\cite{kaliaUltralightDarkMatter2024}.

Combining the three contributions, the total magnetic field noise spectrum of a single levitated ferromagnet can be written as
\begin{equation}
S_{B,1}(\omega)
=
S_{B,1}^\mathrm{ th}(\omega)
+
S_{B,1}^\mathrm{ imp}(\omega)
+
S_{B,1}^\mathrm{ back}(\omega).
\label{eq:single_noise_total}
\end{equation}
This single-particle noise budget provides the reference against which the noise properties of the ferromagnet lattice will be evaluated in the following sections.

\subsection{Collective readout, interaction correction, and noise scaling}

We now extend the single-particle noise analysis to the lattice. In the ideal collective-readout picture, the total signal is obtained by coherently summing the responses of all ferromagnets, while different noise sources scale differently depending on whether they are correlated across the lattice. However, once the dipole-dipole interaction and the finite boundary are taken into account, this simple scaling picture must be refined. As shown in Sec.~IV, the finite boundary mixes the coherent mode with nonuniform modes, and the resulting correction affects different noise channels in qualitatively different ways.

The key distinction is whether the noise source acts directly on the coherent channel alone or drives all lattice modes. A spatially uniform external signal couples only to the coherent mode and acquires an interaction correction only through virtual leakage into nonuniform modes and back, as encoded in the boundary self-energy. Thermal noise behaves differently: since it acts independently on each ferromagnet, it excites all reciprocal-space modes directly. The thermal fluctuations of nonuniform modes can therefore enter the coherent readout channel through boundary-induced mode mixing. This mechanism leads to an additional interaction-dependent correction to the thermal noise, which is absent in the noninteracting picture. By contrast, as will be discussed below, the imprecision and backaction noise remain tied primarily to the coherent readout channel and do not acquire this extra mixing-induced contribution.

\subsubsection{Thermal noise}

We begin with thermal noise. For a single ferromagnet, the thermal torque noise can be represented as an equivalent white magnetic-field noise with spectral density
\begin{equation}
S^\mathrm{th}_{B,1}(\omega)
=
\frac{4k_BTI\gamma}{|\mu|^2}\,\mathbbm 1,
\label{eq:SB1_th_matrix}
\end{equation}
where the noise is assumed to be identical and uncorrelated in the two orthogonal angular directions. In the lattice, thermal noise acts independently on each ferromagnet and can therefore be modeled as a random local magnetic field $\mathbf B_i^\mathrm{noise}(\omega)$. The equation of motion in the reciprocal space reads
\begin{equation}
\chi_{\mathrm{mod}}^{-1}(\omega,\mathbf k) \mathbf u_{\mathbf k}(\omega)
\rev{+}
\mu B_\mathrm{dd} \sum_{\mathbf q\neq \mathbf k}
V_{\mathbf k\mathbf q}
\mathbf u_{\mathbf q}(\omega)
=
-\mu\left[
\mathbf B_\mathrm{ext}(\omega)\delta_{\mathbf k,0}
+
\mathbf B^\mathrm{noise}(\omega,\mathbf k)
\right].
\label{eq:master0_reciprocal_thermal}
\end{equation}

The thermal noise in real space is uncorrelated between different ferromagnets,
\begin{equation}
\braket{
\mathbf B^\mathrm{noise}_{i}(\omega)
\mathbf B^\mathrm{noise}_{j}(\omega')^\dagger
}
=
2\pi\delta(\omega-\omega')\delta_{ij}S^\mathrm{th}_{B,1},
\label{eq:thermal_noise_realspace_corr}
\end{equation}
and therefore remains uncorrelated between different $\mathbf k$ modes after Fourier transformation,
\begin{equation}
\braket{
\mathbf B^\mathrm{noise}(\omega,\mathbf k)
\mathbf B^\mathrm{noise}(\omega',\mathbf k')^\dagger
}
=
\frac{2\pi}{N}\delta(\omega-\omega')\delta_{\mathbf k\mathbf k'}S^\mathrm{th}_{B,1}.
\label{eq:thermal_noise_kspace_corr}
\end{equation}
Thus, thermal noise is white both in real space and in reciprocal space, but unlike an external signal it acts on all lattice modes rather than only on the coherent mode. This is the essential difference between thermal noise and a spatially uniform external field. The external field enters only through the $\mathbf k=0$ channel, and its interaction correction arises because the coherent response leaks into nonuniform modes and then returns. Thermal noise, however, directly drives every $\mathbf k$ mode. As a result, the thermal fluctuations of nonuniform modes can contribute to the coherent readout through the same boundary-induced mode mixing discussed in Sec.~IV. In the coherent-mode-dominance regime, the equations of motion take the form
\begin{equation}
\chi_{\mathrm{mod}}^{-1}(\omega,\mathbf k)\mathbf u_{\mathbf k}
\rev{+}
\mu B_\mathrm{dd}V_{\mathbf k0}\mathbf u_0
=
-\mu \mathbf B^\mathrm{noise}(\omega,\mathbf k),
\qquad
\mathbf k\neq 0,
\label{eq:uk_thermal_pert}
\end{equation}
and
\begin{equation}
\chi_{\mathrm{mod}}^{-1}(\omega,\mathbf 0)\mathbf u_0
\rev{+}
\mu B_\mathrm{dd}\sum_{\mathbf k\neq 0}V_{0\mathbf k}\mathbf u_{\mathbf k}
=
-\mu\left[
\mathbf B_\mathrm{ext}(\omega)
+
\mathbf B^\mathrm{noise}(\omega,0)
\right].
\label{eq:u0_thermal_pert}
\end{equation}
Solving these equations to leading order gives
\begin{equation}
\mathbf u_0
=
-\mu \chi_{\mathrm{coh}}(\omega)
\left[
\mathbf B_\mathrm{ext}(\omega)
+
\mathbf B^\mathrm{noise}_\mathrm{eff} (\omega)
\right],
\label{eq:u0_thermal_solution}
\end{equation}
where
\begin{equation}
\mathbf B^\mathrm{noise}_\mathrm{eff}(\omega)
=
\mathbf B^\mathrm{noise}(\omega,0)
\rev{-}
\mu B_\mathrm{dd}
\sum_{\mathbf k\neq 0}
V_{0\mathbf k}\chi_{\mathrm{mod}}(\omega,\mathbf k)\mathbf B^\mathrm{noise}(\omega,\mathbf k)
\label{eq:Beff_thermal}
\end{equation}
is the effective thermal magnetic field acting on the coherent mode. Compared with the response to an external signal, there is an additional contribution from the thermal noise of all nonuniform modes. The first term in $\mathbf B^\mathrm{noise}_\mathrm{eff}(\omega)$ is the direct coherent contribution, which would remain in the noninteracting limit. The second term is induced entirely by the interaction and the finite boundary, and describes the conversion of thermal fluctuations from nonuniform modes into the coherent readout channel.

Using Eq.~(\ref{eq:thermal_noise_kspace_corr}), the correlation function of the effective thermal field becomes
\begin{align}
&\braket{
\mathbf B^\mathrm{noise}_\mathrm{eff}(\omega)
\mathbf B^\mathrm{noise}_\mathrm{eff}(\omega')^\dagger
}
\nonumber\\
&=
\frac{2\pi}{N}\delta(\omega-\omega')
\left[
\mathbbm 1
+
|\mu|^2B_\mathrm{dd}^2
\sum_{\mathbf k\neq 0}
V_{0\mathbf k}
\chi_{\mathrm{mod}}(\omega,\mathbf k)
\chi_{\mathrm{mod}}^\dagger(\omega,\mathbf k)
V_{\mathbf k0}
\right]
S^\mathrm{th}_{B,1}.
\label{eq:Beff_thermal_corr}
\end{align}
The equivalent thermal magnetic-field noise of the lattice is therefore
\begin{equation}
S^\mathrm{th}_B(\omega)
=
\frac{1}{N} M
S^\mathrm{th}_{B,1},
\label{eq:SBth_lattice_full}
\end{equation}
where
\eq{
M = 
\mathbbm 1
+
|\mu|^2B_\mathrm{dd}^2
\sum_{\mathbf k\neq 0}
V_{0\mathbf k}
\chi_{\mathrm{mod}}(\omega,\mathbf k)
\chi_{\mathrm{mod}}^\dagger(\omega,\mathbf k)
V_{\mathbf k0}
}
is a $2\times 2$ matrix that rescales the thermal noise. Rewriting this in terms of the dimensionless susceptibility defined in Sec.~IV yields
\begin{equation}
M = 
\mathbbm 1
+
\lambda_\mathrm{int}^2
\sum_{\mathbf k\neq 0}
V_{0\mathbf k}
\xi^{-1}(\omega,\mathbf k)
\xi^{-1}(\omega,\mathbf k)^\dagger
V_{\mathbf k0}.
\label{eq:SBth_lattice_dimensionless}
\end{equation}

This result should be contrasted with the deterministic boundary self-energy in Sec.~IV.B. The self-energy depends linearly on the susceptibility of the intermediate modes, whereas the thermal-noise correction depends quadratically on it. Thermal noise is therefore more sensitive to near-singular modes than the coherent susceptibility itself. Even when the deterministic response is only moderately distorted, the thermal noise can already be strongly amplified once the lattice contains modes with anomalously large susceptibility.

\rev{When the near-singular modes must be retained explicitly, the susceptibility of an isolated mode is replaced by the full response matrix of the near-singular subspace. The selected modes are treated by the coupled Green function $G$, while the regular modes outside this subspace retain their perturbative contribution. The thermal noise then becomes
\begin{equation}
\begin{split}
S^\mathrm{th}_B(\omega)
=
\frac{1}{N}
\Bigg\{
\mathbbm 1
+
\lambda_\mathrm{int}^2
\Bigg[&
\sum_{\mathbf s_1,\mathbf s_2,\mathbf s_3\in\mathcal S\setminus\{0\}}
V_{0\mathbf s_1}
G_{\mathbf s_1,\mathbf s_2}(\omega)
G^\dagger_{\mathbf s_2,\mathbf s_3}(\omega)
V_{\mathbf s_3 0}
\\
&+
\sum_{\mathbf k\notin\mathcal S}
V_{0\mathbf k}
\xi^{-1}(\omega,\mathbf k)
\xi^{-1}(\omega,\mathbf k)^\dagger
V_{\mathbf k0}
\Bigg]
\Bigg\}
S^\mathrm{th}_{B,1}.
\end{split}
\label{eq:SBth_lattice_nearsingular}
\end{equation}
Thus, the perturbative and nonperturbative treatments have the same hybrid structure as in Sec.~IV: the near-singular modes are promoted to a coupled response matrix, and the remaining modes supply a regular perturbative remainder. This is also the form implemented in the numerical calculation described in Appendix~\ref{app:finite_lattice_numerics}.}

\begin{figure}
    \centering
    \includegraphics[width=0.75\linewidth]{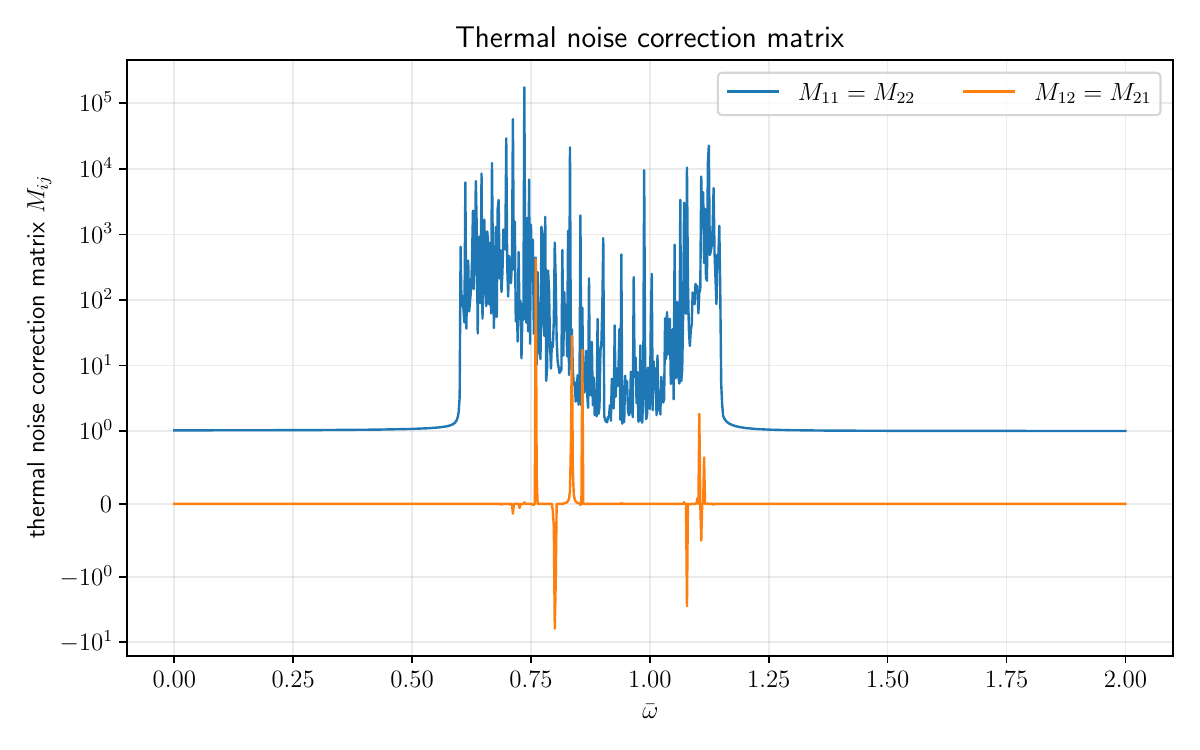}
    \caption{Components of the thermal magnetic-field noise matrix for the finite simple cubic lattice. The vertical axis is shown on a symmetric logarithmic scale that is linear in the interval $[-1,1]$ and logarithmic outside this range. The interaction-induced correction strongly amplifies the thermal noise in the frequency range where near-singular modes appear. As a result, the lattice develops a blind zone in which the observable sensitivity is significantly degraded. The off-diagonal components become nonzero within part of this region, showing that the boundary-induced mode mixing also generates correlations between the two orthogonal angular directions.}
    \label{fig:thermal_noise_matrix_components}
\end{figure}

Numerical results for the thermal-noise matrix are shown in \rev{Fig.}~\ref{fig:thermal_noise_matrix_components}. \rev{Because the correction depends on the square of the susceptibility, every boundary-coupled mode that becomes nearly singular according to Eq.~(\ref{eq:chi_mod_def}) produces a strong thermal-noise enhancement near its eigenfrequency. The frequency support of these near-singular modes therefore defines the blind zone. The analytical small-$\mathbf k$ treatment captures the dominant long-wavelength sector and gives the estimate in Eq.~(\ref{eq:freq_interval}). The blind zone obtained from the full numerical interaction kernel is slightly broader and remains in good overall agreement with this analytical estimate. Within parts of the blind zone, the thermal-noise matrix also develops nonzero off-diagonal components, indicating interaction-induced correlations between the two orthogonal angular directions.}

The thermal-noise result above constitutes the main modification of the ideal collective-readout picture. In the absence of interaction, thermal noise is reduced simply by averaging over independent ferromagnets, achieving a suppression factor of $1/N$. In the interacting finite lattice, however, this $1/N$ scaling is supplemented by a frequency-dependent correction arising from the conversion of nonuniform thermal fluctuations into coherent readout noise. Nevertheless, this blind zone is concentrated only around a small region about the resonance frequency, and for most of the remaining frequency range the thermal noise still remains the favorable $1/N$ scaling. This blind zone can be further compensated for by adopting two distinct resonance frequencies in two experiments, to which we will return in the numerical part.

\subsubsection{Imprecision and backaction noise}

We next consider the readout-induced noise channels, which predominantly comprise imprecision and backaction noise. Unlike thermal noise, these contributions do not originate from independent fluctuations acting on each ferromagnet. Instead, they are tied directly to the collective readout process itself, and therefore couple primarily to the coherent channel. As a result, they do not acquire an additional boundary-induced correction of the thermal-noise type.

The imprecision noise originates from the intrinsic flux noise floor of the SQUID detector. Since this noise enters directly at the readout stage, its equivalent magnetic-field noise is determined by the inverse transfer function from the external magnetic field to the measured collective flux. In the interacting finite lattice, the only modification is therefore that the single-particle susceptibility appearing in the transfer function must be replaced by the corrected coherent susceptibility derived above. For a collective readout aligned with the coherent channel, the total flux can be written as
\begin{equation}
\Phi_{\mathrm{tot}}(\omega)
=
N\,\eta\,\mu\,
\mathbf x^{\mathsf T}
\chi_{\mathrm{coh}}(\omega)\,
\mathbf b\,
B(\omega),
\label{eq:phi_tot_coh}
\end{equation}
so that the collective transfer function becomes
\begin{equation}
\mathcal T_{\mathrm{coh}}(\omega)
=
N\,\eta\,\mu\,
\mathbf x^{\mathsf T}
\chi_{\mathrm{coh}}(\omega)\,
\mathbf b.
\label{eq:Tcoh_def}
\end{equation}
The corresponding imprecision noise is therefore
\begin{equation}
S_B^{\mathrm{imp}}(\omega)
=
\frac{S_{\Phi}^{\mathrm{imp}}(\omega)}
{\left|\mathcal T_{\mathrm{coh}}(\omega)\right|^2}.
\label{eq:SBimp_coh}
\end{equation}
Compared with the ideal noninteracting lattice, no additional contribution from nonuniform modes appears in the numerator. The effect of the interaction is entirely contained in the corrected coherent response entering $\mathcal T_{\mathrm{coh}}(\omega)$.

This point is physically important. The imprecision noise is not a fluctuating force acting on the lattice degrees of freedom, but a limitation of how accurately the collective flux can be read out. It therefore does not excite nonuniform lattice modes independently, and there is no analog of Eq.~(\ref{eq:Beff_thermal}) for the imprecision channel. In the limit where the coherent response reduces to the ideal collective sum, Eq.~(\ref{eq:SBimp_coh}) recovers a $1/N^2$ suppression.

The backaction noise behaves in a different manner. It originates from current fluctuations in the SQUID readout circuit, which generate a fluctuating magnetic field in the pickup coil. Because this magnetic field is produced collectively by the readout circuit and is approximately uniform across the lattice, it couples predominantly to the coherent mode. Its role is therefore analogous to that of an external spatially uniform driving field, except that the field is stochastic rather than deterministic. Consequently, the backaction noise does not receive an extra contribution from independently fluctuating nonuniform modes, and its equivalent magnetic-field spectrum is still determined directly by the readout circuit,
\begin{equation}
S_B^{\mathrm{back}}(\omega)
=
\frac{\eta^2}{|\mu|^2}S_I^{\mathrm{SQUID}}(\omega).
\label{eq:SBback_coh}
\end{equation}
As the backaction magnetic field is correlated across all the ferromagnets, there is no additional suppression due to the number of ferromagnets $N$.

The contrast with thermal noise is therefore clear. Thermal fluctuations act locally on each ferromagnet, and after transforming to reciprocal space they appear as independent noise sources for all $\mathbf k$ modes. This is why boundary-induced mode mixing generates the additional correction in Eqs.~(\ref{eq:SBth_lattice_full})-(\ref{eq:SBth_lattice_nearsingular}), leading to a blind zone. The imprecision and backaction noise, by contrast, are both associated with the collective readout channel itself. The former enters directly as detector noise in the measured flux, and the latter enters as an approximately uniform stochastic magnetic field generated by the pickup circuit. Neither of them produces an additional incoherent injection into nonuniform lattice modes analogous to the thermal-noise channel.

Collecting the above results, the total magnetic-field noise of the lattice can be written schematically as
\begin{equation}
S_B(\omega)
=
S_B^{\mathrm{th}}(\omega)
+
S_B^{\mathrm{imp}}(\omega)
+
S_B^{\mathrm{back}}(\omega),
\label{eq:lattice_noise_total_revised}
\end{equation}
where the thermal contribution is given by the interaction-corrected expressions in Eqs.~(\ref{eq:SBth_lattice_full}) and (\ref{eq:SBth_lattice_nearsingular}), the imprecision noise is obtained from the corrected coherent transfer function through Eq.~(\ref{eq:SBimp_coh}), and the backaction noise remains determined by the collective pickup-circuit fluctuations through Eq.~(\ref{eq:SBback_coh}). In the ideal limit away from the near-singular band, these expressions reduce to the familiar scaling relations
\begin{equation}
S_B^{\mathrm{th}}\sim \frac{1}{N},\qquad
S_B^{\mathrm{imp}}\sim \frac{1}{N^2},\qquad
S_B^{\mathrm{back}}\sim N^0.
\label{eq:lattice_noise_scaling_summary}
\end{equation}
The main new feature of the interacting finite lattice is therefore not a qualitative change in the imprecision or backaction channels, but the strong interaction-induced amplification of the thermal noise near the near-singular sector.

\subsubsection{Readout rebalancing and optimization of the effective coupling}

According to the previous result, the three noise components scale differently. Thermal fluctuations are suppressed through averaging over independent particles, imprecision noise benefits from the enhanced collective signal, while backaction noise remains correlated across the lattice and therefore sets an important limit that must be mitigated through readout optimization. The relative magnitude of the imprecision and backaction contributions depends on the coupling between the pickup coil and the SQUID readout circuit. This coupling is characterized by the flux coupling factor $\eta$ introduced in Eq.~(\ref{eq:flux_def}), which determines both the strength of the signal readout and the magnitude of the current fluctuations fed back to the ferromagnet. As a result, the imprecision and backaction noise contributions exhibit opposite dependences on $\eta$. Increasing the coupling improves the transfer of magnetic flux to the SQUID and therefore reduces the equivalent magnetic field imprecision noise, while at the same time amplifying the magnetic field fluctuations generated by SQUID current noise.

This trade-off can be understood directly from the expressions for the noise contributions. The imprecision noise is obtained by dividing the intrinsic SQUID flux noise by the square of the transfer function defined in Eq.~(\ref{eq:transfer_function}). Since the transfer function is proportional to the coupling factor $\eta$, the equivalent magnetic field noise scales as
\begin{equation}
S_{B}^\mathrm{ imp}(\omega)\propto\eta^{-2}.
\end{equation}
In contrast, the backaction noise originates from SQUID current fluctuations that produce a magnetic field through the pickup circuit. The magnitude of this field is proportional to the same coupling factor, leading to
\begin{equation}
S_{B}^\mathrm{ back}(\omega)\propto\eta^{2}.
\end{equation}
The two contributions therefore vary in opposite directions as the coupling strength is adjusted. Because the imprecision and backaction noise originate from the same quantum fluctuations of the SQUID detector, their spectral densities cannot be reduced simultaneously. The uncertainty relation discussed in Sec.~V.A implies that decreasing the flux noise necessarily increases the associated current fluctuations.

In practice, this balancing can be implemented by introducing a two-stage pickup circuit. One pickup coil with inductance $L_\mathrm{p}$ couples directly to the ferromagnet lattice, while a second coil with inductance $L_\mathrm{i}$ is coupled to the SQUID input coil of inductance $L_\mathrm{s}$. The two pickup coils are connected in series and therefore carry the same current. If the mutual inductance between the intermediate coil and the SQUID input coil is $M_\mathrm{i}=k\sqrt{L_\mathrm{i}L_\mathrm{s}}$, with $k<1$ determined by the circuit geometry, the entire readout chain is equivalent to a single effective coupling in which the flux coupling $\eta$ is rescaled by a factor $M_\mathrm{i}/L_\mathrm{tot}$, where $L_\mathrm{tot}=L_\mathrm{p}+L_\mathrm{i}$. Accordingly, the effective transfer function acquires the replacement
\begin{equation}
\eta \rightarrow \eta_\mathrm{eff}\equiv \eta\,\frac{M_\mathrm{i}}{L_\mathrm{tot}}.
\end{equation}
Since the backaction noise is generated by SQUID current fluctuations fed back through the same circuit, it scales as $\eta_\mathrm{eff}^2$, whereas the imprecision noise, obtained by dividing the SQUID flux noise by the square of the transfer function, scales as $\eta_\mathrm{eff}^{-2}$. The readout optimization problem therefore reduces to choosing the circuit parameters such that $\eta_\mathrm{eff}$ places the two contributions at comparable levels in the frequency range of interest.

For a circular pickup coil of radius $R_\mathrm{p}$ and loop number $N_\mathrm{p}$, the coil inductance is
\begin{equation}
L_\mathrm{p}
=
N_\mathrm{p}^2\mu_0R_\mathrm{p}
\left[
\ln\!\left(\frac{8R_\mathrm{p}}{a_\mathrm{p}}\right)-2
\right],
\end{equation}
with $a_\mathrm{p}$ the wire radius. \rev{The relative readout-noise balance is quantified by}
\begin{equation}
\frac{S_B^\mathrm{ back}(\omega = 0)}{S_B^\mathrm{ imp}(\omega = 0)}
=
 \frac{N^2 \eta_\mathrm{eff}^4}{V_{\theta\theta}^2} \frac{S^\mathrm{SQUID}_I(\omega= 0 )}{S^\mathrm{imp}_\Phi(\omega= 0 )},
\label{eq:back_imp_ratio}
\end{equation}
where $V_{\theta\theta}$ denotes the relevant curvature of the trapping potential. \rev{Varying $N_\mathrm{p}$ or $R_\mathrm{p}$ changes both the direct magnetic coupling and the total inductance of the pickup circuit and thereby tunes $\eta_\mathrm{eff}$. The readout-induced noise is minimized by bringing the ratio in Eq.~(\ref{eq:back_imp_ratio}) as close to unity as the accessible circuit parameters allow. For fixed trapping and SQUID parameters, the factor $N^2$ shifts the balanced value of $\eta_\mathrm{eff}$ downward relative to the single-ferromagnet case, with $\eta_\mathrm{eff}\propto N^{-1/2}$. The direction of optimization is determined by the actual noise balance: when imprecision dominates and the ratio is below unity, $\eta_\mathrm{eff}$ should be increased, whereas a backaction-dominated configuration requires a reduction of $\eta_\mathrm{eff}$.}

\subsection{Secondary imperfections of the lattice}
\label{sec:Secondary imperfections of the lattice}

The analysis above identifies the dominant interaction-induced modification of the lattice noise: the finite boundary mixes the coherent mode with nonuniform modes, and this mixing strongly amplifies the thermal noise in the near-singular band. In the discussions before, we assumed an ideal lattice model with identical ferromagnets and an exact scale hierarchy. In a realistic implementation, however, the lattice is not perfectly ideal. Small fabrication-induced variations among the ferromagnets and the finite spatial extent of the readout geometry introduce additional corrections beyond the ideal lattice model. These effects do not change the basic physical mechanism responsible for the blind zone, but they can modify the detailed structure of the response and the noise spectrum.

In the following, we discuss two such secondary imperfections. The first is fabrication-induced disorder in the local single-particle response, which enters the mode-space dynamics as an additional kernel on top of the boundary correction. The second is the spatial nonuniformity of the readout and backaction profiles, which causes the measured signal and the backaction drive to deviate slightly from a pure projection onto the coherent channel. Both effects can be incorporated naturally into the reciprocal-space framework developed above. Their main impact is to reshape the detailed structure of the response within the blind zone, while leaving the behavior outside that region only weakly affected.

\subsubsection{Fabrication-induced disorder in the dynamical kernel}

We first consider fabrication-induced variations among the ferromagnets. In an ideal lattice, all particles are assumed to have identical properties, so that the single-particle susceptibility is the same on every site. In practice, however, small variations in radius, magnetic moment, moment of inertia, trap curvature, or equilibrium orientation lead to a weak site dependence of the local response. This can be described by writing the single-particle inverse susceptibility on site $i$ as
\begin{equation}
\chi_i^{-1}(\omega)
=
\chi^{-1}(\omega)
+
\delta K_i(\omega),
\label{eq:chiinv_i_disorder}
\end{equation}
where $\chi(\omega)$ is the ideal susceptibility and $\delta K_i(\omega)$ denotes the local fabrication-induced deviation. In real space, the lattice equation of motion then becomes
\begin{equation}
\sum_j
\left[
\chi^{-1}(\omega)\delta_{ij}
+
\delta K_i(\omega)\delta_{ij}
\rev{+}
\mu B_{\mathrm{dd}}T_{ij}
\right]
\delta\mathbf n_j(\omega)
=
-\mu\,\mathbf B_i(\omega),
\label{eq:lattice_eom_disorder_real}
\end{equation}
where $\mathbf B_i$ denotes the total magnetic field acting on site $i$.

After transforming to reciprocal space, the disorder term is no longer diagonal in momentum space. Instead, it generates an additional mode-mixing kernel
\begin{equation}
\delta K_{\mathbf k\mathbf q}(\omega)
\equiv
\frac{1}{N}
\sum_i
e^{-i\mathbf k\cdot\mathbf r_i}
\delta K_i(\omega)
e^{i\mathbf q\cdot\mathbf r_i},
\label{eq:deltaK_kq_def}
\end{equation}
which enters the mode-space equation on exactly the same footing as the boundary correction. The equation of motion therefore takes the form
\begin{equation}
\chi_{\mathrm{mod}}^{-1}(\omega,\mathbf k)\mathbf u_{\mathbf k}
\rev{+}
\mu B_{\mathrm{dd}}
\sum_{\mathbf q}
V_{\mathbf k\mathbf q}\mathbf u_{\mathbf q}
+
\sum_{\mathbf q}
\delta K_{\mathbf k\mathbf q}(\omega)\mathbf u_{\mathbf q}
=
-\mu\,\mathbf B_{\mathbf k}(\omega).
\label{eq:lattice_eom_disorder_k}
\end{equation}
Equivalently, one may regard the disorder correction as simply shifting the mode-mixing kernel according to
\begin{equation}
V_{\mathbf k\mathbf q}
\;\rightarrow\;
V_{\mathbf k\mathbf q}
 \rev{+} \frac{1}{\mu B_\mathrm{dd}}
\delta K_{\mathbf k\mathbf q}(\omega).
\label{eq:V_to_VplusdeltaK}
\end{equation}
All of the formalism developed in Secs.~IV and V.B then remains valid after this replacement. Therefore, the fabrication disorder does not introduce a qualitatively new mechanism, but rather modifies the same leakage processes already induced by the finite boundary. The coherent mode can still couple to nonuniform modes, except that the corresponding mode-mixing amplitudes are now shifted by the additional kernel $\delta K_{\mathbf k\mathbf q}$. As a result, the self-energy correction to the coherent susceptibility and the interaction-induced correction to the thermal noise are both modified in a straightforward way by the replacement in Eq.~(\ref{eq:V_to_VplusdeltaK}).

A rough estimate of the size of this effect can be obtained by introducing a characteristic relative fabrication error $\varepsilon_{\mathrm{dis}}$, representing the typical fractional variation of the microscopic parameters entering the local susceptibility. Since $\delta K_i$ arises from these local variations, one expects its matrix elements to scale as
\begin{equation}
\|\delta K_i\|
\sim
\varepsilon_{\mathrm{dis}}\,
\|\chi^{-1}\|,
\label{eq:deltaKi_scaling}
\end{equation}
and therefore the corresponding mode-space kernel satisfies
\begin{equation}
\|\delta K_{\mathbf k\mathbf q}\|
\sim
\varepsilon_{\mathrm{dis}}\,
\|\xi(\omega,\mathbf k)\|,
\label{eq:deltaKkq_scaling}
\end{equation}
up to geometric factors associated with the Fourier transform. Away from the near-singular region, where the lattice response remains regular, such corrections remain perturbative and lead only to small quantitative shifts in the response. Inside the blind zone, as the \rev{susceptibility} is already strongly enhanced, a relatively weak $\delta K_{\mathbf k\mathbf q}$ can reshape the detailed line shape by shifting, broadening, or splitting the structures generated by the near-singular modes. The important point is that fabrication disorder does not by itself create a new broad noise-dominated band. Its main effect is to modify the detailed structure within the interaction-induced blind zone, where the coherent response is already highly sensitive to mode mixing. Outside that region, the response and noise spectra remain controlled primarily by the ideal kernel and the boundary correction, with disorder contributing only a small perturbative correction. In this sense, the interaction-induced blind zone is robust against moderate fabrication imperfections. Disorder changes its fine structure, but not the basic physical origin of the effect or its overall location in frequency space.

\subsubsection{Nonuniform readout and backaction profiles}

\rev{A finite pickup coil does not sample every lattice site with exactly the same weight. Let $g(\mathbf r_i)$ denote its dimensionless readout profile, normalized so that $N^{-1}\sum_i g(\mathbf r_i)=1$, with the mean single-ferromagnet coupling absorbed into $\eta$. The measured flux is then}
\rev{\begin{equation}
\Phi(\omega)
=\eta\sum_i g(\mathbf r_i)\,\mathbf x^{\mathsf T}\delta\mathbf n_i(\omega)
=\eta\sum_{\mathbf k}g_{-\mathbf k}\,\mathbf x^{\mathsf T}\mathbf u_{\mathbf k}(\omega),
\label{eq:phi_nonuniform_profile}
\end{equation}}
\rev{where the reciprocal-space readout profile is defined by}
\rev{\begin{equation}
g_{\mathbf k}
=\sum_i g(\mathbf r_i)e^{-i\mathbf k\cdot\mathbf r_i}.
\label{eq:gk_def}
\end{equation}}
\rev{Thus $g_0=N$, while a perfectly uniform profile has $g_{\mathbf k}=N\delta_{\mathbf k,0}$. A finite spatial profile gives nonzero $g_{\mathbf k\neq0}$ and, by electromagnetic reciprocity, the same profile determines the nonuniform part of the backaction drive. The full coil geometry determines the detailed coupling. We evaluate $\eta$ and the complete set of profile weights from the unit-current pickup field, as described in Appendix~\ref{app:pickup_coil_scan}.}

\rev{Away from the blind zone, the nonuniform-mode response remains regular and the pickup profile produces only a perturbative correction to the coherent-channel description. Near the blind zone, the same geometric admixture can be enhanced by the near-singular susceptibility and can modify the internal line shape. The spectrum of near-singular modes fixed by the interaction kernel continues to set the blind-zone frequency range. The analogous statement applies to backaction because electromagnetic reciprocity relates the readout weights and the magnetic field generated by current in the same pickup loop.}

\rev{For the benchmark centered single-loop geometry, the radius scan in Appendix~\ref{app:pickup_coil_scan} quantifies this tradeoff directly. The full-mode noncoherent admixture is $\epsilon_{\mathrm{nc}}=0.167$ at $R_{\mathrm p}/L_{\mathrm{lat}}=1$ and decreases to $5.40\times10^{-3}$ at the scan point $R_{\mathrm p}/L_{\mathrm{lat}}=5.05$. Over the same change, the effective coherent coupling is reduced by a factor of $17.9$. The coherent-channel calculation at $\bar\omega=0$ gives a corresponding increase of the total magnetic-field noise power spectral density by a factor of $3.08\times10^2$, because the benchmark remains strongly imprecision dominated. Thus, within this particular circuit and geometry, the stronger coupling at $R_{\mathrm p}=L_{\mathrm{lat}}$ outweighs the larger noncoherent admixture. The optimum radius generally depends on the relative importance of thermal noise, imprecision, backaction, and the direct dynamical response of the finite-profile noncoherent sector.}

Taken together, the two secondary imperfections discussed in this subsection have a common effect: they perturb the ideal coherent-channel picture without changing its basic validity over most of the spectrum. Fabrication disorder modifies the dynamical kernel itself, while finite-profile readout and backaction introduce small nonuniform projections in the measurement and drive channels. Both effects are strongly amplified only in the same near-singular frequency range already identified in the ideal analysis. They therefore alter the detailed internal structure of the blind zone, but not the existence of the blind zone or the overall behavior away from it.

\subsection{Benchmark parameters and numerical noise spectra}

\begin{table}[ht]
    \caption{\rev{Reference benchmark parameters of the ferromagnet lattice and readout apparatus used in the numerical calculations. Parameters marked with an asterisk are directly adopted from Ref.~\cite{kaliaUltralightDarkMatter2024}, including the microscopic properties of each individual ferromagnet, the thermal environment, and the SQUID-related quantities. The remaining parameters are introduced or estimated in the present work to quantify the lattice geometry, the trapping configuration, and the collective readout circuit; together they define the reference theoretical benchmark used here.}}
\begin{ruledtabular}
    \begin{tabular}{ll}
    Parameters & Values \\
    \colrule
        *Ferromagnet radius $R$ & $20\ \upmu\mathrm{ m}$ \\
        *Ferromagnet magnetization $M$ & $7.0\times 10^5\ \mathrm{A/m}$\\
        *Ferromagnet dipole $\mu$ & $-2.3\times 10^{-8}\ \mathrm{A\cdot m^2}$\\
        *Ferromagnet density $\rho$ & $7400\ \mathrm{kg/m^3}$\\
        *Einstein de-Haas frequency $\omega_\mathrm I$ & $3.3\ \mathrm{Hz}$\\
        *Magnetization frequency scale $\omega_\mathrm s$ (Neodymium magnet) & $1.5\times 10^{11} \ \mathrm{Hz}$\\
        *Temperature $T$ & $4.0\ \mathrm{K}$\\
        *Dissipation rate $\gamma$ & $10^{-2} \ \mathrm{Hz}$\\
        *Energy resolution $\kappa$ & $1000\hbar$\\
        Trapping potential $V_{\theta\theta} = V_{\phi\phi}$ & $10^{-13}\ \mathrm{J}$\\
        Ferromagnet number $N$ & $10^6$\\
        Lattice constant $a_0$ & $2000\ \upmu\mathrm{m}$\\
        Pickup coil radius $R_\mathrm{p}$ & $200\ \mathrm{mm}$\\
        *Pickup coil wire radius $a_\mathrm{p}$ & $0.1\ \mathrm{mm}$\\
        *SQUID inductance $L_\mathrm{s}$ & $80\ \mathrm{pH}$\\
        *Intermediate coil inductance $L_\mathrm{i}$ & $1.8\ \upmu\mathrm{H}$\\
        Coupling coefficient $\eta/\sqrt{L_\mathrm s}$ (without rescaling) & $8.0\times 10^{-9} \ \mathrm{\sqrt{J}}$\\
        Rescaled coupling coefficient $\eta_\mathrm{eff}/\sqrt{L_\mathrm s}$ & $2.2\times 10^{-11} \ \mathrm{\sqrt{J}}$
    \end{tabular}
\end{ruledtabular}
\label{tab:params}
\end{table}

\begin{figure}[ht]
    \centering
    \includegraphics[width=0.9\linewidth]{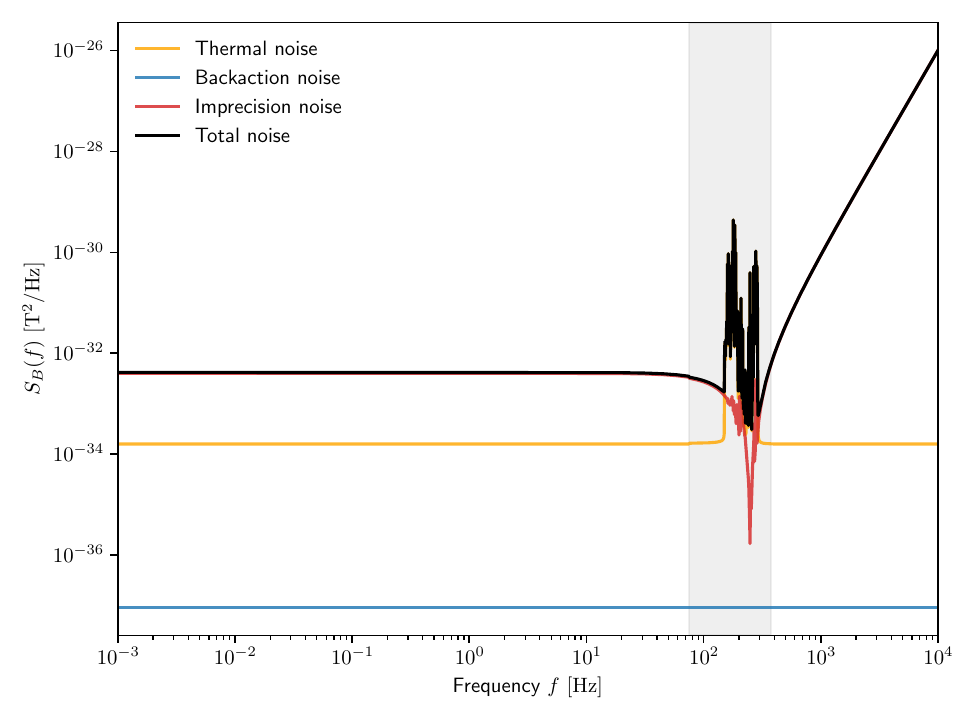}
    \caption{Magnetic-field noise power spectral density $S_{B}(f)$ for the ferromagnet ensemble. The total noise (black line) is shown together with its individual contributions: thermal noise, backaction noise, and imprecision noise. The imprecision noise is evaluated using the corrected coherent susceptibility $\chi_\mathrm{coh}(\omega)$, resulting in a steep rise at high frequencies. We mark the blind zone as the gray area, within which interaction severely reshapes the noise spectrum. At most frequencies, the noise is dominated by the imprecision noise, while at the blind zone the thermal noise sets the sensitivity. For the benchmark configuration, $v_{\theta\theta}=v_{\phi\phi}$, so the two bare libration frequencies coincide and only a single resonance appears.}
    \label{fig:noise_spectrum}
\end{figure}

Having established the general noise budget of the lattice magnetometer, we now specify a representative set of benchmark parameters and evaluate the corresponding noise spectra. \rev{This platform-independent reference configuration is chosen to keep $\lambda_{\mathrm{int}}$ moderate, retain a finite macroscopic lattice size, and allow direct comparison with the existing single-ferromagnet SQUID benchmark. We first identify the dominant scaling relations that govern the parameter dependence and then choose parameters that isolate the collective scaling and interaction-induced limitations. A complete experimental optimization will additionally include technical noise sources and constraints specific to the chosen levitation, trapping, shielding, and readout architecture.}

The most important constraint comes from the interaction parameter \(\lambda_{\mathrm{int}}\). This parameter characterizes the strength of the residual dipole-dipole interaction relative to the trapping curvature, and must remain sufficiently small for the perturbative treatment of the interaction-induced mode mixing to remain valid. For fixed \(\lambda_{\mathrm{int}}\), the trapping curvature is constrained by
\begin{equation}
V_{\theta\theta}
=
\frac{\omega_s}{\gamma_e\lambda_{\mathrm{int}}}
\frac{R^3}{a_0^3}\mu
\propto R^6 a_0^{-3},
\label{eq:Vtt_lambda_scaling}
\end{equation}
where \(R\) is the radius of each ferromagnet and \(a_0\) is the lattice constant. Therefore, once the single-particle design and the target value of \(\lambda_{\mathrm{int}}\) are fixed, \(a_0\) and \(V_{\theta\theta}\) can no longer be varied independently.

Before choosing explicit numerical values, it is useful to understand the resulting scaling of the dominant noise contribution. In the sparse and macroscopic lattice configurations considered here, the leading readout limitation is typically the imprecision noise. Physically, this is because the pickup coil must enclose the full array, so the geometric coupling to each individual ferromagnet becomes weaker as the system size increases. The equivalent magnetic-field imprecision noise scales as
\begin{equation}
S_B^{\mathrm{imp}}
\propto
\frac{V_{\theta\theta}^2}{\left(N\mu\eta_{\mathrm{eff}}\right)^2},
\label{eq:Simp_scaling_basic}
\end{equation}
where \(N\mu\) is the coherent magnetic moment of the lattice, and \(\eta_{\mathrm{eff}}\) denotes the effective readout coupling after including both the geometric pickup factor and the inductive rescaling from the SQUID circuit. As discussed earlier, the coupling of the pickup coil to different momentum modes is determined only by the ratio \(R_{\mathrm p}/L_{\mathrm{lat}}\), rather than by the absolute values of \(R_{\mathrm p}\) and \(L_{\mathrm{lat}}\) separately. One may therefore enlarge the system while preserving the readout mode profile, provided that this ratio is kept fixed. Motivated by this observation, we keep \(R_{\mathrm p}/L_{\mathrm{lat}}\) constant in the scaling analysis, so that changing the overall system size does not alter which lattice modes are preferentially read out. The lattice size is
\begin{equation}
L_{\mathrm{lat}}\sim N^{1/3}a_0.
\end{equation}
For a fixed coil-to-lattice aspect ratio, the effective coupling has the approximate scaling
\begin{equation}
\eta_{\mathrm{eff}}\propto R^3 L_{\mathrm{lat}}^{-\alpha},
\qquad
1\le \alpha \le 2,
\label{eq:eta_eff_scaling}
\end{equation}
where the two limiting cases correspond to different inductive regimes of the readout circuit. When the intermediate-coil inductance dominates over the pickup-coil inductance, \(L_{\mathrm i}\gg L_{\mathrm p}\), one finds approximately \(\alpha=1\). When the pickup-coil inductance dominates, \(L_{\mathrm i}\ll L_{\mathrm p}\), the coupling decreases faster with the system size, giving approximately \(\alpha=2\). More precisely, \(\eta_{\mathrm{eff}}\) also contains a weak logarithmic dependence on \(R_{\mathrm p}\), but this only produces a mild \(\ln(R_{\mathrm p})\) correction and does not affect the leading power-law behavior. We therefore neglect this logarithmic dependence in the following scaling estimate.

Combining Eqs.~\eqref{eq:Vtt_lambda_scaling} and \eqref{eq:eta_eff_scaling}, together with \(N\sim (L_{\mathrm{lat}}/a_0)^3\) and \(\mu\propto R^3\), gives
\begin{equation}
S_B^{\mathrm{imp}}
\propto
L_{\mathrm{lat}}^{2\alpha-6}.
\label{eq:Simp_power_law}
\end{equation}
Thus, along a family of designs with fixed \(\lambda_{\mathrm{int}}\) and fixed \(R_{\mathrm p}/L_{\mathrm{lat}}\), the leading imprecision-noise scaling is controlled primarily by the total lattice size. At this level, the explicit dependence on the microscopic parameters \(R\) and \(a_0\) cancels out. In the two limiting readout regimes,
\begin{equation}
S_B^{\mathrm{imp}}\propto L_{\mathrm{lat}}^{-4}
\quad
(L_{\mathrm i}\gg L_{\mathrm p}),
\qquad
S_B^{\mathrm{imp}}\propto L_{\mathrm{lat}}^{-2}
\quad
(L_{\mathrm i}\ll L_{\mathrm p}).
\end{equation}
Therefore, within this fixed-\(\lambda_{\mathrm{int}}\) scaling trajectory, enlarging the total lattice size improves the imprecision-limited sensitivity, even though the geometric coupling of an individual large pickup coil becomes weaker.

This conclusion should be distinguished from a different type of variation in which \(\lambda_{\mathrm{int}}\) itself is changed. If the trapping curvature and the single-particle parameters are held fixed, increasing the lattice spacing \(a_0\) reduces the dipole-dipole interaction strength and hence lowers \(\lambda_{\mathrm{int}}\). This is useful for suppressing interaction-induced mode mixing, but it also enlarges the physical size of a lattice with fixed \(N\), thereby weakening the overall pickup coupling. This statement refers to the tradeoff encountered when \(a_0\) is used to tune the interaction strength itself. It does not contradict the power law in Eq.~\eqref{eq:Simp_power_law}, which applies after \(\lambda_{\mathrm{int}}\) has already been fixed and the corresponding trapping curvature has been adjusted consistently.

The scaling analysis leads to the following parameter strategy. Since the leading imprecision-noise scaling is insensitive to the microscopic choice of \(R\) and \(a_0\) once \(\lambda_{\mathrm{int}}\) is fixed, we keep the single-ferromagnet parameters the same as those of the existing single-particle setup \cite{kaliaUltralightDarkMatter2024} in order to allow a direct comparison. The remaining design freedom is then used to choose a lattice spacing and trapping curvature that keep \(\lambda_{\mathrm{int}}\) moderate while keeping the total lattice size within a feasible macroscopic range. For this purpose, we fix
\begin{equation}
\lambda_{\mathrm{int}}=0.0665,
\end{equation}
and choose
\begin{equation}
a_0=100R=2000~\upmu\mathrm{m}.
\end{equation}
For a lattice containing \(N=10^6\) ferromagnets, this corresponds to a \(100\times100\times100\) simple-cubic lattice with total linear size
\begin{equation}
L_{\mathrm{lat}}=N^{1/3}a_0=200~\mathrm{mm}.
\end{equation}
This choice should be understood as a compromise between interaction control and experimental scale. A larger \(a_0\), at fixed trapping curvature, would further reduce \(\lambda_{\mathrm{int}}\), but would also enlarge the array for the same \(N\). Conversely, a smaller \(a_0\) would make the device more compact, but would require a larger trapping curvature in order to keep \(\lambda_{\mathrm{int}}\) fixed. The benchmark value above keeps the residual interaction sufficiently weak while avoiding an excessively large lattice.

The same scaling argument also suggests a straightforward route toward further improvement: one may continue to lower the noise by increasing the total number of ferromagnets and correspondingly enlarging the lattice size. In practice, however, a larger lattice also introduces more severe technical challenges for levitation, since the trapping and stabilization of a macroscopic three-dimensional array become increasingly demanding as the system size grows.

For the chosen \(a_0\) and target \(\lambda_{\mathrm{int}}\), we have
\begin{equation}
V_{\theta\theta}=V_{\phi\phi}=10^{-13}~\mathrm{J}.
\end{equation}
This value keeps the angular response relatively large while maintaining the interaction strength in the perturbative regime. \rev{As quantified in Eqs.~(\ref{eq:bias_curvature}) and (\ref{eq:benchmark_bias}), it corresponds to an engineered orientational confinement of approximately $4.3\ \upmu\mathrm T$ if supplied by a magnetic bias field, rather than to residual leakage through the shield.} The equality \(V_{\theta\theta}=V_{\phi\phi}\) is assumed for simplicity, so that the two transverse directions have the same bare trapping curvature.

For the collective readout, we choose the pickup-coil radius to be comparable to the lattice size,
\begin{equation}
R_{\mathrm p}=L_{\mathrm{lat}}=200~\mathrm{mm}.
\end{equation}
This implements the fixed-aspect-ratio assumption adopted in the scaling estimate above and ensures that the coil encloses the full lattice. \rev{The finite-profile calculation and the scan over $R_{\mathrm p}/L_{\mathrm{lat}}$ are presented in Appendix~\ref{app:pickup_coil_scan}.} The effective coupling to the SQUID is written as
\begin{equation}
\eta_{\mathrm{eff}}
=
\eta\frac{M_{\mathrm i}}{L_{\mathrm{tot}}},
\end{equation}
where \(\eta\) is the geometric pickup coupling, \(M_{\mathrm i}\) is the mutual inductance between the input coil and the SQUID, and \(L_{\mathrm{tot}}\) is the total inductance of the pickup and input circuit. \rev{In the present benchmark configuration, imprecision noise is the dominant readout-induced contribution, so Eq.~(\ref{eq:back_imp_ratio}) favors increasing $\eta_\mathrm{eff}$ within the accessible circuit range. Increasing the pickup-coil loop number raises the pickup inductance and reduces the effective coupling for the benchmark parameters. We therefore choose the minimal loop number,}
\begin{equation}
N_{\mathrm p}=1.
\end{equation}
\rev{The two coupling coefficients listed in Table~\ref{tab:params} imply $M_{\mathrm i}/L_{\mathrm{tot}}=2.73\times10^{-3}$ for this reference circuit. The geometric factor $\eta$, the noncoherent admixture, and the resulting imprecision noise retain explicit dependence on the finite pickup profile through Eqs.~(\ref{eq:pickup_reciprocity})--(\ref{eq:noncoherent_admixture}).}

The benchmark parameters used in the numerical calculations are summarized in Table~\ref{tab:params}. Parameters marked with an asterisk are inherited directly from Ref.~\cite{kaliaUltralightDarkMatter2024}, while the lattice spacing, trapping strength, pickup-coil radius, and readout configuration are chosen here according to the scaling considerations and interaction constraints described above.

With these parameters fixed, we evaluate the magnetic-field noise spectrum of the lattice. The thermal contribution is computed using the interaction-corrected expression derived in Sec.~V.B, including the boundary-induced mode mixing and the associated blind zone. The imprecision noise is obtained from the corrected coherent transfer function, while the backaction noise is determined directly by the collective pickup-circuit fluctuations. The resulting spectrum is shown in Fig.~\ref{fig:noise_spectrum}, where the thermal, imprecision, and backaction contributions are plotted separately together with their sum.

The resulting spectrum reflects the scaling picture discussed above. Over most of the accessible frequency range, the total noise is limited by imprecision noise, reflecting the weak effective coupling of the macroscopic pickup coil. The thermal noise remains subdominant away from the interaction-induced blind zone, where it follows the expected collective scaling. Inside the blind zone, however, boundary-mediated coupling to near-singular modes strongly amplifies the thermal contribution and sharply degrades the observable sensitivity. The backaction noise remains parametrically smaller throughout the benchmark configuration and does not set the final sensitivity floor.  A SQUID detector operating with near-quantum-limited noise, or a larger lattice containing $1000$ ferromagnets on each side, would be sufficient to suppress the imprecision noise below the thermal noise over a substantial part of the accessible band. In that regime, the sensitivity would no longer be set by the readout chain, but instead approach the intrinsic thermal limit of the lattice magnetometer.

\section{ULDM Detection}

\subsection{General signal model and SNR formalism}

We now project the magnetic-field sensitivity of the ferromagnet lattice magnetometer onto several well-motivated ultralight dark matter scenarios. Each dark matter candidate considered below induces an effective oscillating magnetic-like field acting on the spins of the ferromagnets. Once this effective field is specified, the corresponding signal-to-noise ratio can be obtained directly from the noise spectrum derived in Sec.~V. For all channels of interest, the dark matter field behaves as a coherently oscillating classical background on laboratory timescales. The signal can therefore be modeled as a narrowband effective magnetic field,
\begin{equation}
\mathbf B_\mathrm{ DM}(t)=\mathbf B_0 \cos(\omega_\mathrm{ DM} t+\varphi),
\label{eq:BDM_time}
\end{equation}
where $\omega_\mathrm{ DM}\simeq m_\mathrm{ DM}$ is set by the dark matter mass, $\mathbf B_0$ is the channel-dependent field amplitude, and $\varphi$ is an unobservable initial phase. Since the dark matter velocity distribution is nonrelativistic, the signal is not perfectly monochromatic but has a finite fractional bandwidth of order $v_\mathrm{ DM}^2$. Equivalently, the signal possesses a finite coherence time
\begin{equation}
t_\mathrm{ coh}\sim \frac{1}{m_\mathrm{ DM}v_\mathrm{ DM}^2},
\label{eq:tcoh_def}
\end{equation}
or, in frequency space, a bandwidth
\begin{equation}
\delta f_\mathrm{ DM}\sim \frac{1}{t_\mathrm{ coh}}.
\label{eq:dfdm_def}
\end{equation}

The magnetic-field sensitivity of the detector is described by the total noise spectral density $S_B(\omega)$ obtained in Sec.~V. For a signal oscillating at frequency $\omega_\mathrm{ DM}$, only the noise within the dark-matter bandwidth contributes appreciably to the measurement. The corresponding signal-to-noise ratio (SNR) may therefore be written as \cite{kaliaUltralightDarkMatter2024}
\begin{equation}
\mathrm{ SNR}
=
\frac{B_\mathrm{ DM}^2}{6}
\sqrt{\mathrm{ Tr}\!\left[S_B^{-2}(\omega_\mathrm{ DM})\right]
\,t_\mathrm{ int}\,\min(t_\mathrm{ int},t_\mathrm{ coh})},
\label{eq:SNR_general}
\end{equation}
where $B_\mathrm{ DM}=|\mathbf B_0|$ denotes the amplitude of the effective magnetic field, and $t_\mathrm{ int}$ is the total integration time. Here the trace is taken over the two transverse response directions of the ferromagnet. The SNR improves linearly with the coherent integration time, but only up to the dark-matter coherence time. For $t_\mathrm{ int}\gg t_\mathrm{ coh}$, the measurement consists of many incoherent time segments and the net sensitivity grows only through the number of such segments. The exclusion reach for a given dark matter model is obtained by requiring that the corresponding signal remain below the detection threshold. Throughout this work we adopt the criterion
\begin{equation}
\mathrm{ SNR}=3,
\label{eq:SNR_threshold}
\end{equation}
which defines the projected upper limit on the relevant coupling parameter at each dark matter mass. Once the channel-dependent relation between $B_\mathrm{ DM}$ and the microscopic coupling is specified, the magnetic-field reach of the detector can therefore be translated directly into a projected sensitivity curve.

\subsection{Axion-electron coupling}

We first consider an axion-like particle coupled to electrons through the derivative interaction
\begin{equation}
\mathcal L_{ae}
=
\frac{g_{ae}}{2m_e}\,
\partial_\mu a\,
\bar\psi_e\gamma^\mu\gamma^5\psi_e .
\label{eq:Lae}
\end{equation}
In the nonrelativistic limit, this interaction reduces to a coupling between the axion gradient and the electron spin, and may therefore be interpreted as an effective oscillating magnetic field acting on the ferromagnet,
\begin{equation}
\mathbf B_{ae}
=
\frac{g_{ae}}{\gamma_e m_e}\,\nabla a .
\label{eq:Bae_general}
\end{equation}
For axion dark matter, the field may be written locally as
\begin{equation}
a(t,\rev{\mathbf r})=a_\mathrm{DM}\cos(\omega_\mathrm{ DM} t-\mathbf k\!\cdot\!\rev{\mathbf r}+\varphi),
\end{equation}
with $\omega_\mathrm{ DM}\simeq m_a$ and $|\mathbf k|=m_a v_\mathrm{ DM}$. Using the relation between the oscillation amplitude and the local dark matter density,
\begin{equation}
\rho_\mathrm{ DM}\simeq \frac{1}{2}m_a^2 a_\mathrm{DM}^2,
\end{equation}
one obtains
\begin{equation}
|\nabla a|
\sim
m_a v_\mathrm{ DM} a_\mathrm{DM}
=
v_\mathrm{ DM}\sqrt{2\rho_\mathrm{ DM}}.
\end{equation}
The effective magnetic field amplitude is therefore
\begin{equation}
B_{ae}
\sim
\frac{g_{ae}}{\gamma_e m_e}\,
v_\mathrm{ DM}\sqrt{2\rho_\mathrm{ DM}} .
\label{eq:Bae_final}
\end{equation}

Several points are worth emphasizing. First, the signal amplitude in Eq.~(\ref{eq:Bae_final}) is set entirely by the dark matter background and does not depend on the ferromagnet lattice itself. In this channel, the lattice does not enhance the signal-generation mechanism. Its role is instead to improve the magnetic-field sensitivity through the collective noise reduction derived in Sec.~V. Accordingly, the projected improvement relative to a single ferromagnet arises solely from the reduced noise floor. Second, because the axion-wind signal is spatially uniform across the scale of the lattice, all ferromagnets experience the same effective magnetic field to an excellent approximation. The response therefore adds coherently across the array, and the noise spectrum obtained in Sec.~V can be applied directly without further geometric suppression.

The projected sensitivity to the axion-electron coupling is obtained by substituting Eq.~(\ref{eq:Bae_final}) into the general SNR relation, Eq.~(\ref{eq:SNR_general}), and imposing the threshold condition Eq.~(\ref{eq:SNR_threshold}). This gives the upper limit on $g_{ae}$ as a function of the axion mass $m_a$.  Compared with the existing single-ferromagnet setup, the lattice improves the projected reach over the full frequency range shown by about 3 orders of magnitude, reflecting the collective reduction of the effective noise.

\subsection{Dark photon kinetic mixing}

We next consider dark-photon dark matter. In the low-energy description relevant here, the dark photon $A'_\mu$ is kinetically mixed with the ordinary photon and behaves as a coherently oscillating background vector field. Its coupling to the electromagnetic sector can be written as
\begin{equation}
\mathcal L_{A'}
=
-\epsilon m_{A'}^2 A_\mu A'^\mu ,
\label{eq:LAp}
\end{equation}
where $\epsilon$ is the kinetic-mixing parameter and $m_{A'}$ is the dark-photon mass. The dark photon field then acts as an effective source for ordinary electromagnetism,
\begin{equation}
J^\mu_\mathrm{ eff}
=
\epsilon m_{A'}^2 A'^\mu .
\label{eq:Jeff_darkphoton}
\end{equation}

Inside a magnetically shielded apparatus, the electromagnetic response to this source is determined by the boundary conditions imposed by the shield. In the low-frequency regime relevant for the present experiment, the induced magnetic response is set parametrically by the characteristic size $L_\mathrm{sh}$ of the shielded region. Solving the induced field under these boundary conditions gives the effective magnetic-field amplitude
\begin{equation}
B_{A'}
\sim
\epsilon \sqrt{2\rho_\mathrm{ DM}}\,m_{A'}L_\mathrm{sh} .
\label{eq:Bdarkphoton}
\end{equation}
As in the axion-electron case, this signal is determined by the dark matter background together with the macroscopic shield geometry; it is not enhanced by the ferromagnet lattice itself. We stress that Eq.~(\ref{eq:Bdarkphoton}) is only an order-of-magnitude estimate of the effective magnetic field. A more precise expression would require a dedicated analysis of the electromagnetic response of the specific shield geometry.

The role of the lattice in the dark-photon channel is therefore again to reduce the detector noise rather than to modify the signal-generation mechanism. Since the induced magnetic-like field is approximately uniform across the lattice for $L_\mathrm{lat}\ll L_\mathrm{sh}$, all ferromagnets respond coherently and the collective readout formalism developed above remains applicable. The projected sensitivity to the kinetic-mixing parameter $\epsilon$ is then obtained similarly as before. The lattice configuration also improves the projected dark-photon sensitivity relative to the single-ferromagnet implementation across the mass range of interest by about 4 orders of magnitude, again due to the collective improvement of the detector noise performance.

\subsection{Axion-photon coupling}
\label{sec:axion_photon}

Finally, we consider the axion coupling to standard model photons,
\begin{equation}
\mathcal L_{a\gamma}
=
-\frac{1}{4}g_{a\gamma}aF_{\mu\nu}\tilde F^{\mu\nu},
\label{eq:Lag}
\end{equation}
where $g_{a\gamma}$ denotes the axion-photon coupling constant. In the presence of a background electromagnetic field, the axion induces an oscillatory electromagnetic response that may be described by an effective current,
\begin{equation}
J^\mu_\mathrm{ eff}
=
g_{a\gamma}\partial_\nu a\,\tilde F^{\mu\nu}.
\label{eq:Jeff_axionphoton}
\end{equation}
Unlike the axion-electron and dark-photon channels discussed above, the resulting signal depends explicitly on the ambient electromagnetic field and therefore on the experimental configuration itself. In the present setup, this background field is generated by the ferromagnet lattice, so that the detector does not merely probe the signal passively, but also participates directly in its generation.

Following \rev{Refs.}~\cite{higgins_maglev_2024,chaudhuri_radio_2015}, the axion-photon-induced magnetic field inside the shielded volume may be expanded in the cavity modes of the shield,
\begin{equation}
\mathbf B_{a\gamma}(\mathbf r)
=
\sum_n c_n \frac{f_n}{f_a}\mathbf B_n(\mathbf r)e^{-i m_a t},
\label{eq:Bag_mode}
\end{equation}
with
\begin{equation}
c_n
=
-\frac{\sqrt{\hbar c^3}\,g_{a\gamma}f_a^2 a_\mathrm{DM}}{f_n^2-f_a^2}
\int dV\,\mathbf E_n^*(\mathbf r)\cdot \mathbf B_0(\mathbf r),
\label{eq:cavity}
\end{equation}
\rev{where $\mathbf E_n$ and $\mathbf B_n$ are the electric and magnetic fields of the cavity modes, $f_n$ is the mode frequency, and $f_a=m_a/2\pi$ in natural units. When $m_a$ is quoted as an energy and $f_a$ as an SI frequency, the corresponding conversion is $f_a=m_a/(2\pi\hbar)$. The axion-field amplitude is}
\begin{equation}
a_\mathrm{DM}=\frac{\sqrt{2\rho_\mathrm{ DM}}}{m_a}.
\end{equation}
The cavity modes are normalized as
\begin{equation}
\int dV\,|\mathbf E_n(\mathbf r)|^2=1.
\end{equation}
The signal is therefore controlled by the overlap integral
\begin{equation}
I_n
=
\int dV\,\mathbf E_n^*(\mathbf r)\cdot \mathbf B_0(\mathbf r).
\label{eq:In_def}
\end{equation}
For the shield dimensions of interest, the relevant cavity frequencies satisfy $f_n\sim c/L_\mathrm{sh}\gg f_a$, so that Eq.~(\ref{eq:cavity}) simplifies to
\begin{equation}
\rev{c_n
\simeq
-\sqrt{\hbar c^3}\,g_{a\gamma}\frac{f_a^2}{f_n^2}a_\mathrm{DM} I_n.}
\label{eq:cn_lowf}
\end{equation}
The problem is thus reduced to estimating the scaling of $I_n$ with the ferromagnet number $N$.

The background magnetic field may be decomposed as
\rev{\begin{equation}
\mathbf B_0=\mu_0\left(\mathbf H_0+\mathbf M\right),
\end{equation}}
where $\mathbf M$ is the magnetization of the ferromagnets. Since there is no free current, one has $\nabla\times \mathbf H_0=0$, and hence
\begin{equation}
\mathbf H_0=\nabla\Psi_0
\end{equation}
for a magnetic scalar potential $\Psi_0$. The overlap integral then separates into two pieces,
\rev{\begin{equation}
I_n
=
\mu_0\int dV\,\mathbf E_n^*\cdot \nabla\Psi_0
+
\mu_0\int dV\,\mathbf E_n^*\cdot \mathbf M.
\end{equation}}
Using integration by parts and $\nabla\cdot \mathbf E_n=0$ inside the cavity volume, the first term reduces to a surface contribution,
\rev{\begin{equation}
I_n
=
\mu_0\int d\mathbf S\cdot \mathbf E_n^*\,\Psi_0
+
\mu_0\int dV\,\mathbf E_n^*\cdot \mathbf M.
\label{eq:In_split}
\end{equation}}

We first consider the surface term. The scalar potential is generated by the array of magnetic dipoles and modified by the shield. Since both the shield size and the minimum distance from the ferromagnets to the shield are much larger than the size of each individual ferromagnet, the shield-induced correction to the scalar potential is subleading for the scaling argument, and one may approximate
\begin{equation}
\Psi_0(\mathbf r)
=
\sum_i
\frac{\boldsymbol\mu\cdot(\mathbf r-\mathbf r_i)}
{4\pi |\mathbf r-\mathbf r_i|^3},
\label{eq:Psi0_exact}
\end{equation}
where $\mathbf r_i$ denotes the position of the $i$th ferromagnet and all dipoles are assumed aligned with the same magnetic moment $\boldsymbol\mu$. On the shield surface, where $|\mathbf r|\sim L_\mathrm{sh}\gg |\mathbf r_i|$, this may be expanded about the geometric center of the array. Choosing the origin such that
\begin{equation}
\sum_i \mathbf r_i=0,
\end{equation}
the dipole expansion gives
\begin{equation}
\Psi_0(\mathbf r)\big|_\mathrm{ shield}
=
\frac{N\,\boldsymbol\mu\cdot \mathbf r}{4\pi |\mathbf r|^3}
\left[
1+\mathcal O\!\left(\frac{L_\mathrm{lat}^2}{L^2_\mathrm{sh}}\right)
\right].
\label{eq:Psi0_surface}
\end{equation}
Thus, to leading order, the surface contribution is identical to that of a single dipole carrying the effective magnetic moment
\begin{equation}
\boldsymbol\mu_\mathrm{ eff}=\sum_i \boldsymbol\mu_i=N\boldsymbol\mu.
\label{eq:mueff}
\end{equation}

We next consider the magnetization term in Eq.~(\ref{eq:In_split}). Since each ferromagnet is much smaller than the shield and the relevant low-frequency cavity modes vary only on the scale of $\rev{L_\mathrm{sh}}$, the electric field of a given cavity mode is approximately constant across each individual ferromagnet. The volume integral may therefore be approximated as
\rev{\begin{equation}
\mu_0\int dV\,\mathbf E_n^*(\mathbf r)\cdot \mathbf M
\approx
\mu_0\boldsymbol\mu\cdot \sum_i \mathbf E_n^*(\mathbf r_i).
\end{equation}}
If the array is centrally symmetric and satisfies $L_\mathrm{lat}\ll L_\mathrm{sh}$, the cavity field may be expanded about the array center,
\begin{equation}
\mathbf E_n(\mathbf r_i)
=
\mathbf E_n(0)
+
r_i^a\partial_a \mathbf E_n(0)
+
\frac{1}{2}r_i^a r_i^b \partial_a\partial_b \mathbf E_n(0)
+\cdots .
\end{equation}
The linear term vanishes after summing over the array because $\sum_i \mathbf r_i=0$, leaving
\begin{equation}
\sum_i \mathbf E_n^*(\mathbf r_i)
=
N\,\mathbf E_n^*(0)
+
\mathcal O\!\left(N\,L_\mathrm{lat}^2 \partial^2 \mathbf E_n^*(0)\right).
\end{equation}
Since the cavity-mode variation scale is set by $L_\mathrm{sh}$, one has parametrically $\partial^2 \mathbf E_n\sim \mathbf E_n/L_\mathrm{sh}^2$, and therefore
\rev{\begin{equation}
\mu_0\int dV\,\mathbf E_n^*\cdot \mathbf M
\approx
N\mu_0\boldsymbol\mu\cdot \mathbf E_n^*(0)
\left[
1+\mathcal O\!\left(\frac{L_\mathrm{lat}^2}{L_\mathrm{sh}^2}\right)
\right].
\label{eq:magterm_scaling}
\end{equation}}

Equations~(\ref{eq:Psi0_surface}) and (\ref{eq:magterm_scaling}) show that both contributions to $I_n$ scale linearly with $N$ as long as the array remains small compared with the spatial variation scale of the relevant cavity modes. The overlap integral therefore takes the form
\begin{equation}
I_n
=
N\, I_n^{(1)}
\left[
1+\mathcal O\!\left(\frac{L_\mathrm{lat}^2}{L_\mathrm{sh}^2}\right)
\right],
\label{eq:In_scaling}
\end{equation}
where $I_n^{(1)}$ is the corresponding single-ferromagnet overlap. Substituting Eq.~(\ref{eq:In_scaling}) into Eq.~(\ref{eq:cn_lowf}) then implies that the cavity-mode coefficients scale as
\begin{equation}
c_n=N\, c_n^{(1)}
\left[
1+\mathcal O\!\left(\frac{L_\mathrm{lat}^2}{L_\mathrm{sh}^2}\right)
\right].
\end{equation}
The axion-induced magnetic field experienced by the lattice is therefore enhanced by the same factor,
\begin{equation}
B_{a\gamma}
\sim
N B_{a\gamma}^{(1)},
\label{eq:Bag_N_scaling}
\end{equation}
up to an overall geometric factor of order unity that depends on the detailed array geometry and the cavity-mode structure.

Using the single-ferromagnet estimate of Refs.~\cite{higgins_maglev_2024,kaliaUltralightDarkMatter2024}, one finally obtains
\begin{equation}
\rev{B_{a\gamma}
\sim
\mathcal O(0.1)\,N\,
g_{a\gamma}\sqrt{2\hbar c\mu_0\rho_\mathrm{ DM}}
\frac{|\mu|}{L_\mathrm{sh}^2}.}
\label{eq:Bag_lattice}
\end{equation}
\rev{Equation~(\ref{eq:Bag_lattice}) gives an order-of-magnitude estimate whose coefficient is set by the shield geometry, the array placement, and the cavity-mode overlap. The factor denoted by $\mathcal O(0.1)$ is inherited from the representative single-ferromagnet estimate. The robust content of the analytical calculation is the possible coherent enhancement at the level of $N$ when the relevant modes vary slowly across the array. The representative numerical calculation below tests this relative enhancement against a single ferromagnet in the same shield, while the absolute normalization remains fixed by the geometry-dependent overlap in Eq.~(\ref{eq:Bag_lattice}).}

\rev{The enhancement in Eq.~(\ref{eq:Bag_lattice}) arises from the low-frequency cavity modes that vary only on the scale of the shield size $L_\mathrm{sh}$.} For these modes, the lattice is effectively point-like, and both contributions to the overlap integral add coherently up to corrections of order $L_\mathrm{lat}^2/L_\mathrm{sh}^2$. \rev{Higher-frequency cavity modes entering the mode sum in Eq.~(\ref{eq:Bag_mode}), especially in the polarization term of Eq.~(\ref{eq:In_split}), can vary across the lattice. Different ferromagnets then sample different phases of the mode function, and the contributions to $\sum_i \mathbf E_n(\mathbf r_i)$ can partially cancel rather than scale linearly with $N$. At the level of the present order-of-magnitude treatment, we retain the low-frequency contribution for which the array is unresolved and write}
\begin{equation}
B_{a\gamma}\propto N .
\end{equation}
\rev{We test this scaling in a representative cylindrical ideal-conductor geometry, as detailed in Appendix~\ref{app:axion_photon_cylinder}. For a cylinder of diameter and height $1~\mathrm m$, with the benchmark $N=10^6$ lattice displaced by $0.15~\mathrm m$ from the cylinder axis, the coherent magnetic-field amplitude is enhanced by $9.9673\times10^5$ relative to an identical single ferromagnet at the same location. Equivalently, the enhancement per ferromagnet is $0.99673$, only $0.327\%$ below exact $N$ scaling. This result supports the theoretical scaling estimate for this specified geometry, while the vanishing coherent response of the centered symmetry-control configuration illustrates why the absolute response is not geometry independent.}

\rev{The projected sensitivity to $g_{a\gamma}$ is obtained by inserting the representative normalization in Eq.~(\ref{eq:Bag_lattice}) into the general SNR formula, Eq.~(\ref{eq:SNR_general}), and solving for the value of $g_{a\gamma}$ that saturates the threshold condition Eq.~(\ref{eq:SNR_threshold}) at each axion mass. In the axion-photon channel, the lattice benefits from the collective reduction of the effective noise and may also receive the coherent signal enhancement described above. With the representative geometry factor adopted in Eq.~(\ref{eq:Bag_lattice}), the benchmark curve improves the single-ferromagnet projection by about seven orders of magnitude. This order-of-magnitude improvement is tied to the geometry assumption entering the overlap coefficient.}

\subsection{Sensitivity projections}

\begin{figure*}[t]
    \begin{tabular}{ccc}
    \includegraphics[width=0.33\linewidth]{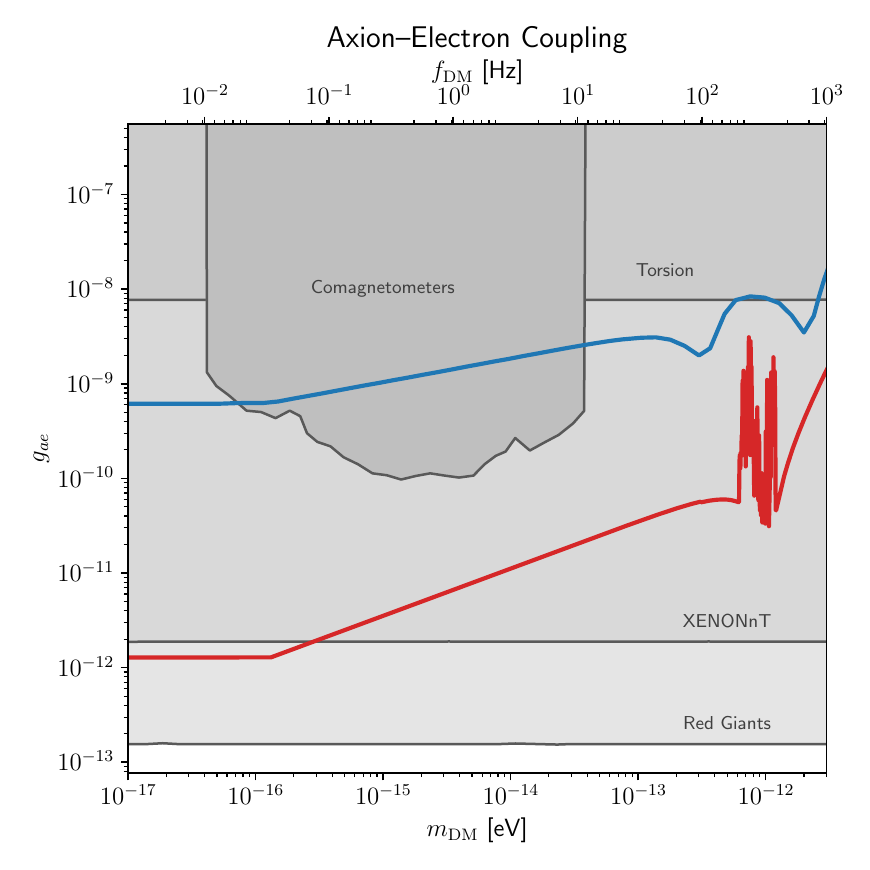} & 
    \includegraphics[width=0.33\linewidth]{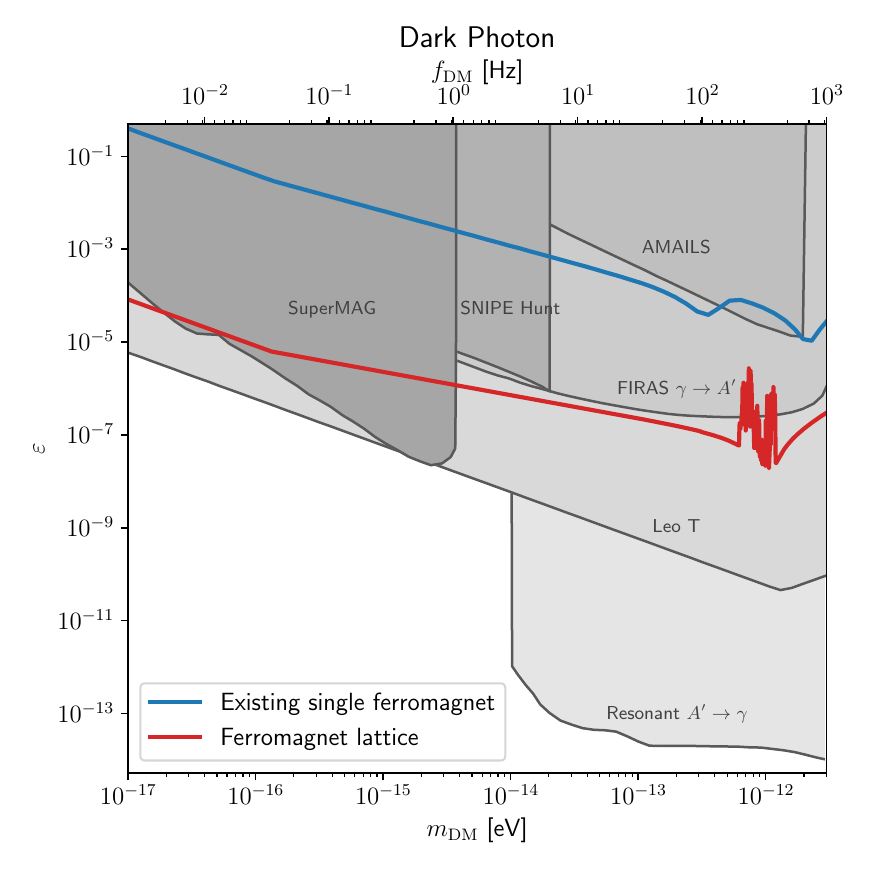} &
    \includegraphics[width=0.33\linewidth]{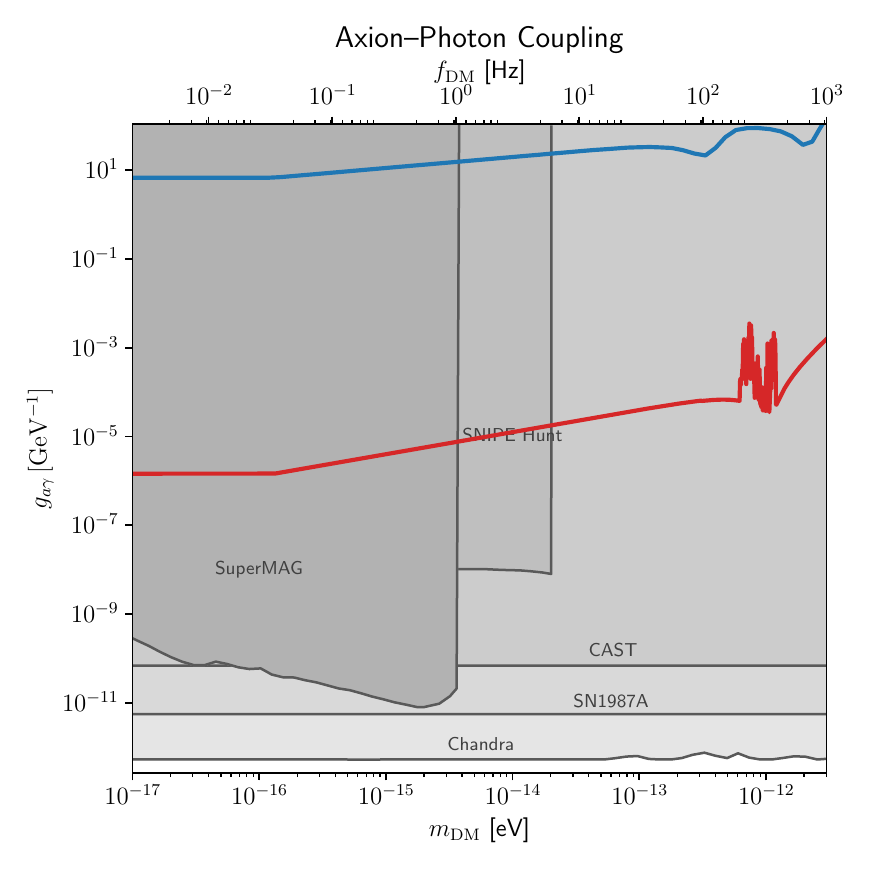}    \\
        (a)&(b)&(c)
    \end{tabular}
    \caption{\rev{Projected sensitivities derived for the reference ferromagnet-lattice benchmark are shown for axion-electron, dark-photon, and axion-photon couplings. The solid red curves show the projections obtained in this work. Shaded regions indicate existing constraints from the literature \cite{caputo_dark_2021-1,AxionLimits}, and the blue curves denote the single-ferromagnet benchmark of Ref.~\cite{kaliaUltralightDarkMatter2024}. We take $L_\mathrm{sh}=1\ \mathrm{m}$ and otherwise use the single-ferromagnet parameters specified in the text. The lattice improves the projected reach in all three cases under these assumptions. The axion-photon curve additionally uses the representative, geometry-dependent normalization in Eq.~(\ref{eq:Bag_lattice}) and is an order-of-magnitude projection.}}
    \label{fig:DMlimits}
\end{figure*}

We now combine the benchmark noise spectrum obtained in Sec.~V with the effective magnetic fields derived above and convert the magnetic-field reach into projected sensitivities to the corresponding ULDM couplings. In all cases, we assume a local dark matter density $\rho_\mathrm{ DM}$, an integration time
\begin{equation}
t_\mathrm{ int}=1~\mathrm{ yr},
\end{equation}
and the detection threshold $\mathrm{ SNR}=3$. The resulting exclusion curves are obtained by evaluating Eq.~(\ref{eq:SNR_general}) at each dark matter mass and solving for the coupling value that saturates Eq.~(\ref{eq:SNR_threshold}). The results are shown in \rev{Fig.}~\ref{fig:DMlimits}.

For the axion-electron and dark-photon channels, the improvement relative to a single levitated ferromagnet originates entirely from the reduced noise floor of the lattice magnetometer. Since the effective fields in Eqs.~(\ref{eq:Bae_final}) and (\ref{eq:Bdarkphoton}) are not enhanced by the lattice itself, the projected reach tracks directly the improvement in magnetic-field sensitivity derived in Sec.~V. In both cases, the gain is most significant in the low-frequency regime where the total noise is thermally limited and the collective readout suppresses the imprecision contribution most efficiently.

\rev{For the representative normalization in Eq.~(\ref{eq:Bag_lattice}), the axion-photon channel benefits more strongly from increasing ferromagnet number because the lattice lowers the noise floor and also enhances the signal through the collective ambient field. Its improvement relative to the single-ferromagnet baseline is therefore more pronounced than in the axion-electron and dark-photon channels. Its quantitative reach remains conditioned on the assumed electromagnetic overlap.}

\rev{Overall, the ferromagnet lattice improves the projected ULDM sensitivity in all three reference scenarios considered here.} The common gain comes from collective readout and the resulting reduction of the effective noise floor. \rev{With the representative electromagnetic normalization, the axion-photon case shows the largest relative improvement over the single-ferromagnet baseline because the signal itself also scales favorably with the number of ferromagnets. The robust conclusion is the collective improvement common to the three channels; the absolute axion-photon curve remains an order-of-magnitude estimate. The projections shown here use the current SQUID noise benchmark. If the SQUID performance could be improved to the quantum limit, the projected sensitivities would be further enhanced by approximately two orders of magnitude. Under the remaining benchmark assumptions, the axion-electron and dark-photon projections could then approach or exceed existing bounds in part of the accessible frequency range.}

\section{Conclusion}

In this work, we investigated a ferromagnet lattice as a platform for ULDM detection. By replacing a single levitated ferromagnet with a coherently read-out array, the detector achieves improved sensitivity without significantly modifying the intrinsic single-particle dynamics. The central idea is to exploit collective response and readout to enhance the effective signal-to-noise ratio while retaining a well-controlled microscopic description.

We developed a theoretical framework that separates the problem into three layers. At the single-particle level, the rotational dynamics are described by a well-defined susceptibility in the fully trapped regime. At the lattice level, dipole-dipole interactions modify this response through a momentum-dependent kernel, while finite-size effects introduce boundary-induced mode mixing. These effects can be systematically incorporated as a self-energy correction to the coherent ($k=0$) channel. This formulation makes it possible to identify the conditions under which the collective mode dominates, as well as the regimes where near-singular modes lead to strong deviations and define the blind-zone structure of the detector response.

On top of this dynamical framework, we analyzed the noise properties of the lattice detector. Thermal, imprecision, and backaction noise scale differently with the number of ferromagnets, and their interplay determines the final sensitivity. Collective readout reduces the effective noise floor, while interaction-induced mode mixing leads to a frequency-dependent modification of the thermal noise spectrum. Away from the blind-zone regime, the detector performance is well captured by an effective coherent-channel description, which provides a transparent connection between microscopic dynamics and macroscopic sensitivity.

\rev{Using this framework, we derived benchmark-dependent projected sensitivities for axion-electron, dark-photon, and axion-photon couplings. Under the reference assumptions, the lattice configuration improves the reach relative to the single-ferromagnet implementation, reflecting the collective reduction of the effective noise. In the axion-photon channel, the signal may additionally benefit from an order-$N$ coherent enhancement associated with the lattice-induced electromagnetic background. A representative calculation for a $1~\mathrm m$ cylindrical ideal-conductor shield gives an enhancement of $9.9673\times10^5$ for the $N=10^6$ benchmark, supporting the coherent scaling estimate. The absolute normalization of the projected curve nevertheless remains dependent on the shield geometry and array placement.}

\rev{The projections presented here are based on current SQUID noise benchmarks and on the reference theoretical configuration specified in Table~\ref{tab:params}. If the SQUID performance can be improved to the quantum limit, the noise spectrum would be further reduced by three orders of magnitude, corresponding to an additional improvement of approximately two orders of magnitude in the ULDM limit. Under the remaining benchmark assumptions, the axion-electron and dark-photon channels could approach or extend beyond existing bounds over part of the accessible frequency range. Near-quantum-limited SQUID amplification has already been demonstrated experimentally, with microwave SQUID amplifiers developed for ADMX reaching noise temperatures within a factor of about 1.4 of the quantum limit \cite{asztalos_squid-based_2010-2}. This limit is the fundamental added-noise bound for phase-preserving linear amplification, corresponding to one-half quantum of input-referred added noise in the high-gain regime \cite{clerk_introduction_2010}.}

Several directions for further improvement are suggested by the present analysis. First, the simple pickup-coil geometry considered here leads to a coupling that decreases with increasing lattice size, which becomes a limiting factor for large arrays. Optimized coil designs, or alternative readout schemes that are less sensitive to the spatial extent of the lattice, could significantly enhance the effective coupling and enable further scaling of the system size. Second, compared with the single-particle case, the lattice configuration requires the trapping potential to exceed the interaction scale, which limits the possibility of improving sensitivity by reducing the trap stiffness. If the effective interaction scale can be suppressed through dynamical control, geometric design, or other engineering approaches, the sensitivity could be substantially improved beyond the regime explored in this work.

\rev{Overall, the ferromagnet lattice provides a scalable detector architecture for precision sensing of ultralight dark matter. The collective-response and noise framework developed here is independent of a specific implementation, whereas the numerical reach is conditional on the reference trapping and readout assumptions. Translating this benchmark into an experimental design will require architecture-specific studies of trapping, angular stability, magnetic shielding, readout coupling, and technical noise.}

\clearpage
\appendix

\section{\texorpdfstring{\rev{Finite-lattice numerical calculations}}{Finite-lattice numerical calculations}}
\label{app:finite_lattice_numerics}

\subsection{\texorpdfstring{\rev{Open-boundary interaction kernel}}{Open-boundary interaction kernel}}

\rev{We describe here the numerical procedure used for the finite simple-cubic lattice in Secs.~IV and V. Let the lattice contain $N_\ell^3$ sites, with integer coordinates $\mathbf n=(n_x,n_y,n_z)$ and $0\leq n_\alpha<N_\ell$. In units of the lattice constant, the two-dimensional transverse dipolar kernel for a displacement $\mathbf d=\mathbf n-\mathbf m$ is}
\rev{\begin{equation}
T_{ab}(\mathbf d)
=
\frac{d^2\delta_{ab}-3d_a d_b}{|\mathbf d|^5},
\qquad
a,b\in\{x,y\},
\label{eq:app_realspace_kernel}
\end{equation}}
\rev{with $T_{ab}(\mathbf 0)=0$. Its action on a lattice field is the open-boundary convolution}
\rev{\begin{equation}
(T^{\mathrm{finite}}u)_a(\mathbf n)
=
\sum_{\mathbf m\in D_{N_\ell}}
\sum_b
T_{ab}(\mathbf n-\mathbf m)u_b(\mathbf m),
\label{eq:app_open_convolution}
\end{equation}}
\rev{where $D_{N_\ell}$ denotes the finite cube. We evaluate Eq.~(\ref{eq:app_open_convolution}) by embedding both the kernel and the input field in a zero-padded array and performing the convolution with fast Fourier transforms. Choosing the padding length $P\geq2N_\ell-1$ removes the circular wraparound and is therefore equivalent to the direct open-boundary sum, up to floating-point precision. The benchmark calculation uses $N_\ell=100$, corresponding to $N=10^6$, and $P=256$. This reduces the memory requirement from that of a dense $N\times N$ matrix to a small number of $P^3$ arrays.}

\rev{The diagonal reference kernel is defined operationally as $T^{\mathrm{diag}}(\mathbf k)=T^{\mathrm{finite}}_{\mathbf k\mathbf k}$. It can be evaluated without constructing the full reciprocal-space matrix. For the cubic window, the number of ordered site pairs with displacement $\mathbf d$ gives}
\rev{\begin{equation}
T^{\mathrm{diag}}_{ab}(\mathbf k)
=
\sum_{\mathbf d\neq0}
\left[
\prod_{\alpha=x,y,z}
\left(1-\frac{|d_\alpha|}{N_\ell}\right)
\right]
T_{ab}(\mathbf d)e^{-i\mathbf k\cdot\mathbf d},
\label{eq:app_weighted_diagonal}
\end{equation}}
\rev{where $-(N_\ell-1)\leq d_\alpha\leq N_\ell-1$. The boundary matrix is then defined by}
\rev{\begin{equation}
V_{\mathbf k\mathbf q}
=
T^{\mathrm{finite}}_{\mathbf k\mathbf q}
-\delta_{\mathbf k\mathbf q}T^{\mathrm{diag}}(\mathbf k),
\label{eq:app_boundary_definition}
\end{equation}}
\rev{so that $V_{\mathbf k\mathbf k}=0$ by construction. To obtain $V_{0\mathbf k}$, we apply the open-boundary convolution to the uniform mode and Fourier transform the result over the finite cube. Matrix elements $V_{\mathbf s\mathbf s'}$ required within a selected subspace are computed only when needed: a plane wave $\mathbf s'$ is used as the input to Eq.~(\ref{eq:app_open_convolution}), and the output is projected onto $\mathbf s$. Thus, all $10^6$ diagonal modes and all $V_{0\mathbf k}$ are retained, whereas the full dense matrix $V_{\mathbf k\mathbf q}$ is never stored.}

\subsection{\texorpdfstring{\rev{Finite-lattice static angular Hessian}}{Finite-lattice static angular Hessian}}
\label{app:static_angular_hessian}

\rev{For the finite-lattice stability check in Sec.~IV.D, we form the Hessian of the total static energy with respect to the two transverse angular displacements of every ferromagnet, evaluated about the uniformly aligned reference configuration. The open boundary generates a nonzero first derivative of the dipolar energy at some sites. We cancel these static torques by local linear compensation so that the reference configuration is a stationary point; because the compensation is linear in the angular displacement, it does not modify the Hessian or its eigenvalues.}

\rev{The action of the full $2N_\ell^3\times2N_\ell^3$ Hessian is evaluated without constructing a dense matrix. The open-boundary dipolar sum is implemented by the same zero-padded FFT convolution described above, with no real-space cutoff, and the lowest eigenvalues are obtained with an iterative symmetric eigensolver. Normalizing the Hessian to the isotropic trapping curvature, the minimum eigenvalues for $N_\ell=20$, $40$, $60$, $80$, and $100$ are $0.46911$, $0.45580$, $0.45156$, $0.44945$, and $0.44818$, respectively. The benchmark value at $N_\ell=100$ is positive and corresponds to $4.48\times10^{-14}\,\mathrm{J}$. It lies $1.19\%$ above the long-wavelength bulk estimate $1-(8\pi/3)\lambda_{\mathrm{int}}=0.44289$. The lowest finite-lattice eigenspace is localized predominantly on the outer faces, so the finite-open result is a direct stability check rather than an identification of the softest finite-lattice mode with a bulk plane wave.}

\rev{For validation, the FFT Hessian action and its lowest spectrum were compared with direct dense calculations for small lattices, and the energy curvature was independently checked by finite differences. The benchmark eigenvalue was also unchanged, at the quoted precision, under tighter eigensolver tolerances and a larger zero-padding array. These checks confirm that the positive minimum in Eq.~(\ref{eq:finite_angular_stability}) is stable against the matrix-free implementation and the padding choice.}

\subsection{\texorpdfstring{\rev{Hybrid treatment of near-singular modes}}{Hybrid treatment of near-singular modes}}

\rev{The analytical discussion in the main text uses the oscillation-dominated approximation within the fully trapped regime and neglects the antisymmetric precessional term in Eq.~(\ref{eq:chi}). This approximation is sufficient away from resonance. In the numerical inversion, the full susceptibility in Eq.~(\ref{eq:chi}) is restored near resonance in order to obtain the correct response. The coefficient of the antisymmetric term is fixed entirely by the benchmark parameters specified in the text and Table~\ref{tab:params}. With the resulting dimensionless inverse susceptibility $\xi(\bar\omega,\mathbf k)$, we first evaluate the perturbative self-energy over every nonzero momentum mode,}
\rev{\begin{equation}
\sigma_{\mathrm{bnd}}^{(0)}(\bar\omega)
=
\lambda_{\mathrm{int}}^2
\sum_{\mathbf k\neq0}
V_{0\mathbf k}
\xi^{-1}(\bar\omega,\mathbf k)
V_{\mathbf k0}.
\label{eq:app_sigma_initial}
\end{equation}}
\rev{This expression is computed in vectorized form over all $N_\ell^3-1$ nonuniform modes. A coupled subspace treatment is activated only when $\left|\mathrm{Tr}\,\sigma_{\mathrm{bnd}}^{(0)}/2\right|>c_1$. Within an activated frequency point, the modes are ranked using the coupling-weighted singularity criterion}
\rev{\begin{equation}
\left|\det\xi(\bar\omega,\mathbf k)\right|
<
c_2 w_{\mathbf k},
\qquad
w_{\mathbf k}
=
\|V_{0\mathbf k}\|_{\mathrm F}^2.
\label{eq:app_mode_selection}
\end{equation}}
\rev{The weight $w_{\mathbf k}$ prevents a nearly singular but negligibly coupled mode from entering the explicit subspace. For the results shown in the main text, $c_1=0.1$, $c_2=0.05$, and no more than $64$ modes are retained at any frequency. These quantities are numerical selection parameters rather than physical detector parameters.}

\rev{The selected nonzero modes are combined with the coherent mode into $\mathcal S$, so that $0\in\mathcal S$. On the nonzero sector $\mathcal S\setminus\{0\}$, we construct the $2(|\mathcal S|-1)\times2(|\mathcal S|-1)$ block matrix}
\rev{\begin{equation}
\left[G_{\mathcal S}^{-1}(\bar\omega)\right]_{\mathbf s\mathbf s'}
=
\xi(\bar\omega,\mathbf s)\delta_{\mathbf s\mathbf s'}
-\lambda_{\mathrm{int}}V_{\mathbf s\mathbf s'},
\qquad
\mathbf s,\mathbf s'\in\mathcal S\setminus\{0\},
\label{eq:app_subspace_green}
\end{equation}}
\rev{using the on-demand plane-wave projections described above. After inverting this matrix, the self-energy is evaluated as the sum of the coupled near-singular contribution and the perturbative remainder,}
\rev{\begin{equation}
\begin{split}
\sigma_{\mathrm{bnd}}(\bar\omega)
=
\lambda_{\mathrm{int}}^2
\sum_{\mathbf s,\mathbf s'\in\mathcal S\setminus\{0\}}
V_{0\mathbf s}
\left[G_{\mathcal S}(\bar\omega)\right]_{\mathbf s\mathbf s'}
V_{\mathbf s'0}
+
\lambda_{\mathrm{int}}^2
\sum_{\mathbf k\notin\mathcal S}
V_{0\mathbf k}
\xi^{-1}(\bar\omega,\mathbf k)
V_{\mathbf k0}.
\end{split}
\label{eq:app_hybrid_sigma}
\end{equation}}
\rev{The thermal-noise correction is evaluated with the same partition. Defining the block row $B_{\mathcal S}$ by concatenating $V_{0\mathbf s}$ for $\mathbf s\in\mathcal S\setminus\{0\}$, the matrix multiplying the single-particle thermal noise is}
\rev{\begin{equation}
\begin{split}
M(\bar\omega)
=
\mathbbm 1
&+
\lambda_{\mathrm{int}}^2
B_{\mathcal S}
G_{\mathcal S}G_{\mathcal S}^\dagger
B_{\mathcal S}^\dagger
\\
&+
\lambda_{\mathrm{int}}^2
\sum_{\mathbf k\notin\mathcal S}
V_{0\mathbf k}
\xi^{-1}(\bar\omega,\mathbf k)
\xi^{-1}(\bar\omega,\mathbf k)^\dagger
V_{\mathbf k0}.
\end{split}
\label{eq:app_hybrid_thermal}
\end{equation}}
\rev{Equations~(\ref{eq:app_hybrid_sigma}) and (\ref{eq:app_hybrid_thermal}) reduce to the perturbative expressions when $\mathcal S\setminus\{0\}$ is empty. In the implementation, the large zero-padded arrays are stored at reduced precision, while the selected block matrices, their inversion, and the accumulated $2\times2$ response matrices are evaluated at double precision.}

\subsection{\texorpdfstring{\rev{Near-singular spectrum and resolution of the blind-zone structure}}{Near-singular spectrum and resolution of the blind-zone structure}}

\rev{The benchmark scan uses $\lambda_{\mathrm{int}}=0.0665$ and $1001$ uniformly spaced values of $\bar\omega$ between $0$ and $2$, with spacing $\Delta\bar\omega=0.002$. The explicit near-singular subspace is used at $58$ frequency points. Between $2$ and $30$ modes are selected at each of these points, remaining below the imposed cap of $64$. The selected modes occur over}
\rev{\begin{equation}
0.678\leq\bar\omega\leq1.128.
\label{eq:app_selected_band}
\end{equation}}
\rev{For the same interaction strength, the long-wavelength result in Eq.~(\ref{eq:freq_interval}) gives}
\rev{\begin{equation}
0.666\lesssim\bar\omega\lesssim1.131.
\label{eq:app_analytic_band_numeric}
\end{equation}}
\rev{For the full numerical kernel, the near-singular frequencies are obtained from the eigenvalues of $\xi(\bar\omega,\mathbf k)$ for all momentum modes through Eq.~(\ref{eq:chi_mod_def}). Their frequency support determines the blind-zone range independently of the frequency grid used to display the susceptibility and noise curves. The close agreement between Eqs.~(\ref{eq:app_selected_band}) and (\ref{eq:app_analytic_band_numeric}) provides a numerical validation of the analytical small-$\mathbf k$ method: the leading long-wavelength form of the dipolar kernel captures the dominant near-singular modes and gives an accurate estimate of the blind-zone range.}

\rev{Resolving the individual resonances of all $N_\ell^3=10^6$ momentum modes would require a much denser or adaptive frequency grid. The present $1001$-point grid is sufficient for a qualitative display of the response inside the blind zone, which is excluded from the projected sensitivity. Frequency resolution, working precision, and the numerical subspace-selection parameters determine the displayed fine structure, including how many individual peaks are resolved, while leaving the near-singular spectrum of the converged interaction kernel unchanged. Finite-lattice geometry and boundary coupling determine the discrete peak positions and weights through the allowed modes and $V_{0\mathbf k}$. The leading small-$\mathbf k$ dipolar kernel is independent of these numerical choices and controls the existence and leading frequency range of the blind zone, whereas its internal line shape remains geometry dependent.}

\section{\texorpdfstring{\rev{Pickup-coil geometry and radius scan}}{Pickup-coil geometry and radius scan}}
\label{app:pickup_coil_scan}

\subsection{\texorpdfstring{\rev{Finite-lattice coupling profile}}{Finite-lattice coupling profile}}

\rev{The benchmark pickup is a single circular loop of radius $R_{\mathrm p}$ in the $yz$ plane. The loop is centered on the geometric center of the cubic lattice and its normal is parallel to $\mathbf x$, so that it reads out the $\mathbf x$-polarized angular response. The $N=N_\ell^3$ lattice sites are placed at}
\rev{\begin{equation}
\mathbf r_{\mathbf n}
=
a_0
\left[
\mathbf n-\frac{N_\ell-1}{2}(1,1,1)
\right],
\qquad
0\leq n_\alpha<N_\ell .
\label{eq:pickup_lattice_coordinates}
\end{equation}}
\rev{For the benchmark, $N_\ell=100$, $a_0=2~\mathrm{mm}$, and $L_{\mathrm{lat}}=N_\ell a_0=200~\mathrm{mm}$. The smallest loop in the scan has $R_{\mathrm p}=L_{\mathrm{lat}}$ and therefore encloses the entire $yz$ cross section of the lattice; the geometric minimum for enclosure is $\sqrt{2}(N_\ell-1)a_0/2\simeq0.70L_{\mathrm{lat}}$.}

\rev{The spatial coupling follows directly from reciprocity. Let $b_{\mathrm p,x}^{(1)}(\mathbf r)$ be the $x$ component of the magnetic field per unit current generated by the pickup loop. For $\rho=(y^2+z^2)^{1/2}$, the Biot--Savart line integral gives}
\rev{\begin{equation}
b_{\mathrm p,x}^{(1)}(x,\rho;R_{\mathrm p})
=
\frac{\mu_0}{4\pi}
\int_0^{2\pi}
\frac{
R_{\mathrm p}^2-R_{\mathrm p}\rho\cos\varphi
}{
\left(
x^2+R_{\mathrm p}^2+\rho^2
-2R_{\mathrm p}\rho\cos\varphi
\right)^{3/2}
}
\,d\varphi .
\label{eq:pickup_ring_profile}
\end{equation}}
\rev{A transverse change $\delta\boldsymbol\mu_i$ of the magnetic moment at site $i$ generates the flux}
\rev{\begin{equation}
\delta\Phi_i
=
\delta\boldsymbol\mu_i\cdot
\mathbf b_{\mathrm p}^{(1)}(\mathbf r_i).
\label{eq:pickup_reciprocity}
\end{equation}}
\rev{Thus, the same unit-current field that determines the pickup profile also determines the reciprocal backaction field. Writing $b_i=b_{\mathrm p,x}^{(1)}(\mathbf r_i;R_{\mathrm p})$, we absorb the site-averaged coupling into $\eta$ and normalize the real-space profile introduced in Eq.~(\ref{eq:gk_def}) according to}
\rev{\begin{equation}
\eta(R_{\mathrm p})
=
\frac{|\mu|}{N}\sum_i b_i,
\qquad
g(\mathbf r_i;R_{\mathrm p})
=
\frac{|\mu|b_i}{\eta(R_{\mathrm p})}.
\label{eq:pickup_profile_normalization}
\end{equation}}
\rev{The measured flux is therefore given directly by Eq.~(\ref{eq:phi_nonuniform_profile}). For a spatially uniform angular displacement, $\mathbf u_0=\delta\mathbf n$ and $g_0=N$, so the coherent flux is $N\eta\,\mathbf x^{\mathsf T}\delta\mathbf n$. The quantity $\eta$ is the per-ferromagnet geometric coupling used in the main text and Table~\ref{tab:params}.}

\rev{For the single-turn loop, the pickup inductance is approximated by}
\rev{\begin{equation}
L_{\mathrm p}(R_{\mathrm p})
=
\mu_0R_{\mathrm p}
\left[
\ln\!\left(\frac{8R_{\mathrm p}}{a_{\mathrm p}}\right)-2
\right].
\label{eq:pickup_inductance_scan}
\end{equation}}
\rev{After the pickup loop is connected in series with the intermediate coil, the per-ferromagnet effective coupling to the SQUID is}
\rev{\begin{equation}
\eta_{\mathrm{eff}}(R_{\mathrm p})
=
\eta(R_{\mathrm p})
\frac{M_{\mathrm i}}{L_{\mathrm i}+L_{\mathrm p}(R_{\mathrm p})}.
\label{eq:pickup_effective_mode_coupling}
\end{equation}}
\rev{The coefficient multiplying mode $\mathbf k$ after the circuit is $\eta_{\mathrm{eff}}g_{-\mathbf k}$, and the coherent coefficient is $N\eta_{\mathrm{eff}}$. Equations~(\ref{eq:pickup_ring_profile})--(\ref{eq:pickup_effective_mode_coupling}) include both the decrease of the direct field with increasing radius and the simultaneous increase of the pickup inductance.}

\subsection{\texorpdfstring{\rev{Numerical radius scan and full-mode admixture}}{Numerical radius scan and full-mode admixture}}

\rev{The line integral in Eq.~(\ref{eq:pickup_ring_profile}) is evaluated on $512$ uniformly spaced azimuthal points. Axial symmetry allows sites with the same pair $(x,\rho^2)$ to be grouped before the integral is performed. If $b_c$ is the coupling of a symmetry class and $n_c$ is its site multiplicity, the coherent sum and the total real-space weight are evaluated as}
\rev{\begin{equation}
\sum_i b_i=\sum_c n_c b_c,
\qquad
\sum_i |b_i|^2=\sum_c n_c |b_c|^2.
\label{eq:pickup_class_sums}
\end{equation}}
\rev{The scan contains $21$ equally spaced values of $R_{\mathrm p}/L_{\mathrm{lat}}$ between $1$ and $10$. The circuit parameters are $a_{\mathrm p}=0.1~\mathrm{mm}$, $L_{\mathrm i}=1.8~\upmu\mathrm H$, $M_{\mathrm i}=1.02\times10^{-8}~\mathrm H$, and $L_{\mathrm s}=80~\mathrm{pH}$, together with the magnetic moment and lattice parameters in Table~\ref{tab:params}. Repeating the calculation at $R_{\mathrm p}/L_{\mathrm{lat}}=1$ with $256$, $512$, and $1024$ azimuthal points leaves $\eta$, $\eta_{\mathrm{eff}}$, and the mode-admixture measure below unchanged at the reported precision. The $512$-point quadrature is therefore sufficient for the radius scan.}

\rev{The complete noncoherent admixture is defined by}
\rev{\begin{equation}
\epsilon_{\mathrm{nc}}
=
\left(
\frac{\sum_{\mathbf k\neq0}|g_{\mathbf k}|^2}
{|g_{\mathbf 0}|^2}
\right)^{1/2}.
\label{eq:noncoherent_admixture}
\end{equation}}
\rev{With the normalization in Eq.~(\ref{eq:pickup_profile_normalization}), Parseval's identity gives the complete admixture directly from the real-space weights,}
\rev{\begin{equation}
\epsilon_{\mathrm{nc}}^2
=
\frac{N\sum_i|b_i|^2}
{\left|\sum_i b_i\right|^2}
-1.
\label{eq:pickup_parseval_admixture}
\end{equation}}
\rev{Equations~(\ref{eq:pickup_class_sums}) and (\ref{eq:pickup_parseval_admixture}) therefore include the complete set of $N-1$ nonzero momentum modes without a mode truncation.}

\rev{At $R_{\mathrm p}/L_{\mathrm{lat}}=1$, the calculation gives $L_{\mathrm p}=1.93~\upmu\mathrm H$ and $M_{\mathrm i}/(L_{\mathrm i}+L_{\mathrm p})=2.73\times10^{-3}$. Restoring the benchmark magnetic moment gives}
\rev{\begin{equation}
\frac{\eta}{\sqrt{L_{\mathrm s}}}
=
7.86\times10^{-9}\sqrt{\mathrm J},
\qquad
\frac{\eta_{\mathrm{eff}}}{\sqrt{L_{\mathrm s}}}
=
2.15\times10^{-11}\sqrt{\mathrm J},
\label{eq:pickup_benchmark_couplings}
\end{equation}}
\rev{which reproduce, after rounding, the two coupling coefficients in Table~\ref{tab:params}. The full-mode admixture is $\epsilon_{\mathrm{nc}}=0.167$ at this radius. At the scan point $R_{\mathrm p}/L_{\mathrm{lat}}=5.05$, $\epsilon_{\mathrm{nc}}=5.40\times10^{-3}$ and $\eta_{\mathrm{eff}}/\sqrt{L_{\mathrm s}}=1.20\times10^{-12}\sqrt{\mathrm J}$. At $R_{\mathrm p}/L_{\mathrm{lat}}=10$, the corresponding values are $1.37\times10^{-3}$ and $3.06\times10^{-13}\sqrt{\mathrm J}$. A larger loop therefore approaches a purer coherent-mode projection, but at the price of a substantially weaker coherent coupling.}

\subsection{\texorpdfstring{\rev{Coherent-channel noise as a function of pickup radius}}{Coherent-channel noise as a function of pickup radius}}

\rev{To quantify the consequence for the benchmark noise, we evaluate the $1,1$ component of the magnetic-field noise matrix at $\bar\omega=0$. The thermal contribution uses the finite-lattice correction derived in Sec.~V.B,}
\rev{\begin{equation}
S_{B,11}^{\mathrm{th}}(0)
=
\frac{M_{11}(0)}{N}S_{B,1}^{\mathrm{th}},
\label{eq:pickup_thermal_radius_scan}
\end{equation}}
\rev{and is independent of the pickup radius. At each radius, the coherent imprecision and backaction contributions are recalculated with the effective coupling in Eq.~(\ref{eq:pickup_effective_mode_coupling}), so that}
\rev{\begin{equation}
S_B^{\mathrm{imp}}\propto\eta_{\mathrm{eff}}^{-2},
\qquad
S_B^{\mathrm{back}}\propto\eta_{\mathrm{eff}}^{2},
\qquad
S_B^{\mathrm{tot}}
=
S_B^{\mathrm{th}}+S_B^{\mathrm{imp}}+S_B^{\mathrm{back}}.
\label{eq:pickup_noise_radius_scaling}
\end{equation}}

\rev{For $R_{\mathrm p}/L_{\mathrm{lat}}=1$, the calculation gives $S_B^{\mathrm{th}}=1.70\times10^{-34}~\mathrm{T^2/Hz}$, $S_B^{\mathrm{imp}}=4.18\times10^{-33}~\mathrm{T^2/Hz}$, $S_B^{\mathrm{back}}=9.20\times10^{-38}~\mathrm{T^2/Hz}$, and $S_B^{\mathrm{tot}}=4.35\times10^{-33}~\mathrm{T^2/Hz}$. The imprecision contribution accounts for $96.1\%$ of the total. Increasing the radius to $R_{\mathrm p}/L_{\mathrm{lat}}=5.05$ and $10$ raises the total power spectral density to $1.34\times10^{-30}~\mathrm{T^2/Hz}$ and $2.05\times10^{-29}~\mathrm{T^2/Hz}$, respectively. Thus, throughout the scanned range, the reduction in noncoherent admixture obtained with a larger loop is outweighed by the loss of effective coupling in this imprecision-dominated reference circuit.}

\begin{figure*}[t]
    \centering
    \includegraphics[width=0.48\textwidth]{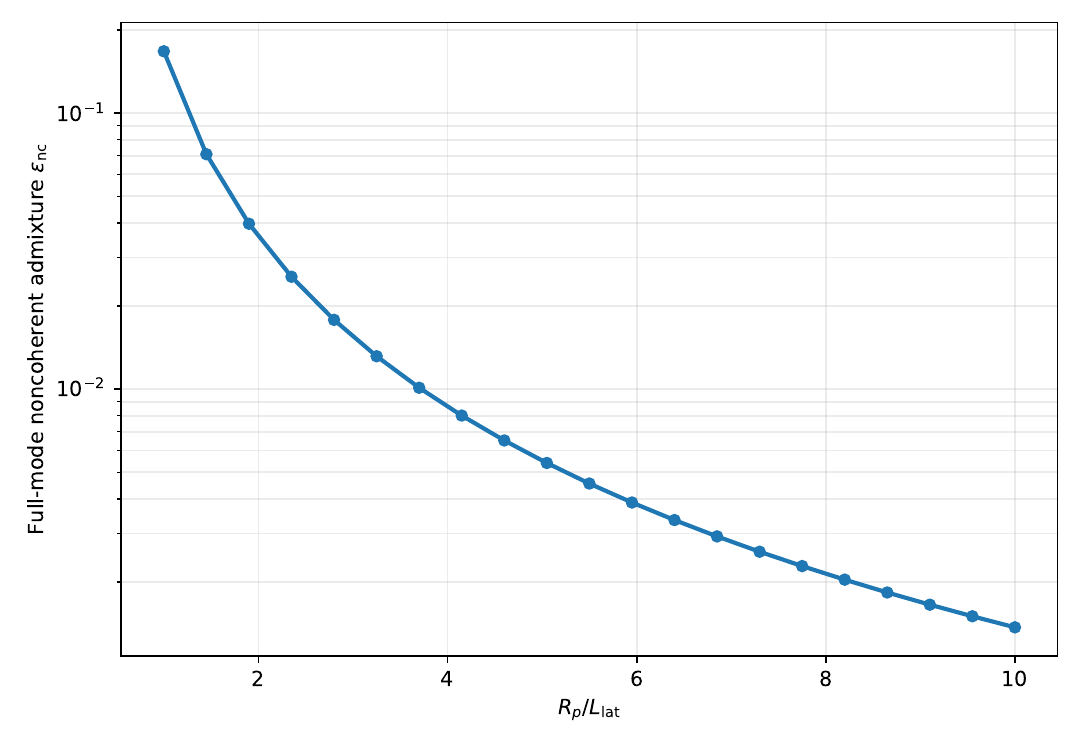}
    \hfill
    \includegraphics[width=0.48\textwidth]{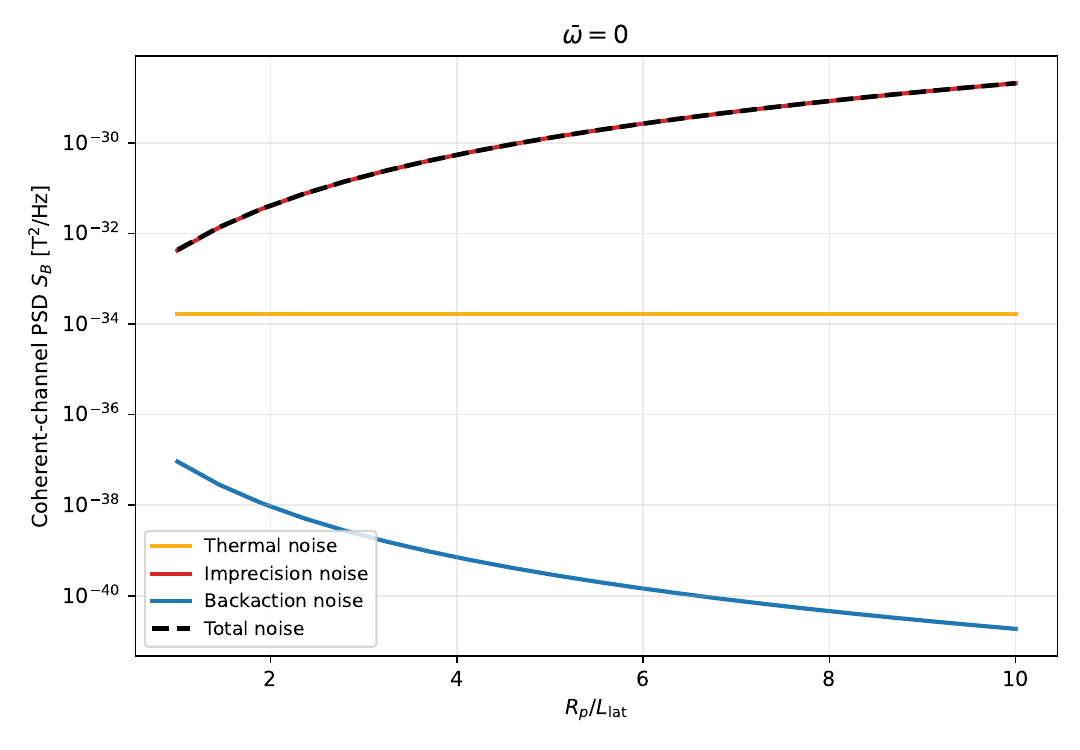}
    \caption{\rev{Pickup-radius scan for the centered single-turn circular loop. Left: full-mode noncoherent admixture obtained from Eq.~(\ref{eq:pickup_parseval_admixture}). Right: coherent-channel magnetic-field noise power spectral density at $\bar\omega=0$. The thermal, imprecision, and backaction contributions are shown separately, together with their sum. The total nearly overlaps the imprecision curve over the displayed range.}}
    \label{fig:pickup_radius_scan}
\end{figure*}

\rev{The noise calculation in this subsection is a coherent-channel scan. The factor $M_{11}$ includes thermal fluctuations of nonzero-momentum modes that enter the coherent channel through boundary-induced dynamical mixing, while $\epsilon_{\mathrm{nc}}$ quantifies the complete geometrical mode admixture of the pickup profile. Figure~\ref{fig:pickup_radius_scan} evaluates the imprecision and backaction terms using their coherent-channel projections. This approximation is adequate for the stated comparison away from the blind zone when the noncoherent response remains perturbative. A detailed calculation near the blind zone would additionally include the direct readout and backaction projections associated with $g_{\mathbf k\neq0}$.}

\clearpage

\section{\texorpdfstring{\rev{Representative cylindrical-shield calculation of the axion-photon enhancement}}{Representative cylindrical-shield calculation of the axion-photon enhancement}}
\label{app:axion_photon_cylinder}

\subsection{\texorpdfstring{\rev{Geometry and coherent observable}}{Geometry and coherent observable}}

\rev{We numerically test the order-$N$ scaling in Sec.~\ref{sec:axion_photon} for one specified electromagnetic geometry. The shield is a closed cylinder of radius $0.500~\mathrm m$ and height $1.000~\mathrm m$, with its axis along $\mathbf z$. We impose ideal-conductor boundary conditions on the axion-induced fields and take the no-trapped-flux branch for the static background,}
\rev{\begin{equation}
\hat{\mathbf n}\times\mathbf E_a=0,
\qquad
\hat{\mathbf n}\cdot\mathbf B_a=0,
\qquad
\hat{\mathbf n}\cdot\mathbf B_0=0.
\label{eq:app_cylinder_boundary}
\end{equation}}
\rev{The benchmark simple-cubic lattice has $N_\ell=100$ sites along each direction, lattice constant $a_0=2~\mathrm{mm}$, and magnetic moments aligned with $+\mathbf z$. Its occupied span is $(N_\ell-1)a_0=0.198~\mathrm m$, and its centroid is displaced from the cylinder axis by $d=0.150~\mathrm m$ along $\mathbf y$. Thus, the full $N=N_\ell^3=10^6$ array remains well inside the shield.}

\rev{The coherent drive of the uniformly responding lattice is characterized by the signed site average}
\rev{\begin{equation}
\overline B_{a,x}^{(N)}
=
\frac{1}{N}\sum_{i=1}^{N}B_{a,x}(\mathbf r_i).
\label{eq:app_axion_coherent_average}
\end{equation}}
\rev{As a single-ferromagnet reference, we use an identical uniformly magnetized sphere at the same lattice centroid, in the same cylinder, with the same orientation and $B_x$ readout. The complex enhancement and the normalized coherent factor are then defined as}
\rev{\begin{equation}
\mathcal E_N
=
\frac{\overline B_{a,x}^{(N)}}{B_{a,x}^{(1)}},
\qquad
C_N=\frac{\mathcal E_N}{N}.
\label{eq:app_axion_enhancement_definition}
\end{equation}}
\rev{This off-axis configuration is required for a nonzero value of the chosen coherent observable. If the same lattice is centered in the cylinder, inversion symmetry makes both the lattice average and the centered single-ferromagnet $B_x$ response vanish exactly, leaving an undefined $0/0$ enhancement ratio. This centered result provides a symmetry control and also demonstrates explicitly that a nonzero absolute response is geometry dependent.}

\subsection{\texorpdfstring{\rev{Electromagnetic calculation and result}}{Electromagnetic calculation and result}}

\rev{The static field and the axion-induced response are evaluated for finite uniformly magnetized spheres rather than point sources. The dynamic field is expanded in the analytic transverse-electric and transverse-magnetic modes of the cylindrical ideal-conductor cavity, while the static boundary-reaction field is evaluated with a separate cylindrical-harmonic expansion. The source overlaps and the coherent readout are summed over the exact Cartesian coordinates of all lattice sites, so no continuum approximation to the $100^3$ array is made. The free-space part of the coherent $B_x$ transfer is odd under pair exchange and cancels exactly for both the lattice and the single reference. We therefore remove this part analytically and evaluate only the smooth boundary-reaction remainder with a Riesz-smoothed cylindrical-mode sum. This separation provides a convergent representation without changing the physical observable.}

\rev{The final calculation retains $105460$ cavity modes up to $12~\mathrm{GHz}$ and uses a second-order Riesz weight; a fourth-order weight is used as an independent convergence check. The axion-frequency points cover $2.418\times10^{-3}$--$241.8~\mathrm{Hz}$, corresponding to $m_a=10^{-17}$--$10^{-12}~\mathrm{eV}$. The lowest empty-cylinder mode is at $2.295\times10^8~\mathrm{Hz}$, so the full frequency-dependent denominator changes by at most $1.11\times10^{-12}$ over the mass interval of the sensitivity projection. The coherent enhancement is consequently frequency independent at the precision relevant here.}

\begin{figure}[h!]
    \centering
    \includegraphics[width=0.92\linewidth]{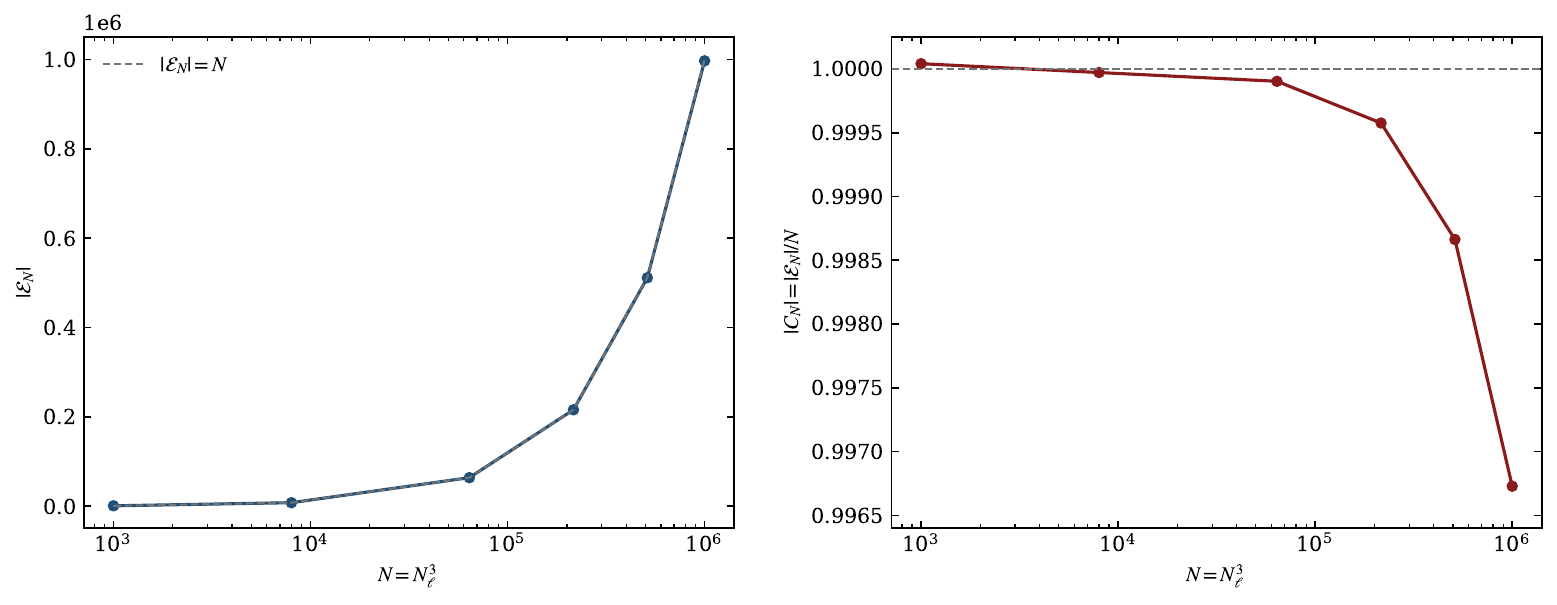}
    \caption{\rev{Axion-photon coherent enhancement for the off-axis cylindrical-shield geometry. Left: magnitude of the field enhancement $|\mathcal E_N|$ relative to an identical single ferromagnet at the same position. Right: normalized coherent factor $|C_N|=|\mathcal E_N|/N$; the dashed line denotes exact order-$N$ scaling. The full $N=10^6$ result is $|C_N|=0.996729$.}}
    \label{fig:axion_photon_enhancement_N}
\end{figure}

\rev{For the full benchmark lattice, the calculation gives}
\rev{\begin{equation}
\mathcal E_{10^6}=9.967290\times10^5,
\qquad
C_{10^6}=0.996729,
\qquad
\arg\mathcal E_{10^6}=0.
\label{eq:app_axion_enhancement_result}
\end{equation}}
\rev{The finite-size correction is therefore a $0.3271\%$ reduction from exact $N$ scaling. Figure~\ref{fig:axion_photon_enhancement_N} shows the scan over $N_\ell=10$, $20$, $40$, $60$, $80$, and $100$. The normalized factor remains within $0.33\%$ of unity throughout the scan, directly supporting the coherent order-$N$ estimate for this geometry.}

\rev{Refining the dynamic cylindrical-mode truncation changes $C_N$ by $0.230\%$, while independently refining the static boundary-reaction expansion changes it by $0.0205\%$. Changing the Riesz order from two to four changes $C_N$ by $1.96\times10^{-3}\%$. A one-site calculation also reproduces the single reference mode by mode and gives $\mathcal E_1=1$. These checks establish convergence at better than the $1\%$ level. Because the common modal normalization cancels in Eq.~(\ref{eq:app_axion_enhancement_definition}), this calculation tests the relative coherent scaling and its finite-size correction; the absolute $\mathcal O(0.1)$ coefficient in Eq.~(\ref{eq:Bag_lattice}) remains set by the geometry-dependent normalization.}

\clearpage

\begin{acknowledgments}
    \textit{Acknowledgments}--This work is in part supported by the National Natural Science Foundation of China (Grant No. 62505007).
\end{acknowledgments}


%

\end{document}